\documentclass[aps,prc,superscriptaddress,10pt,showpacs,notitlepage,  
nofootinbib
]{revtex4-1}
\usepackage{bm}
\usepackage{amsmath,amssymb,amsthm}
\usepackage{latexsym,graphicx,color,subfigure}
\usepackage{enumerate}
\usepackage{amssymb}
\usepackage{hyperref}
\usepackage{mathtools}

\usepackage{graphicx}
\graphicspath{ {images/} }

\usepackage{color}

\newcommand{\grad}{\mathrm{grad}}

\newcommand{\bulk}{\mathrm{bulk}}

\newcommand{\be}{\begin{equation}}
\newcommand{\ee}{\end{equation}} 
\newcommand{\beq}{\begin{eqnarray}}
\newcommand{\eeq}{\end{eqnarray}}

\newcommand{\bea}{\begin{eqnarray}}
\newcommand{\eea}{\end{eqnarray}}

\def\tc{\textcolor{red}}
\def\tc2{\textcolor{blue}}
 
\renewcommand{\vec}[1]{\boldsymbol{#1}}

\allowdisplaybreaks[3]
\def\simge{\mathrel{
   \rlap{\raise 0.511ex \hbox{$>$}}{\lower 0.511ex \hbox{$\sim$}}}}
\def\simle{\mathrel{
   \rlap{\raise 0.511ex \hbox{$<$}}{\lower 0.511ex \hbox{$\sim$}}}}
\def\bigs{\mathrel{
   \rlap{\raise 0.531ex \hbox{$>$}}{\lower 0.531ex \hbox{$<$}}}}

\usepackage[normalem]{ulem}

\renewcommand\sout{\bgroup \color{blue} \ULdepth=-.5ex \ULset}

\begin{document}
\title{Domain walls in neutron $^{3}P_{2}$ superfluids in neutron stars}
\author{Shigehiro Yasui}
\email{yasuis@keio.jp}
\affiliation{Department of Physics $\&$ Research and Education Center for Natural Sciences,\\ Keio University,Hiyoshi 4-1-1, Yokohama, Kanagawa 223-8521, Japan}
\author{Muneto Nitta}
\email{nitta(at)phys-h.keio.ac.jp}
\affiliation{Department of Physics $\&$ Research and Education Center for Natural Sciences,\\ Keio University,Hiyoshi 4-1-1, Yokohama, Kanagawa 223-8521, Japan}
\date{\today}
\begin{abstract}
We work out domain walls in neutron $^{3}P_{2}$ superfluids realized in the core of neutron stars.
Adopting the Ginzburg-Landau (GL) theory as a bosonic low-energy effective theory,
we consider configurations of domain walls interpolating ground states, {\it i.e.}, the uniaxial nematic (UN), D$_{2}$-biaxial nematic (D$_{2}$-BN), and D$_{4}$-biaxial nematic (D$_{4}$-BN) phases in the presence of zero, small and large magnetic fields, respectively.
We solve the Euler-Lagrange equation from the GL free energy density, and calculate  surface energy densities of the domain walls.
We find that one extra Nambu-Goldstone mode is localized in the vicinity of 
a domain wall in the UN phase while a U(1) symmetry restores 
 in the vicinity of  one type of domain wall in the D$_{2}$-BN phase and 
 all domain walls in the D$_{4}$-BN phase. 
Considering a pile of domain walls in the neutron stars,
we find that the most stable configurations are
domain walls perpendicular to the magnetic fields piled up in the direction along the magnetic fields
 in the D$_{2}$-BN and D$_{4}$-BN phases. 
We estimate the energy released from the deconstruction of the domain walls in the edge of a neutron star, and show that it can reach an astrophysical scale such as glitches in neutron stars. 
\end{abstract}
%\pacs{ 21.65.Qr, 11.27+d, 74.25.Uv, 03.65.Ta}
\maketitle
%\newpage

%\tableofcontents

\section{Introduction}

Domain walls or kinks are solitonic objects separating 
two discrete vacua or ground states of a system~\cite{manton_sutcliffe_2004,vachaspati_2006,rajaraman1982solitons}
and play important roles in various subjects of physics from 
condensed matter physics~\cite{volovik} to cosmology~\cite{Vilenkin:2000jqa} and supersymmetric field theories~\cite{shifman_yung_2009}.
They are often created in phase transitions associated with 
symmetry breakings~\cite{Kibble:1976sj}.  
In cosmology, if they appear at a phase transition in the early Universe, 
then the so-called domain wall problem occurs~\cite{Vilenkin:2000jqa}:
the domain wall energy dominates Universe to make it collapse. 
In helium superfluids, such domain walls are created in a similar manner, 
thereby simulating cosmological phase transitions~\cite{Zurek:1985qw}. 
Here, we focus on domain walls in neutron stars, 
more precisely those in nuclear matter.

Neutron stars are compact stars under extreme conditions, thereby serving as astrophysical laboratories for studying nuclear matter at high density, under rapid rotation and with a strong magnetic field 
(see Refs.~\cite{Graber:2016imq,Baym:2017whm} for recent reviews). 
The recent progresses in astrophysical observations promote us to study the neutron stars more precisely, 
such as
the recent reports on massive neutron stars whose masses are almost twice as large as the solar mass~\cite{Demorest:2010bx,Antoniadis1233232} and
the gravitational waves from a binary neutron star merger~\cite{TheLIGOScientific:2017qsa}.

Inside neutron stars, one of the most important key ingredients for understanding the inner structure is neutron superfluidity and proton superconductivity (see Refs.~\cite{Chamel2017,Haskell:2017lkl,Sedrakian:2018ydt} for recent reviews).
Since the superfluid and superconducting components can alter excitation modes at low energy from the normal phase, their existence can affect several properties in neutron stars, such as neutrino emissivities and specific heats relevant to the long relaxation time after in the sudden speed-up events (glitches) of neutron stars~\cite{Baym1969,Pines1972,Takatsuka:1988kx}, and the enhancement of neutrino emission around 
the critical point of the superfluid transition~\cite{Yakovlev:2000jp,Potekhin:2015qsa,Yakovlev:1999sk,Heinke:2010cr,Shternin2011,Page:2010aw}.
Glitches in pulsars may also be explained by quantized vortices in superfluids~\cite{reichley,Anderson:1975zze}.
The neutron superfluids are realized by the attraction between two neutrons at the low density in the $^1S_0$ channel. This channel becomes, however, repulsive in the high density regime.\footnote{Although the $^1S_0$ superfluidity at low density was proposed in Ref.~\cite{Migdal:1960}, it was shown in Ref.~\cite{1966ApJ...145..834W} that this channel turns to be repulsive due to the strong short-range repulsion at higher densities.}
Instead,
at higher density, neutron $^3P_2$ superfluids, in which neutron pairs possess the total angular momentum $J=2$ with spin-triplet and $P$-wave, become more relevant~\cite{Tabakin:1968zz,Hoffberg:1970vqj,Tamagaki1970,Hoffberg:1970vqj,Takatsuka1971,Takatsuka1972,Fujita1972,Richardson:1972xn,Amundsen:1984qc,Takatsuka:1992ga,Baldo:1992kzz,Elgaroy:1996hp,Khodel:1998hn,Baldo:1998ca,Khodel:2000qw,Zverev:2003ak,Maurizio:2014qsa,Bogner:2009bt,Srinivas:2016kir}\footnote{It is noted that the interaction in the $^{3}P_{0}$ and $^{3}P_{1}$ channels are repulsive at high density, and hence they are irrelevant to the formation of superfluidity~\cite{Dean:2002zx}.}.
The $^3P_2$ interaction originates from a strong spin-orbit ($LS$) force at large scattering energy, and thus the neutron $^3P_2$ superfluids are expected to be realized in the high density regions in the inner cores of neutron stars. 
The neutron $^3P_2$ superfluids can survive in the neutron stars with strong magnetic fields, such as in the magnetars with the magnetic field $10^{15}$-$10^{18}$ G,
because the spin $\uparrow\uparrow$ or $\downarrow\downarrow$ pairs in the spin-triplet pairing cannot be broken by the Zeeman effects and hence the neutron $^3P_2$ superfluids are tolerant against the strong magnetic field.\footnote{The origin of the strong magnetic fields in neutrons stars or in magnetars is still an open problem although there are many theoretical works: spin-dependent interactions~\cite{Brownell1969,RICE1969637,Silverstein:1969zz,Haensel:1996ss}, pion domain walls~\cite{Eto:2012qd,Hashimoto:2014sha}, spin polarizations in quark-matter in the neutron star core~\cite{Tatsumi:1999ab,Nakano:2003rd,Ohnishi:2006hs} and so on. It may be worthwhile to mention that a negative result for the generation of strong magnetic fields was recently announced in a study in terms of the nuclear many-body calculations~\cite{Bordbar:2008zz}.
}
The possible existence of neutron $^{3}P_{2}$ superfluids inside the neutron stars are pursued in astrophysical observations.
It has been recently pointed out that the rapid cooling of the neutron star in Cassiopeia A may be explained by the enhancement of neutrino emissivities which is caused by the formation and dissociation of neutron $^3P_2$ Cooper pairs~\cite{Heinke:2010cr,Shternin2011,Page:2010aw}.
In the theoretical studies, it is known that neutron $^3P_2$ superfluids have rich structures in the condensates due to a variety of combinations of spin-triplet and $P$-wave angular momentum in the Cooper pairs. 
The superfluid states with $J=2$ are classified into nematic, cyclic, and ferromagnetic phases~\cite{Mermin:1974zz}, 
among which the nematic phase is the ground state in the weak coupling limit of $^3P_2$ superfluids~\cite{Fujita1972,Richardson:1972xn,Sauls:1978lna,Muzikar:1980as,Sauls:1982ie,Vulovic:1984kc,Masuda:2015jka,Masuda:2016vak}.
The nematic phase is continuously degenerated 
and consists of the three subphases:
the uniaxial nematic (UN), and dihedral-two and dihedral-four biaxial nematic (D$_{2}$-BN and D$_{4}$-BN) phases according to the continuous/discrete symmetries of the $^{3}P_{2}$ order parameter: the U(1) symmetry in the UN phase, and the D$_{2}$ and D$_{4}$ symmetries in the D$_{2}$-BN and D$_{4}$-BN phases, respectively.\footnote{See e.g. Appendix~B in Refs.~\cite{Yasui:2019unp,Yasui:2019pgb} for more information on the definitions of the UN, D$_{2}$-BN, and D$_{4}$-BN phases.}

In terms of fermion dynamics, all these phases are accompanied by the Bogoliubov quasiparticles.
The phase diagram on the plane spanned by the temperature and the magnetic field was drawn by solving the 
Bogoliubov de-Gennes (BdG) equation self-consistently with the Fermi liquid corrections, and it was shown that the UN phase exists at zero magnetic field, and the D$_{2}$-BN and D$_{4}$-BN phases appear in the weak and strong magnetic fields, respectively~\cite{Mizushima:2016fbn}.
There are the first and second-order transitions at the boundary between the D$_{2}$-BN  and D$_{4}$-BN phases, and those two transitions meet at the (tri)ctirical endpoint (CEP).
The existence of the CEP is important because the fluctuations can significantly affect the thermodynamical properties and the transport coefficients in neutron stars.
The Bogoliubov quasiparticles in neutron $^{3}P_{2}$ superfluids are protected topologically against the perturbation.
In terms of the general classifications,
the nematic phase in the neutron $^{3}P_{2}$ superfluids is a class-DIII topological superconductor in the periodic table,  inducing Majorana fermions on the edge of the superfluids~\cite{Mizushima:2016fbn}.
The cyclic and ferromagnetic phases 
are nonunitary states, in which the time-reversal symmetry is broken, and they serve to host Weyl fermions in the bulk~\cite{Mizushima:2016fbn,Mizushima:2017pma}.  

The $^3P_2$ superfluids allow also bosonic excitations as collective modes~\cite{Bedaque:2003wj,Leinson:2011wf,Leinson:2012pn,Leinson:2013si,Bedaque:2012bs,Bedaque:2013rya,Bedaque:2013fja,Bedaque:2014zta,Leinson:2009nu,Leinson:2010yf,Leinson:2010pk,Leinson:2010ru,Leinson:2011jr},
which are considered to be relevant to cooling process by neutrino emissions from neutron stars.\footnote{It is discussed that the cooling process is related not only to low-energy excitation modes but also to quantum vortices~\cite{Shahabasyan:2011zz}.}
Bosonic excitations can be best discussed in terms of
the Ginzburg-Landau (GL) theory as a bosonic effective theory around the CEP from the normal phase to the superfluid phase~\cite{Fujita1972,Richardson:1972xn,Sauls:1978lna,Muzikar:1980as,Sauls:1982ie,Vulovic:1984kc,Masuda:2015jka,Chatterjee:2016gpm,Masuda:2016vak,Yasui:2018tcr,Yasui:2019tgc,Yasui:2019unp,Yasui:2019pgb}. 
The GL equation can be obtained by a systematic expansion of the functional with respect to the order parameter field and magnetic field, where the fermionic degrees of freedom are integrated out.
We notice that, in the GL expansion up to the fourth order in terms of the order parameter, the ground state cannot be determined uniquely, because there exists a continuous degeneracy among the UN, D$_{2}$-BN and D$_{4}$-BN phases.\footnote{At the fourth order, there happens to exist an SO(5) symmetry in the potential term. This is an extended symmetry, which is absent in the original Hamiltonian. It is known that, in this case, the spontaneous breaking eventually generate a quasi-Nambu-Goldstone (NG) modes. However, such NG modes should be regarded as being irrelevant to the excitations in the true ground state~\cite{Uchino:2010pf}.
This is nothing but the origin of the continuous degeneracy.
}
Instead, the ground state is determined uniquely in the GL expansion up to the sixth order~\cite{Masuda:2015jka}.
However this is still not sufficient for the expansion order, because it is stable only locally and there exists the instability for a large value of the order parameter in the variational calculation.
It was found recently that the eighth-order terms of the condensates ensure the stability of the ground state~\cite{Yasui:2019unp}.
As a by-product, it was also shown that the eighth-order terms in the GL equation induce the CEP in the phase diagram, although the position of the CEP in the GL theory is different from that in the BdG equation~\cite{Mizushima:2016fbn}.
With the GL expansion in Ref.~\cite{Yasui:2019unp}, we can adopt the GL equation to investigate the position dependence of the order parameter in the nonuniform system,  
and topological defects can be discussed.
So far, the theoretical studies have been conducted in depth: spontaneously magnetized vortices~\cite{Muzikar:1980as,Sauls:1982ie,Fujita1972,Masuda:2015jka},
solitonic excitations on a vortex~\cite{Chatterjee:2016gpm},  
and half-quantized non-Abelian vortices~\cite{Masuda:2016vak}, and
topological defects on the boundary of $^3P_2$ superfluids~\cite{Yasui:2019pgb}. 
Interestingly, the topological states in the $^{3}P_{2}$ share common properties in the condensed matter systems, such as $D$-wave superconductors~\cite{Mermin:1974zz}, $P$-wave superfluidity in $^{3}$He liquid~\cite{vollhardt2013superfluid,volovik}, chiral $P$-wave superconductivity e.g. in Sr$_2$RuO$_4$~\cite{RevModPhys.75.657}, spin-2 Bose-Einstein condensates~\cite{2010arXiv1001.2072K}, and so on.
For instance, topological defects on the boundary of the spin-2 Bose-Einstein condensate have been discussed~\cite{2019arXiv190702216C}, 
as similar to those in $^3P_2$ superfluids in Ref.~\cite{Yasui:2019pgb}.\footnote{See Ref.~\cite{Urbanski_2017} as a recent review for the topological defects in the boundary in liquid crystals.}

In the present study, we consider domain walls, {\it i.e.}, kinks or one-dimensional solitons, which connect two different vacua in the bulk phase in the neutron $^{3}P_{2}$ superfluids.
They are also called textures in analogy to those in crystal liquids, because  the order parameter changes not only in the amplitude but also in the {\it directions}~\cite{Mermin:1979zz,Urbanski_2017}.
In the case of $^{3}P_{2}$, orientations of domain walls are supplied by the spin and the angular momentum.
They can be created via a phase transition and stay as quasistable states when the lifetime is longer than the typical time scale in the systems.\footnote{If the configurations of the solitons are protected topologically, then they keep to be the stable states against perturbation.}
In the literature, the studies of domain walls have been extensively accomplished in the $^{3}$He superfluidity~\cite{PhysRevB.14.118,PhysRevB.16.182,PhysRevB.17.1088,PhysRevLett.38.557,PhysRevB.18.3197} (see also Refs.~\cite{vollhardt2013superfluid,volovik} and the references therein)\footnote{The domain walls in in the $^{3}$He superfluidity were called  also the composite soliton instead of textures in Refs.~\cite{PhysRevB.14.118,PhysRevB.16.182,PhysRevB.17.1088,PhysRevLett.38.557}.}. 
In analogy to the $^{3}$He superfluids, we can expect that domain walls exist also in the neutron $^{3}P_{2}$ superfluids, because there are degenerate vacua which are separated by a barrier potential.
Because the domain walls are the excited states which do not appear in the ground state, the domain walls can be produced through defect formations at a phase transition~\cite{Kibble:1976sj,Zurek:1985qw}.
By adopting the GL theory,
we consider domain walls in the bulk UN, D$_{2}$-BN, and D$_{4}$-BN phases in the zero, weak, and strong magnetic fields, respectively.
We obtain the spatial configurations of the domain walls by solving the Euler-Lagrange (EL) equation from the GL effective potential.
With those solutions, we estimate the energy of domain walls, {\it i.e.}, the surface energy density, for supposing several different configurations and directions.
We show that the domain wall configurations pass thorough different phases, and unbroken symmetries inside the domain walls are different from those in the bulk.
As a result, for instance, one Nambu-Goldstone mode is trapped inside the domain wall in the bulk UN phase.
We also find that the domain walls piled along the magnetic fields are more stable than those in the other directions.
Such domain walls are likely to exist as quasistable states in the neutron stars.
As a simple situation, we consider that a pile of domain walls may be deconstructed by moving to the north or south pole of the neutron star, and in the end they can release huge energy which can be detectable in the astrophysical observations.

The paper is organized as the followings.
In Sec.~\ref{sec:formalism}, we introduce the GL equation as the low-energy effective theory of the neutron $^{3}P_{2}$ superfluids and show the EL equations for describing the domain walls.
In Sec.~\ref{sec:numerical_results}, we show the numerical results of the configurations of the domain walls in the one-dimensional directions for solving the EL equations by adopting appropriate boundary conditions.
We also estimate the surface energy density for each domain wall and show that the domain walls piled along the magnetic field exist at the most stable states.
In Sec.~\ref{sec:discussion}, we will consider the situation that a pile of domain walls exist in the neutron stars, and they can release a huge energy as the astrophysical phenomena.
The final section is devoted to our conclusion.
In Appendix~\ref{sec:EL_appendix}, we present the explicit forms of the EL equations for the domain walls.
In Appendix~\ref{sec:plots_configuration_DW}, we show the numerical results of the configurations of the domain walls.

\section{Formalism}
\label{sec:formalism}

\subsection{Ginzburg-Landau equation}

The condensate of the neutron $^{3}P_{2}$ superfluidity 
can be expressed by a symmetric and traceless three-by-three dimensional tensor $A$ as an order parameter 
which triggers the symmetry breaking.
The components of $A$ are denoted by $A^{ab}$ with the indices $a=1$, $2$, $3$ for the spin and $b=1$, $2$, $3$ for three-dimensional momentum degrees of freedom.
The GL theory is introduced by integrating out the neutron degrees of freedom as a loop expansion, supposing the small coupling strength in the $^{3}P_{2}$ interaction for two neutrons~\cite{Fujita1972,Richardson:1972xn,Sauls:1978lna,Muzikar:1980as,Sauls:1982ie,Vulovic:1984kc,Masuda:2015jka,Masuda:2016vak,Yasui:2018tcr,Yasui:2019unp}.
The GL equation is valid in the region in which the temperature $T$ is close to the critical temperature $ T_{c0}$, $|1-T/T_{c0}| \ll 1$, where
the value of $T_{c0}$ is determined at zero magnetic field.
The concrete form of the GL free energy reads 
\begin{eqnarray}
  f[{A}] = f_{0} + f_{\grad}[A] + f_{8}^{(0)}[{A}] + f_{2}^{(\le4)}[{A}] + f_{4}^{(\le2)}[{A}] + {\cal O}(B^{m}{A}^{n})_{m+n\ge7},
\label{eq:eff_pot_coefficient02_f}
\end{eqnarray}
as an expansion in terms of the condensate $A$ and the magnetic field $\vec{B}$ with the magnitude $B=|\vec{B}|$.
Each term is explained as follows.
The first term $f_{0}$ is the sum of the free part and the spin-magnetic coupling term, given by
\begin{eqnarray}
f_{0}
=
%%1
  - T
    \int \frac{\mathrm{d}^{3}\vec{p}}{(2\pi)^{3}}
    \ln \Bigl(
             \bigl( 1+e^{-\xi_{\vec{p}}^{-}/T} \bigr)
             \bigl( 1+e^{-\xi_{\vec{p}}^{+}/T} \bigr)
         \Bigr),
\label{eq:eff_pot_free_magneticfield}
\end{eqnarray}
with $\displaystyle \xi_{\vec{p}}^{\pm} = \xi_{\vec{p}} \pm |\vec{\mu}_{n}||\vec{B}|$ and $\displaystyle \xi_{\vec{p}}=\vec{p}^{2}/(2m)-\mu$ for the neutron three-dimensional momentum $\vec{p}$, the neutron mass $m$ and the neutron chemical potential $\mu$.
The bare magnetic moment of a neutron is $\vec{\mu}_{n}=-(\gamma_{n}/2)\vec{\sigma}$ 
with the gyromagnetic ratio $\gamma_{n}=1.2 \times 10^{-13}$ MeV/T (in natural units, $\hbar = c=1$) and the Pauli matrices for the neutron spin $\vec{\sigma}$.
The following terms include the condensate $A$:
$f_{\grad}[{A}]$ is the gradient term,
$f_{8}^{(0)}[{A}]$ consists of the terms including the field $A$ up to the eighth order with no magnetic field,
$f_{2}^{(\le4)}[{A}]$ consists of the terms including the field $A$ up to the second order with the magnetic field up to $|\vec{B}|^{4}$, 
and $f_{4}^{(\le2)}[{A}]$ consists of the terms including the field $A$ up to the fourth order with the magnetic field up to $|\vec{B}|^{2}$.
Their explicit expression can be given as follows:
\begin{eqnarray}
 f_{\grad}[{A}]
=
%%4
 K^{(0)}
 \sum_{a,i,j=1}^{3}
  \Bigl(
        \nabla_{i} {A}^{ja\ast}
        \nabla_{i} {A}^{aj}
     + \nabla_{i} {A}^{ia\ast}
        \nabla_{j} {A}^{aj}
     + \nabla_{i} {A}^{ja\ast}
        \nabla_{j} {A}^{ai}
  \Bigr),
\label{eq:eff_pot_kin}
\end{eqnarray}
for the gradient term and 
\begin{widetext}
\begin{eqnarray}
 f_{8}^{(0)}[{A}]
&=&
%%2
 \alpha^{(0)}
   \bigl(\mathrm{tr} {A}^{\ast} {A} \bigr)
\nonumber \\ &&
%%5
+ \beta^{(0)}
   \Bigl(
        \bigl(\mathrm{tr} \, {A}^{\ast} {A} \bigr)^{2}
      - \bigl(\mathrm{tr} \, {A}^{\ast 2} {A}^{2} \bigr)
   \Bigr)
\nonumber \\ &&
%%6
+ \gamma^{(0)}
   \Bigl(
         - 3  \bigl(\mathrm{tr} \, {A}^{\ast} {A} \bigr) \bigl(\mathrm{tr} \, {A}^{2} \bigr) \bigl(\mathrm{tr} \, {A}^{\ast 2} \bigr)
        + 4 \bigl(\mathrm{tr} \, {A}^{\ast} {A} \bigr)^{3}
        + 6 \bigl(\mathrm{tr} \, {A}^{\ast} {A} \bigr) \bigl(\mathrm{tr} \, {A}^{\ast 2} {A}^{2} \bigr)
      + 12 \bigl(\mathrm{tr} \, {A}^{\ast} {A} \bigr) \bigl(\mathrm{tr} \, {A}^{\ast} {A} {A}^{\ast} {A} \bigr)
              \nonumber \\ && \hspace{3em} %%
         - 6 \bigl(\mathrm{tr} \, {A}^{\ast 2} \bigr) \bigl(\mathrm{tr} \, {A}^{\ast} {A}^{3} \bigr)
         - 6 \bigl(\mathrm{tr} \, {A}^{2} \bigr) \bigl(\mathrm{tr} \, {A}^{\ast 3} {A} \bigr)
       - 12 \bigl(\mathrm{tr} \, {A}^{\ast 3} {A}^{3} \bigr)
      + 12 \bigl(\mathrm{tr} \, {A}^{\ast 2} {A}^{2} {A}^{\ast} {A} \bigr)
        + 8 \bigl(\mathrm{tr} \, {A}^{\ast} {A} {A}^{\ast} {A} {A}^{\ast} {A} \bigr)
   \Bigr)
\nonumber \\ &&
%%8
 + \delta^{(0)}
\Bigl(
       \bigl( \mathrm{tr}\,A^{\ast 2} \bigr)^{2} \bigl( \mathrm{tr}\, A^{2} \bigr)^{2}
 + 2 \bigl( \mathrm{tr}\,A^{\ast 2} \bigr)^{2} \bigl( \mathrm{tr}\, A^{4} \bigr)
  - 8 \bigl( \mathrm{tr}\,A^{\ast 2} \bigr) \bigl( \mathrm{tr}\,A^{\ast}AA^{\ast}A \bigr) \bigl( \mathrm{tr}\,A^{2} \bigr)
  - 8 \bigl( \mathrm{tr}\,A^{\ast 2} \bigr) \bigl( \mathrm{tr}\,A^{\ast}A \bigr)^{2} \bigl( \mathrm{tr}\,A^{2} \bigr)
       \nonumber \\ && \hspace{3em}
 - 32 \bigl( \mathrm{tr}\,A^{\ast 2} \bigr) \bigl( \mathrm{tr}\,A^{\ast}A \bigr) \bigl( \mathrm{tr}\,A^{\ast}A^{3} \bigr)
 - 32 \bigl( \mathrm{tr}\,A^{\ast 2} \bigr) \bigl( \mathrm{tr}\,A^{\ast}AA^{\ast}A^{3} \bigr)
 - 16 \bigl( \mathrm{tr}\,A^{\ast 2} \bigr) \bigl( \mathrm{tr}\,A^{\ast}A^{2}A^{\ast}A^{2} \bigr)
       \nonumber \\ && \hspace{3em}
  + 2 \bigl( \mathrm{tr}\,A^{\ast 4} \bigr) \bigl( \mathrm{tr}\,A^{2} \bigr)^{2}
  + 4 \bigl( \mathrm{tr}\,A^{\ast 4} \bigr) \bigl( \mathrm{tr}\,A^{4} \bigr)
  - 32 \bigl( \mathrm{tr}\,A^{\ast 3}A \bigr) \bigl( \mathrm{tr}\,A^{\ast}A \bigr) \bigl( \mathrm{tr}\,A^{2} \bigr)
       \nonumber \\ && \hspace{3em}
  - 64 \bigl( \mathrm{tr}\,A^{\ast 3}A \bigr) \bigl( \mathrm{tr}\,A^{\ast}A^{3} \bigr)
  - 32 \bigl( \mathrm{tr}\,A^{\ast 3}AA^{\ast}A \bigr) \bigl( \mathrm{tr}\,A^{2} \bigr)
  - 64 \bigl( \mathrm{tr}\,A^{\ast 3}A^{2}A^{\ast}A^{2} \bigr)
  - 64 \bigl( \mathrm{tr}\,A^{\ast 3}A^{3} \bigr) \bigl( \mathrm{tr}\,A^{\ast}A \bigr)
       \nonumber \\ && \hspace{3em}
  - 64 \bigl( \mathrm{tr}\,A^{\ast 2}AA^{\ast 2}A^{3} \bigr)
  - 64 \bigl( \mathrm{tr}\,A^{\ast 2}AA^{\ast}A^{2} \bigr) \bigl( \mathrm{tr}\,A^{\ast}A \bigr)
 + 16 \bigl( \mathrm{tr}\,A^{\ast 2}A^{2} \bigr)^{2}
 + 32 \bigl( \mathrm{tr}\,A^{\ast 2}A^{2} \bigr) \bigl( \mathrm{tr}\,A^{\ast}A \bigr)^{2}
       \nonumber \\ && \hspace{3em}
 + 32 \bigl( \mathrm{tr}\,A^{\ast 2}A^{2} \bigr) \bigl( \mathrm{tr}\,A^{\ast}AA^{\ast}A \bigr)
 + 64 \bigl( \mathrm{tr}\,A^{\ast 2}A^{2}A^{\ast 2}A^{2} \bigr)
  -16 \bigl( \mathrm{tr}\,A^{\ast 2}AA^{\ast 2}A \bigr) \bigl( \mathrm{tr}\,A^{2} \bigr)
   + 8 \bigl( \mathrm{tr}\,A^{\ast}A \bigr)^{4}
       \nonumber \\ && \hspace{3em}
 + 48 \bigl( \mathrm{tr}\,A^{\ast}A \bigr)^{2} \bigl( \mathrm{tr}\,A^{\ast}AA^{\ast}A \bigr)
 +192 \bigl( \mathrm{tr}\,A^{\ast}A \bigr) \bigl( \mathrm{tr}\,A^{\ast}AA^{\ast 2}A^{2} \bigr)
 + 64 \bigl( \mathrm{tr}\,A^{\ast}A \bigr) \bigl( \mathrm{tr}\,A^{\ast}AA^{\ast}AA^{\ast}A \bigr)
       \nonumber \\ && \hspace{3em}
  -128 \bigl( \mathrm{tr}\,A^{\ast}AA^{\ast 3}A^{3} \bigr)
 + 64 \bigl( \mathrm{tr}\,A^{\ast}AA^{\ast 2}AA^{\ast}A^{2} \bigr)
 + 24 \bigl( \mathrm{tr}\,A^{\ast}AA^{\ast}A \bigr)^{2}
 +128 \bigl( \mathrm{tr}\,A^{\ast}AA^{\ast}AA^{\ast 2}A^{2} \bigr)
       \nonumber \\ && \hspace{3em}
 + 48 \bigl( \mathrm{tr}\,A^{\ast}AA^{\ast}AA^{\ast}AA^{\ast}A \bigr)
\Bigr),
\nonumber
%\label{eq:eff_pot_w0_coefficient02_f}
\\
%%%%
   F_{2}^{(\le4)}[{A}]
&=&
%3
      \beta^{(2)}
      \vec{B}^{t} {A}^{\ast} {A} \vec{B}
+ \beta^{(4)}
   |\vec{B}|^{2}
   \vec{B}^{t} {A}^{\ast} {A} \vec{B},
\nonumber
%\label{eq:eff_pot_B4w2_coefficient02_f}
\\
%%%%
   f_{4}^{(\le2)}[{A}]
&=&
  \gamma^{(2)}
  \Bigl(
       - 2 \, |\vec{B}|^{2} \bigl(\mathrm{tr} \, {A}^{2} \bigr) \bigl(\mathrm{tr} \, {A}^{\ast 2} \bigr)
       - 4 \, |\vec{B}|^{2} \bigl(\mathrm{tr} \, {A}^{\ast} {A} \bigr)^{2}
      + 4 \, |\vec{B}|^{2} \bigl(\mathrm{tr} \, {A}^{\ast} {A} {A}^{\ast} {A} \bigr)
      + 8 \, |\vec{B}|^{2} \bigl(\mathrm{tr} \, {A}^{\ast 2} {A}^{2} \bigr)
            \nonumber \\ && \hspace{2em}
        + \vec{B}^{t} {A}^{2} \vec{B} \bigl(\mathrm{tr} \, {A}^{\ast 2} \bigr)
       - 8 \, \vec{B}^{t} {A}^{\ast} {A} \vec{B} \bigl(\mathrm{tr} \, {A}^{\ast} {A} \bigr)
         + \vec{B}^{t} {A}^{\ast 2} \vec{B} \bigl(\mathrm{tr} \, {A}^{2} \bigr)
      + 2 \, \vec{B}^{t} {A} {A}^{\ast 2} {A} \vec{B}
            \nonumber \\ && \hspace{2em}
      + 2 \, \vec{B}^{t} {A}^{\ast} {A}^{2} {A}^{\ast} \vec{B}
       - 8 \, \vec{B}^{t} {A}^{\ast} {A} {A}^{\ast} {A} \vec{B}
       - 8 \, \vec{B}^{t} {A}^{\ast 2} {A}^{2} \vec{B}
  \Bigr),
%\label{eq:eff_pot_B2w4_coefficient02_f}
\label{eq:eff_pot_f}
\end{eqnarray}
\end{widetext}
for the potential and interaction terms.
The trace ($\mathrm{tr}$) is taken over the indices of spin and three-dimensional momentum in $A$.
The coefficients are given by
\begin{eqnarray}
  K^{(0)}
&=&
   \frac{7 \, \zeta(3)N(0) p_{F}^{4}}{240m^{2}(\pi T_{c0})^{2}},
\nonumber \\ %%
   \alpha^{(0)}
&=&
   \frac{N(0)p_{F}^{2}}{3} 
   \frac{T-T_{c0}}{T_{c0}},
\nonumber \\ %%
  \beta^{(0)}
&=&
   \frac{7\,\zeta(3)N(0)p_{F}^{4}}{60\,(\pi T_{c0})^{2}},
\nonumber \\ %%
  \gamma^{(0)}
&=&
   - \frac{31\,\zeta(5)N(0)p_{F}^{6}}{13440\,(\pi T_{c0})^{4}},
\nonumber \\ %%
  \delta^{(0)}
&=&
  \frac{127\,\zeta(7)N(0)p_{F}^{8}}{387072\,(\pi T_{c0})^{6}},
\nonumber \\ %%
   \beta^{(2)}
&=&
    \frac{7\,\zeta(3)N(0)p_{F}^{2}\gamma_{n}^{2}}{48(1+F_{0}^{a})^{2}(\pi T_{c0})^{2}},
\nonumber \\ %%
   \beta^{(4)}
&=&
    - \frac{31\,\zeta(5)N(0)p_{F}^{2}\gamma_{n}^{4}}{768(1+F_{0}^{a})^{4}(\pi T_{c0})^{4}},
\nonumber \\ %%
  \gamma^{(2)}
&=&
  \frac{31\,\zeta(5)N(0)p_{F}^{4}\gamma_{n}^{2}}{3840(1+F_{0}^{a})^{2}(\pi T_{c0})^{4}}.
\label{eq:eff_pot_coefficient0_parameters_FL_f}
\end{eqnarray}
We denote $N(0)=m\,p_{F}/(2\pi^{2})$ for the state-number density at the Fermi surface and $|\vec{\mu}_{n}^{\ast}|=(\gamma_{n}/2)/(1+F_{0}^{a})$ for the magnitude of the magnetic momentum of a neutron modified by the Landau parameter $F_{0}^{a}$.\footnote{The interaction Hamiltonian between the neutron and the magnetic field ($\vec{B}$) is modified to $-\vec{\mu}_{n}^{\ast}\!\cdot\!\vec{B}$.}
Notice that $\vec{\mu}_{n}^{\ast}$ is different from the bare magnetic moment of the neutron $\vec{\mu}_{n}$ due to the Fermi-liquid correction.
We notice that the Landau parameter stems from the Hartree-Fock approximation which are not taken into account explicitly in the present mean-field approximation.
Finally, in all the expressions, $\zeta(n)$ is the zeta function.

We comment the physical meanings of each term in Eq.~\eqref{eq:eff_pot_f}.
The $\alpha^{(0)}$ term is the leading order, and the $\beta^{(0)}$ term is the next-to-leading order.
However, those terms are not sufficient to determine uniquely the ground state, because the free energies in the UN, D$_{2}$-BN, and D$_{4}$-BN phases are degenerate at this order.
The degeneracy is resolved by the $\gamma^{(0)}$ term which leads to only the local stability of the ground state, but the global stability is lost at this order.
Finally, the global stability is restored by the $\delta^{(0)}$ term which was recently calculated in Ref.~\cite{Yasui:2019unp}.
Therefore, the eighth order is minimally required to have the globally stable and unique ground state.
Concerning the magnetic field, the $\beta^{(2)}$ term is the leading order, and the $\beta^{(4)}$ and $\gamma^{(2)}$ terms are the higher-order corrections.
The effect of the latter terms was examined for investigating the phase diagrams in strong magnetic fields in magnetars, and it was shown that those terms supply the change of the transition line in the phase diagram by a few percentages at most~\cite{Yasui:2018tcr}.

%%%%%%%%%%%%%%%%%%%%%%%%%%%%%%
\begin{figure}[tb]
\begin{center}
\includegraphics[scale=0.22]{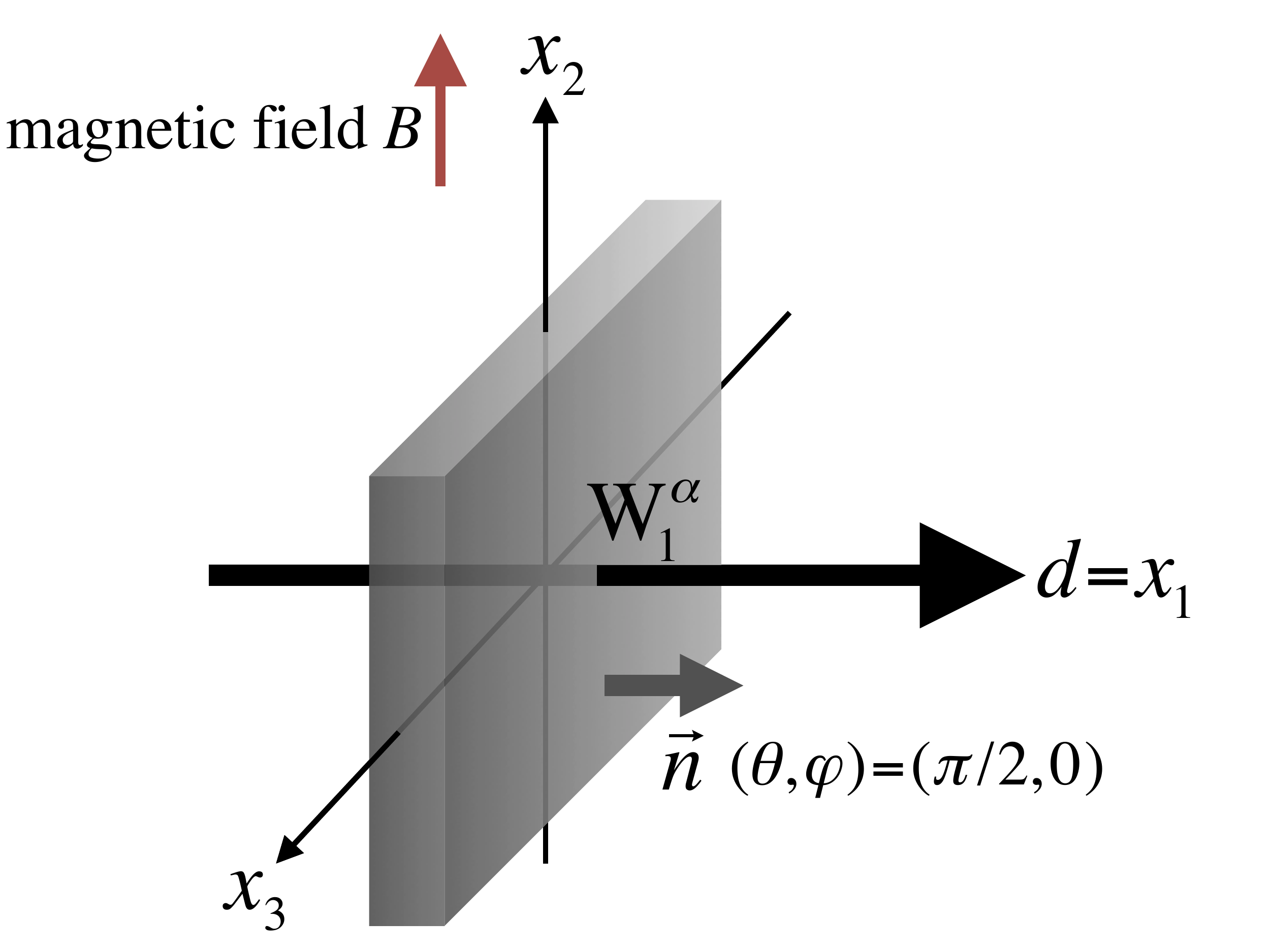}
\hspace{0.5em}
\includegraphics[scale=0.22]{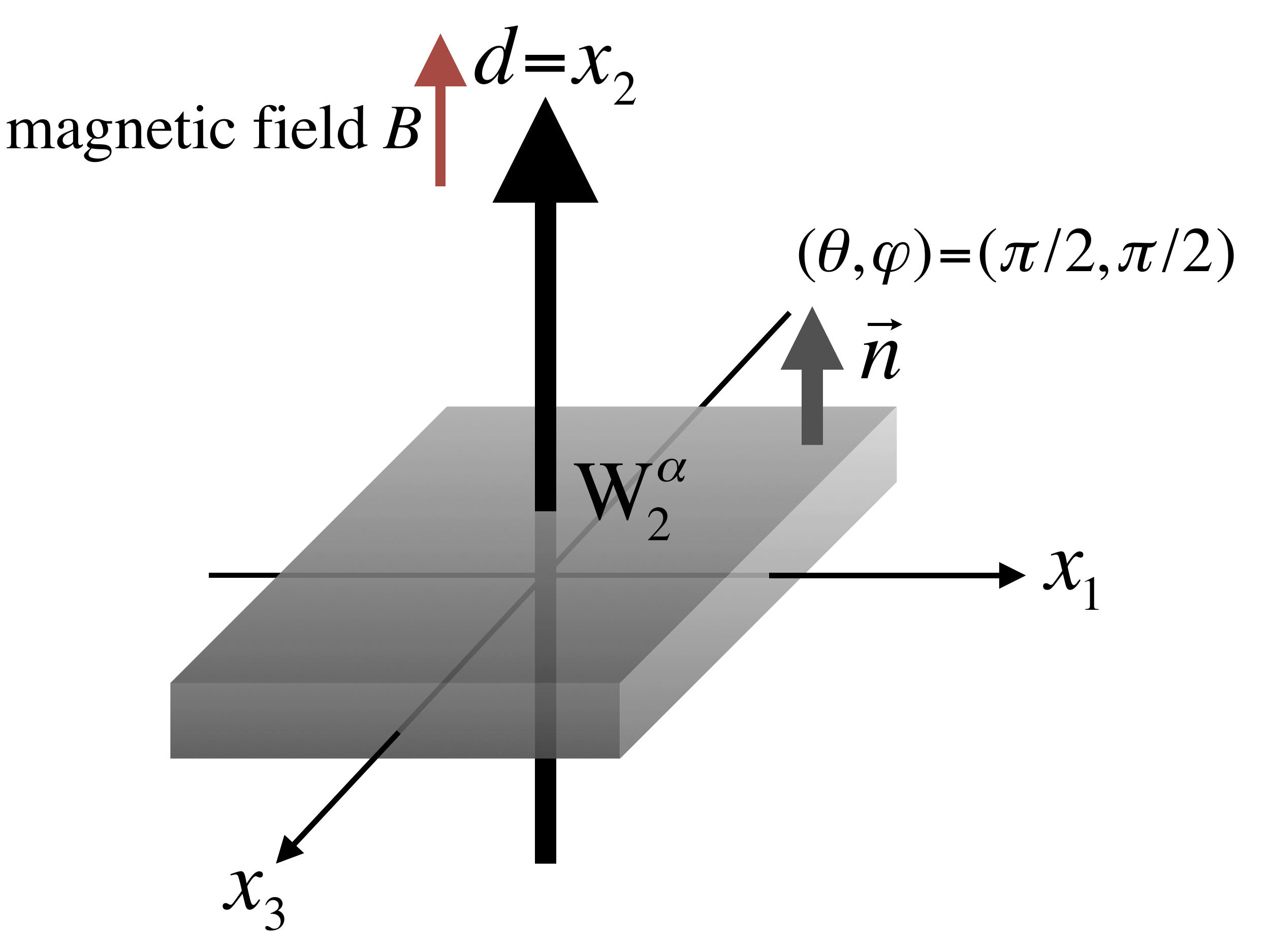}
\hspace{0.5em}
\includegraphics[scale=0.22]{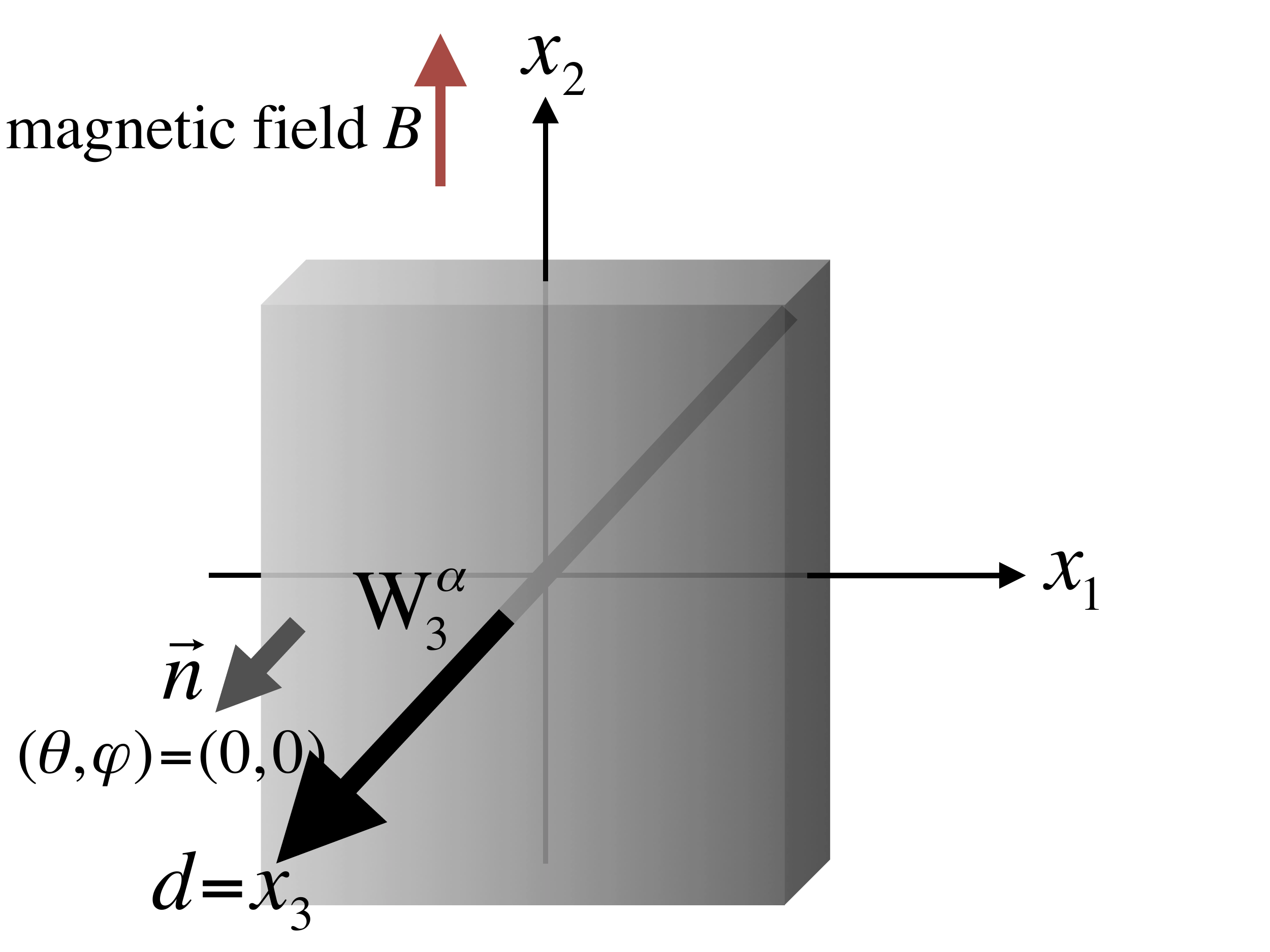}
\caption{The directions of the domain walls $W_{i}^{\alpha}$ ($i=1$, $2$, $3$) for several types $\alpha=1$, $2$, $3$, and $13$ are displayed. Left: $W_{1}^{\alpha}$ ($x_{1}$ direction). Middle: $W_{2}^{\alpha}$ ($x_{2}$ direction). Right: $W_{3}^{\alpha}$ ($x_{3}$ direction). The corresponding polar angles $(\theta,\varphi)$ are shown. The magnetic field is applied along the $x_{2}$ direction.}
\label{fig:Fig_190602_x}
\end{center}
\end{figure}
%%%%%%%%%%%%%%%%%%%%%%%%%%%%%%

\subsection{Euler-Lagrange equation of domain walls}

We consider domain walls as nonuniform solutions in the GL equation.
The domain wall is an extended object in the two-dimensional space, and it is regarded as a one-dimensional solution, {\it i.e.}, a kink along the direction perpendicular to the plane of the domain wall.
We assume that the geometrical shape of the surface is flat by neglecting the curvature of the domain walls.
For the direction of the domain wall, we denote the normal vector perpendicular to the surface: $\vec{n}=(x_{1},x_{2},x_{3})/|\vec{x}|=(n_{1},n_{2},n_{3})$ with the coordinate $\vec{x}=(x_{1},x_{2},x_{3})$ as shown in Fig.~\ref{fig:Fig_190602_x}.
We use a polar-angle parametrization $n_{1}=\sin\theta\cos\varphi$, $n_{2}=\sin\theta\sin\varphi$, and $n_{3}=\cos\theta$ with the angles $\theta$ and $\varphi$.
We also introduce the coordinate $d$ ($-\infty < d < \infty$) in the direction along $\vec{n}$ perpendicular to the domain wall.\footnote{It should be noticed that $d$ defined in the present discussion is different from $d$ used in the previous study~\cite{Yasui:2019pgb}.}
Then the condensate $A$ can be expressed by a matrix whose components are functions of $d$ and $\vec{n}$:
\begin{eqnarray}
   A(d;\vec{n})
=
\left(
\begin{array}{ccc}
 -F_{1}(d;\vec{n}) & G_{3}(d;\vec{n}) & G_{2}(d;\vec{n}) \\
 G_{3}(d;\vec{n}) & -F_{2}(d;\vec{n}) & G_{1}(d;\vec{n}) \\
 G_{2}(d;\vec{n}) & G_{1}(d;\vec{n}) & F_{1}(d;\vec{n})+F_{2}(d;\vec{n})  
\end{array}
\right),
\label{eq:A_matrix}
\end{eqnarray}
where $F_{\alpha}(d;\vec{n})$ ($\alpha=1$, $2$) are the diagonal components and $G_{\beta}(d;\vec{n})$ ($\beta=1$, $2$, $3$) are off-diagonal components with $F_{\alpha}(d;\vec{n})$ and $G_{\beta}(d;\vec{n})$ being complex numbers in general.
We consider that static domain walls, neglecting the dynamical fluctuation on the surface.
Thus, we regard that $d$ is only the coordinate on which the configurations in $A$ depend and that the angles $\theta$ and $\varphi$ are simply the external parameters for fixing the direction of the domain wall.
In the following discussion, setting $\vec{n}$ as a constant vector, we will investigate the angles which are favored to minimize the energy of the domain walls, for which the magnetic field is applied to the $x_2$ direction.
For the short expression, instead of the above notations, we will use the simpler notations $A(d)=A(d;\vec{n})$, $F_{\alpha}(d)=F_{\alpha}(d;\vec{n})$, and $G_{\beta}(d)=G_{\beta}(d;\vec{n})$ by omitting the normal vector $\vec{n}$. 
With the above setup, we express the gradient terms in Eq.~\eqref{eq:eff_pot_kin} as
\begin{eqnarray}
   {f}_{\grad}[A]
&=&
   \frac{{K}^{(0)}}{4}
   \biggl(
           \bigl(
                    2
                 - \sin^{2}\theta \sin^{2}\varphi
           \bigr)
           \bigl(\nabla_{\!d}F_{1}\bigr)^{2}
        + \bigl(
                    2
                  - \sin^{2}\theta \cos^{2}\varphi
           \bigr)
           \bigl(\nabla_{\!d}F_{2}\bigr)^{2}
        + \bigl(
                    1
                 + 2\cos^{2}\theta
           \bigr)
           \bigl(\nabla_{\!d} F_{1}\bigr) \bigl(\nabla_{\!d}F_{2}\bigr)
   \nonumber \\ && \hspace{1em} %%
+ 
           2\cos\theta\sin\theta\sin\varphi \bigl(\nabla_{\!d}F_{1}\bigr) \bigl(\nabla_{\!d}G_{1}\bigr)
        + 2\cos\theta\sin\theta\cos\varphi \bigl(\nabla_{\!d}F_{2}\bigr) \bigl(\nabla_{\!d}G_{2}\bigr)
           \nonumber \\ && \hspace{1em} %%
         - 2\sin^{2}\theta\cos\varphi\sin\varphi \bigl(\nabla_{\!d}F_{1}+\nabla_{\!d}F_{2}\bigr) \bigl(\nabla_{\!d}G_{3}\bigr)
   \nonumber \\ && \hspace{1em} %%
+ 
           \bigl(
                   2
                 - \sin^{2}\theta \cos^{2}\varphi
           \bigr)
           \bigl(\nabla_{\!d}G_{1}\bigr)^{2}
        + \bigl(
                    2
                  - \sin^{2}\theta \sin^{2}\varphi
           \bigr)
           \bigl(\nabla_{\!d}G_{2}\bigr)^{2}
        + \bigl(
                    1
                 + \sin^{2}\theta
           \bigr)
           \bigl(\nabla_{\!d}G_{3}\bigr)^{2}
           \nonumber \\ && \hspace{1em} %%
        + 2\sin^{2}\theta\cos\varphi\sin\varphi \bigl(\nabla_{\!d}G_{1}\bigr) \bigl(\nabla_{\!d}G_{2}\bigr)
        + 2\cos\theta\sin\theta \bigl(\cos\varphi \nabla_{\!d}G_{1}+\sin\varphi \nabla_{\!d}G_{2}\bigr) \nabla_{\!d}G_{3}
   \biggr),
\label{eq:kin_eff_pot}
\end{eqnarray}
with $\nabla_{\!d}=\partial/\partial d$.
We notice that the derivatives with respect to $\theta$ and $\varphi$ are absent because $\vec{n}$ is a constant vector.
By adopting the stationary condition with respect to $F_{\alpha}(d)$ and $G_{\beta}(d)$ in Eq.~\eqref{eq:eff_pot_coefficient02_f} together with Eq.~\eqref{eq:kin_eff_pot},
we then obtain the EL equations for $A$,
\begin{eqnarray}
   -\nabla_{\!d} \frac{\delta {f}[A]}{\delta (\nabla_{\!d}F_{\alpha})}
   + \frac{\delta {f}[A]}{\delta F_{\alpha}} &=& 0,
   \label{eq:EL_f} \\ %%
   -\nabla_{\!d} \frac{\delta {f}[A]}{\delta (\nabla_{\!d}G_{\beta})}
   + \frac{\delta {f}[A]}{\delta G_{\beta}} &=& 0,
   \label{eq:EL_g}
\end{eqnarray}
which provide the solutions of the domain walls.\footnote{The concrete expressions of the left hand sides are presented in detail in Appendix~\ref{sec:EL_appendix}.}
As for the boundary condition for Eqs.~\eqref{eq:EL_f} and \eqref{eq:EL_g},
 we require that the domain wall approaches the bulk state at $d \rightarrow \pm\infty$.
This means that the condensate values in $A$ should satisfy $F_{\alpha}(d)\rightarrow F_{\alpha}^{\bulk}$ and $G_{\beta}(d)\rightarrow 0$,
where $F_{\alpha}^{\bulk}$ are the values in the ground state in the bulk space.
We will consider that the values $F_{\alpha}^{\bulk}$ for $d \rightarrow \infty$ are not necessarily the same as those for $d\rightarrow-\infty$, because the domain walls are supposed to connect degenerate vacua that are considered to be different to each other.

In the following discussion, we restrict the configuration in the order parameter \eqref{eq:A_matrix} to the diagonal form by setting the off-diagonal components to be zero.\footnote{We call an attention to the assumption that the off-diagonal components are set to be zero in the present analysis. In this case, the two different degenerate vacua can be distinguished. However, those two degenerate states can be connected by a symmetry transformation once the off-diagonal components are taken into account. See more discussions below.}
In this setting, we introduce the dimensionless forms by expressing the order parameter $A(d)$ by
\begin{eqnarray}
   A(d) = \frac{T_{c0}}{p_{F}} \tilde{A}(\tilde{d}),
\label{eq:dimensionless_quantities_1}
\end{eqnarray}
with $\tilde{A}(\tilde{d})$ is the dimensionless function parametrized by
\begin{eqnarray}
  \tilde{A}(\tilde{d})
=
\left(
\begin{array}{ccc}
 -f_{1}(\tilde{d}) & 0 & 0 \\
 0 & -f_{2}(\tilde{d}) & 0 \\
 0 & 0 & f_{1}(\tilde{d})+f_{2}(\tilde{d})  
\end{array}
\right)
=
\left(
\begin{array}{ccc}
 -f_{1}(\tilde{d}) & 0 & 0 \\
 0 & -f_{2}(\tilde{d}) & 0 \\
 0 & 0 & -f_{3}(\tilde{d})
\end{array}
\right),
\label{eq:A_matrix_diagonal}
\end{eqnarray}
with $d=(p_{F}/(mT_{c0}))\tilde{d}$ for the dimensionless coordinate $\tilde{d}$ ($-\infty < \tilde{d} < \infty$).
$f_{i}(\tilde{d})$ ($i=1$, $2$) is related to $F_{i}(d)$ through $F_{i}(d)=(T_{c0}/p_{F})f_{i}(\tilde{d})$.
In the last equation in Eq.~\eqref{eq:A_matrix_diagonal},
we have introduced $f_{3}(\tilde{d}) \equiv -f_{1}(\tilde{d})-f_{2}(\tilde{d})$ for convenience of the calculation.
Furthermore, we parametrize $f_{i}(\tilde{d})$ ($i=1$, $2$, $3$) by
\begin{eqnarray}
   f_{1}(\tilde{d}) &=& \biggl( \frac{\cos\phi(\tilde{d})}{\sqrt{2}} - \frac{\sin\phi(\tilde{d})}{\sqrt{6}} \biggr) f_{0}(\tilde{d}), \nonumber \\
   f_{2}(\tilde{d}) &=& \sqrt{\frac{2}{3}} \bigl( \sin\phi(\tilde{d}) \bigr) f_{0}(\tilde{d}), \nonumber \\
   f_{3}(\tilde{d}) &=& \biggl( -\frac{\cos\phi(\tilde{d})}{\sqrt{2}} - \frac{\sin\phi(\tilde{d})}{\sqrt{6}} \biggr) f_{0}(\tilde{d}),
   \label{eq:alpha_f}
\end{eqnarray}
where $f_{0}(\tilde{d})$ and $\phi(\tilde{d})$ are introduced for the {\it amplitude} and the {\it angle} as functions of $\tilde{d}$ (for a fixed $\vec{n}$)
 as a new parametrization of the condensate $A$.
The range of the values are constrained to $f_{0}(\tilde{d}) \ge 0$ and $-\pi \le \phi(\tilde{d}) \le \pi$.
In the following discussions, we will use either the three-dimensional vector $\vec{f}(\tilde{d})\equiv(f_{1}(\tilde{d}),f_{2}(\tilde{d}),f_{3}(\tilde{d}))$ or the polar parametrization $f_{0}(\tilde{d})$ and $\phi(\tilde{d})$ to express the order parameter $A$.
We also introduce the dimensionless forms:
\begin{eqnarray}
 f[A] = N(0)T_{c0}^{2}\tilde{f}[\tilde{A}], \quad
 x_{i} = \frac{p_{F}}{mT_{c0}}\tilde{x}_{i} \quad (i=1,2,3), \quad
 t=\frac{T}{T_{c0}}, \quad
 \vec{B} = \frac{(1+F_{0}^{a})T_{c0}}{\gamma_{n}} \vec{b},
\label{eq:dimensionless_quantities_2}
\end{eqnarray}
for the thermodynamical potential, the coordinate in the real space, the temperature, and the magnetic field, respectively.

Before going to numerics, 
let us emphasize that the domain walls that we are considering are 
described only by the diagonal components in $A(d;\vec{n})$. 
Because of this restriction, configuration may be unstable or metastable 
in the full analysis including  of the off-diagonal components.
See Sec.~\ref{sec:discussion} for a discussion in more detail.

\section{Numerical results}
\label{sec:numerical_results}

First, we present the UN, D$_{2}$-BN, and D$_{4}$-BN phases in the phase diagram in the bulk space.
In the next, with noting that there exist several minima in the effective potential in each phase, we give the definitions of the domain walls connecting the different minima.
Finally, we show the solution of the configurations of the domain walls, estimate their surface energy density, and discuss that multiple domain walls which are piled along the magnetic field can stably exist.

\subsection{Phase diagram in bulk space}

%%%%%%%%%%%%%%%%%%%%%%%%%%%%%%
\begin{figure}[tb]
\begin{center}
\includegraphics[scale=0.23]{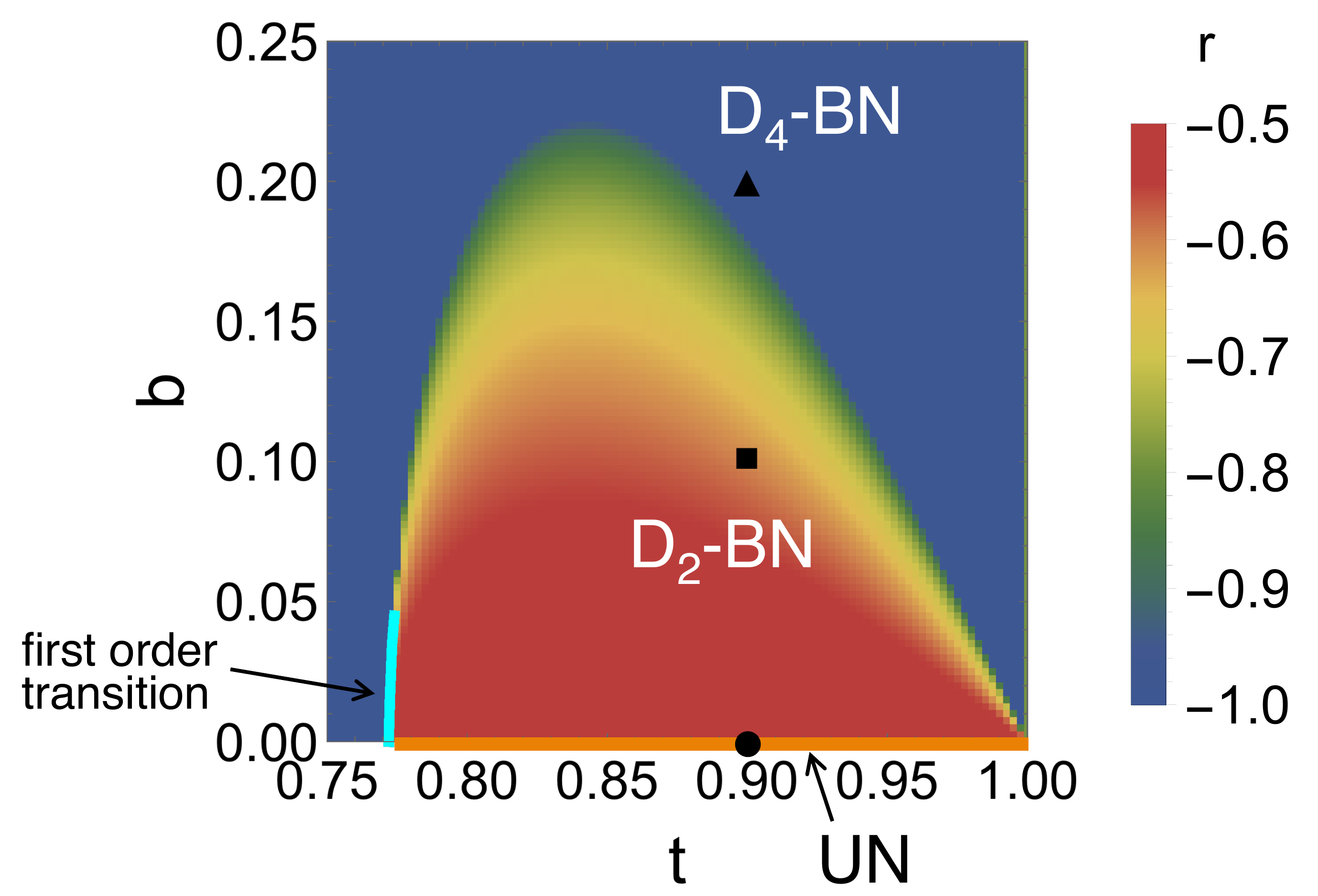}
\caption{We show the phase diagram on the plane spanned by the dimensionless temperature $t$ and the dimensionless magnetic field $b$. We will consider the three examples for the bulk phases at $t=0.9$: $b=0$ for the bulk UN phase (circle), $b=0.1$ for the bulk D$_{2}$-BN phase (square), and $b=0.2$ for the bulk D$_{4}$-BN phase (triangle). The phase boundary by the cyan line indicates the first-order phase transition~\cite{Yasui:2019unp}. The other phase boundaries are the second-order phase transition.}
\label{fig:phase_diagram}
\end{center}
\end{figure}
%%%%%%%%%%%%%%%%%%%%%%%%%%%%%%

In the bulk space,
the order parameter \eqref{eq:A_matrix} can be diagonalized as
\begin{eqnarray}
 A \rightarrow U_{\mathrm{diag}}AU_{\mathrm{diag}}^{-1}
 = A_{0}
\left(
\begin{array}{ccc}
  r & 0 & 0 \\
  0 & -1-r & 0 \\
  0 & 0 & 1
\end{array}
\right),
 \label{eq:A_matrix_0}
\end{eqnarray}
with an appropriate symmetry transformation $U_{\mathrm{diag}}$ of $\mathrm{U}(1) \times \mathrm{SO}(3)$,
where $A_{0} \ge 0$ is the amplitude and $r$ is a real parameter which can be restricted to $-1 \le r \le -1/2$ without loss of generality by the $\mathrm{U}(1)\times \mathrm{SO}(3)$ symmetry.
The different values of $r$ induce the different symmetries in the ground state: the UN phase for $r=-1/2$, the D$_{2}$-BN phase for $-1<r<-1/2$ , and the D$_{4}$-BN phase for $r=-1$.
Substituting Eq.~\eqref{eq:A_matrix_0} into Eq.~\eqref{eq:eff_pot_coefficient02_f}, we perform the variational calculation with respect $A_{0}$
and $r$ and obtain the phase diagram as shown in Fig.~\ref{fig:phase_diagram}.
Roughly, there exist the UN phase at zero magnetic field, the D$_{2}$-BN at weak magnetic field, and D$_{4}$-BN phases at strong magnetic field.\footnote{We may notice that the D$_{4}$-BN phase is also extended at low temperature and small magnetic field. There, the first-order phase transition exists at small magnetic field, and there appears the CEP at the meeting point between the first-order and second-order phase transitions. The first-order phase transition and the CEP is induced by the eighth-order term ($\delta^{(0)}$ term) in the GL equation~\cite{Yasui:2019unp}.
}

\begin{figure}%[tb]
\begin{center}
\includegraphics[scale=0.25]{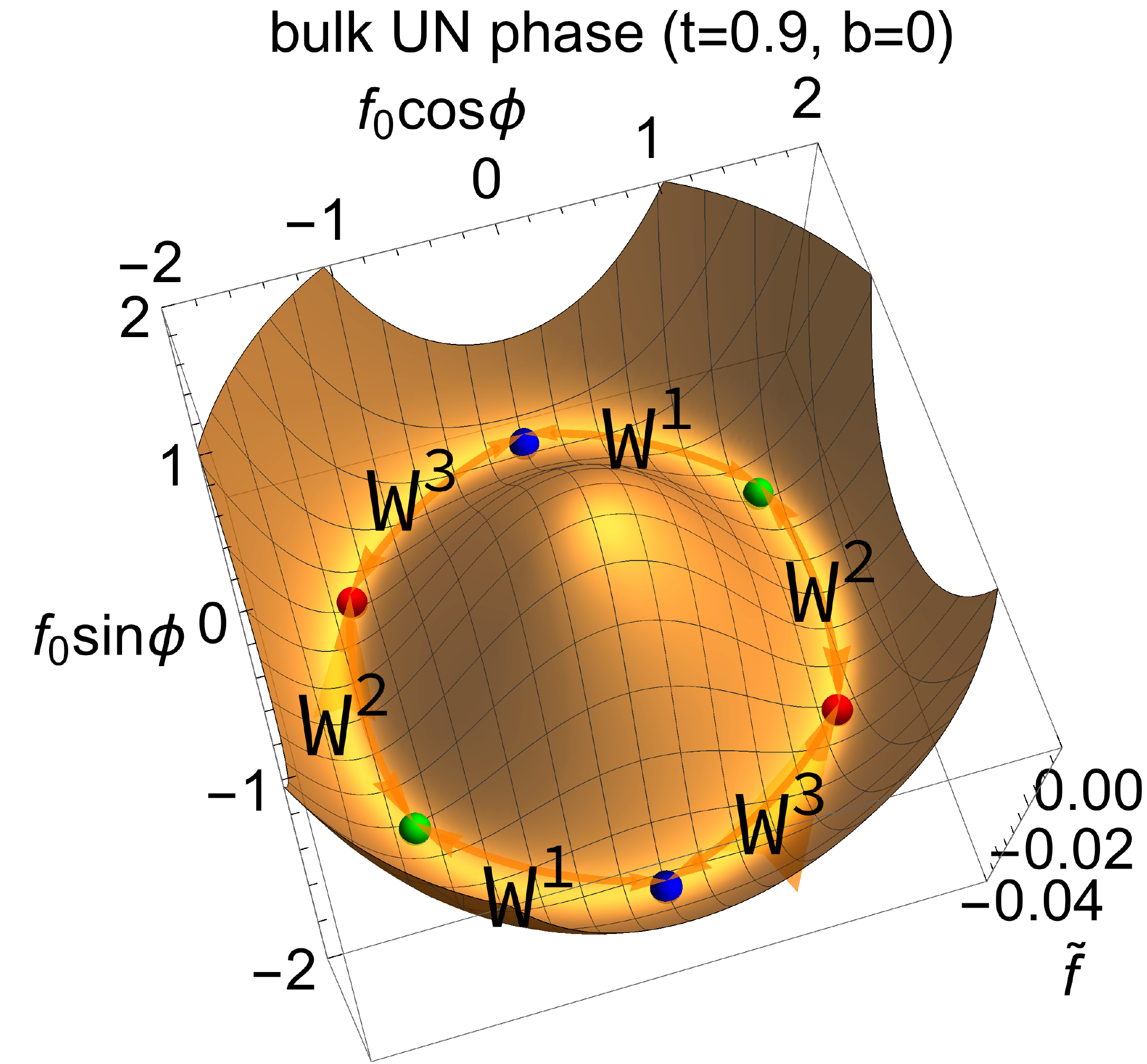}
\hspace{1em}
\includegraphics[scale=0.25]{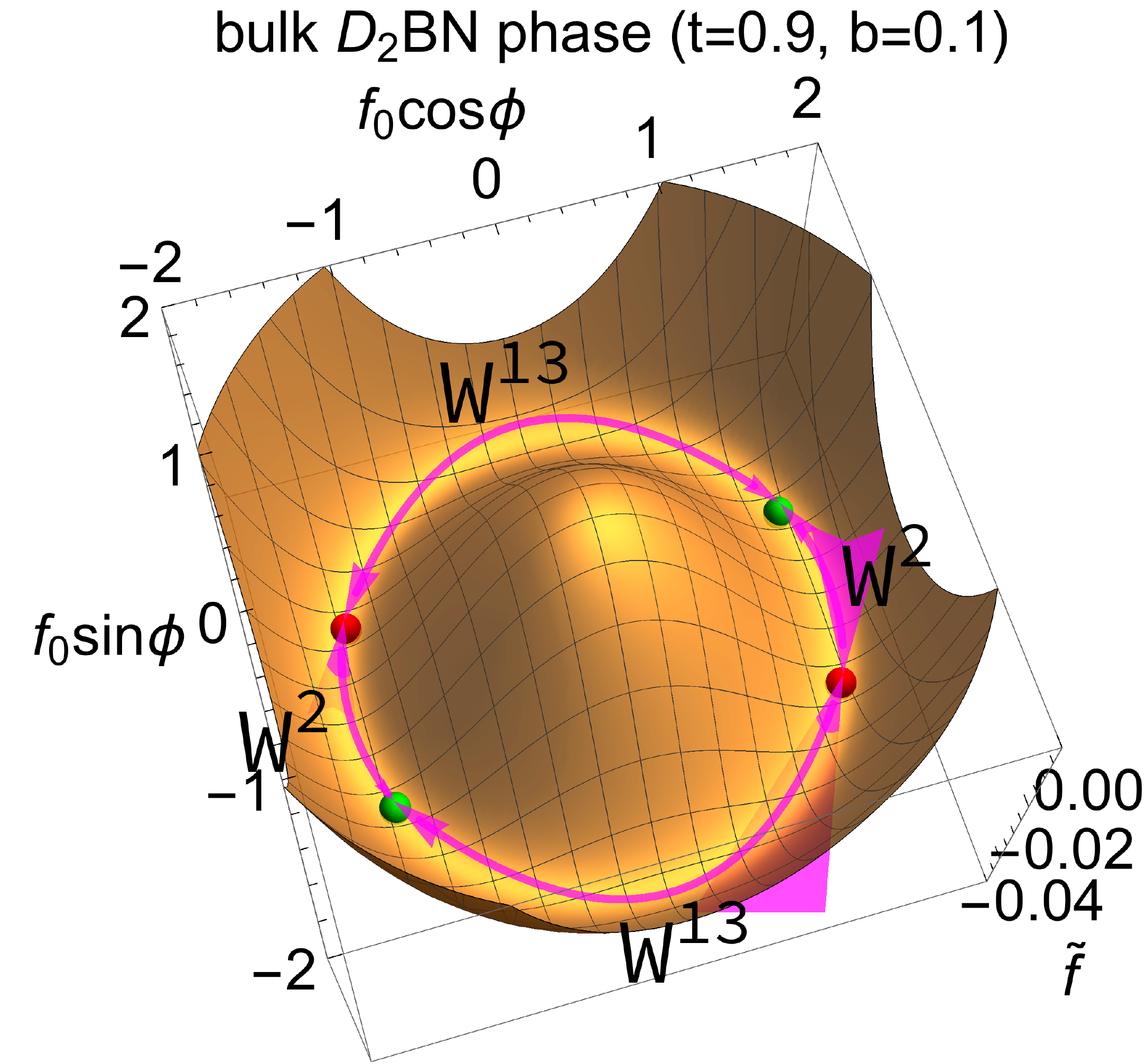}
\hspace{1em}
\includegraphics[scale=0.25]{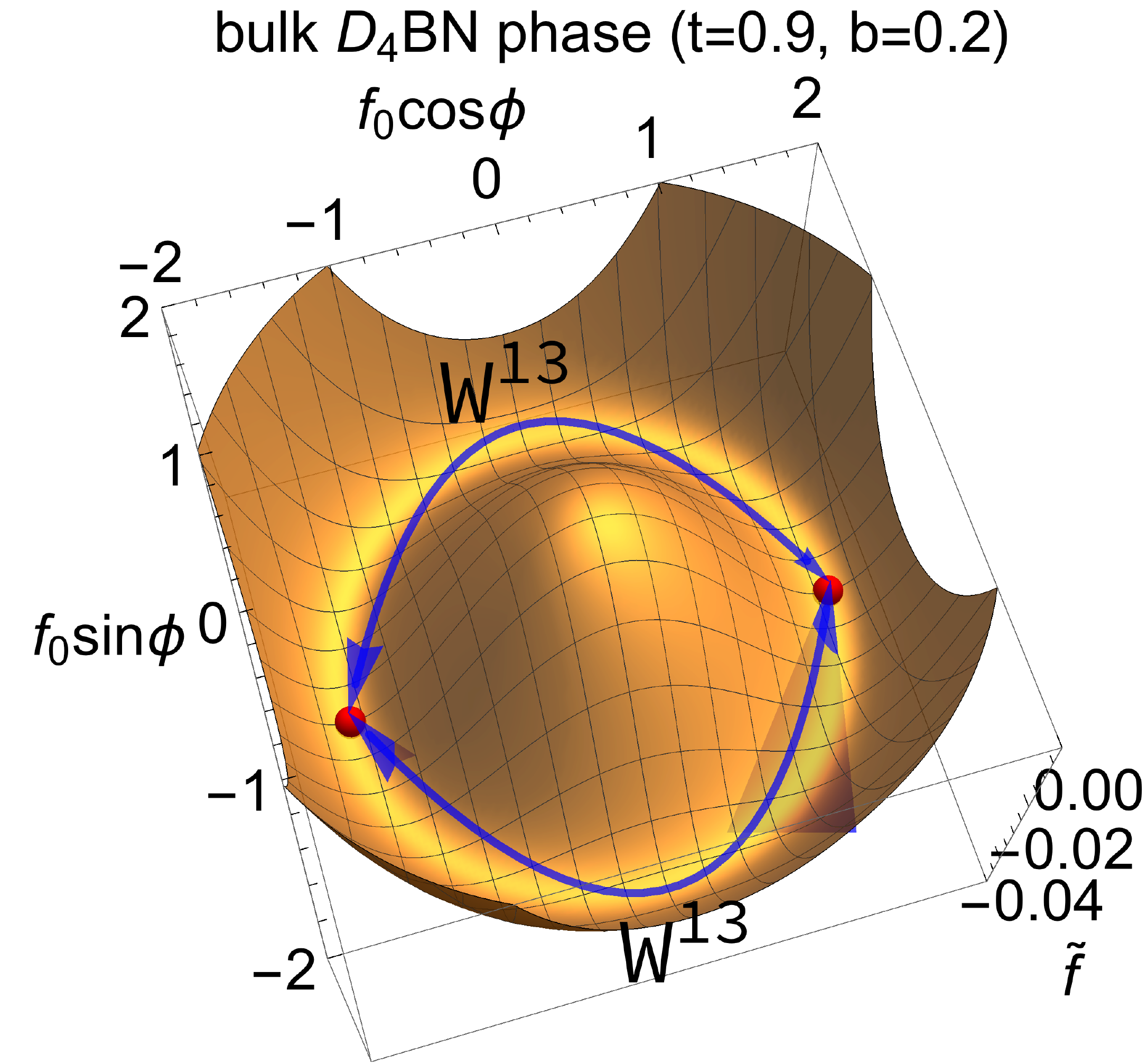}
\\
\vspace{1em}
\includegraphics[scale=0.28]{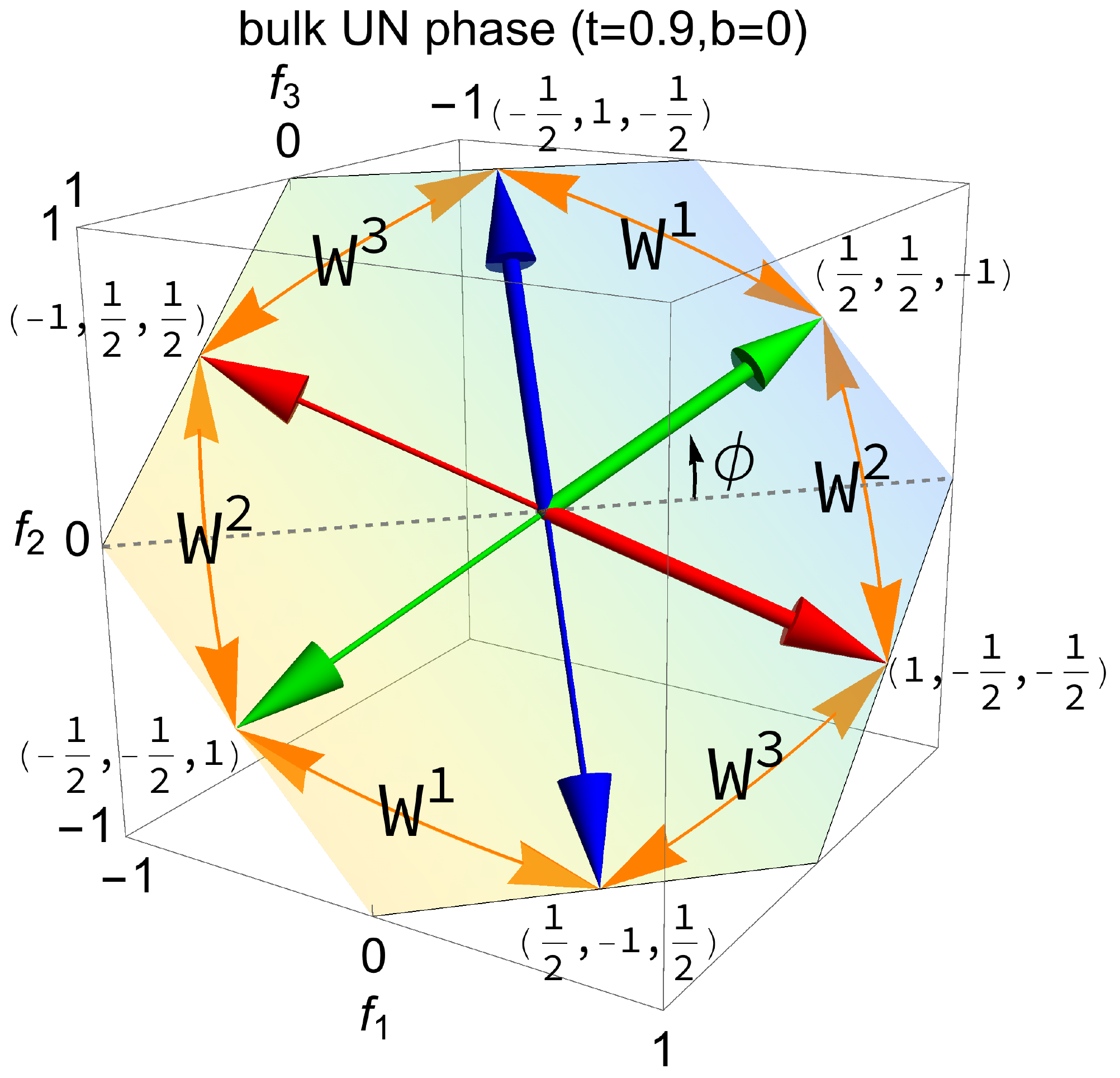}
\hspace{1em}
\includegraphics[scale=0.28]{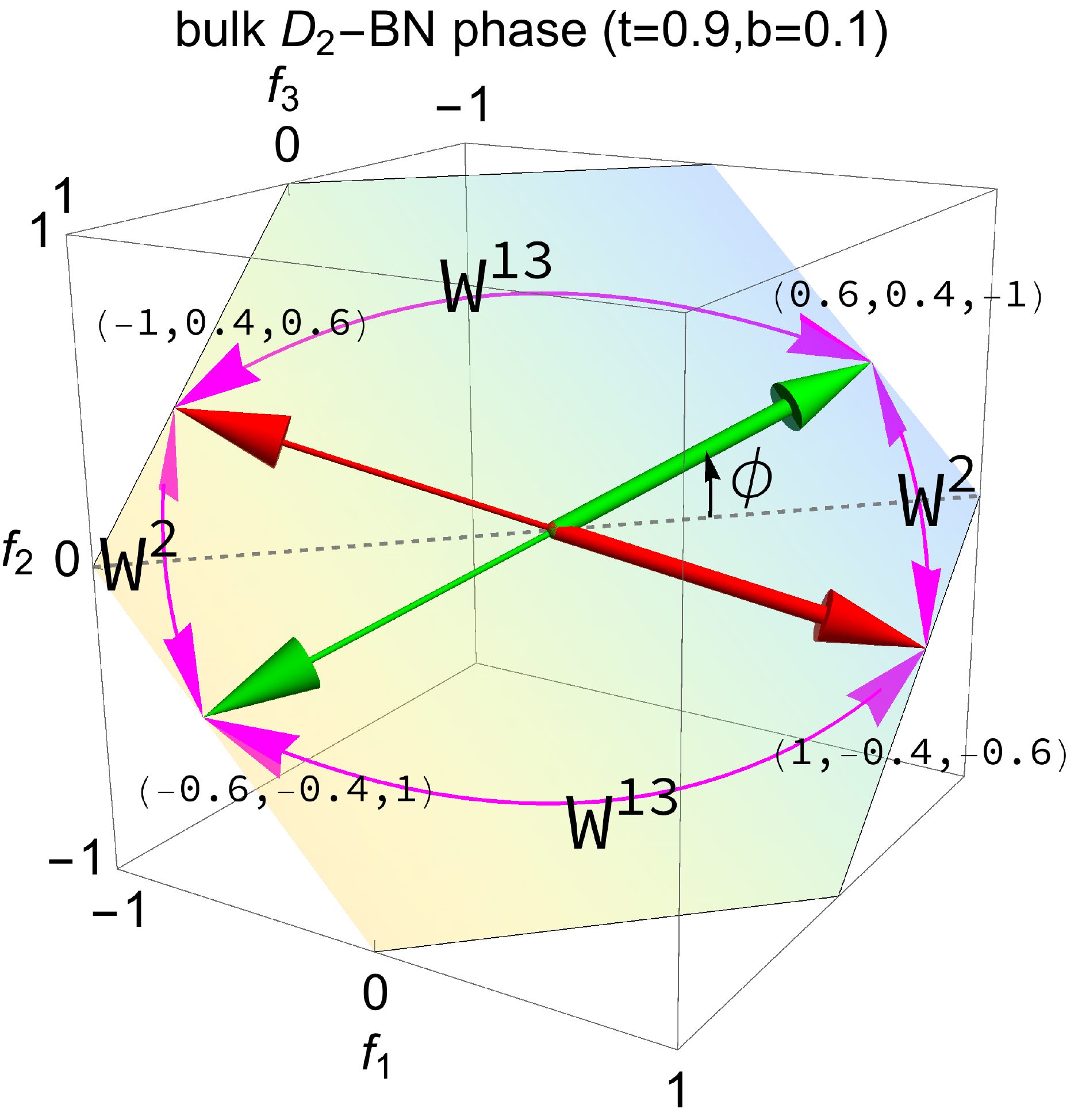}
\hspace{1em}
\includegraphics[scale=0.28]{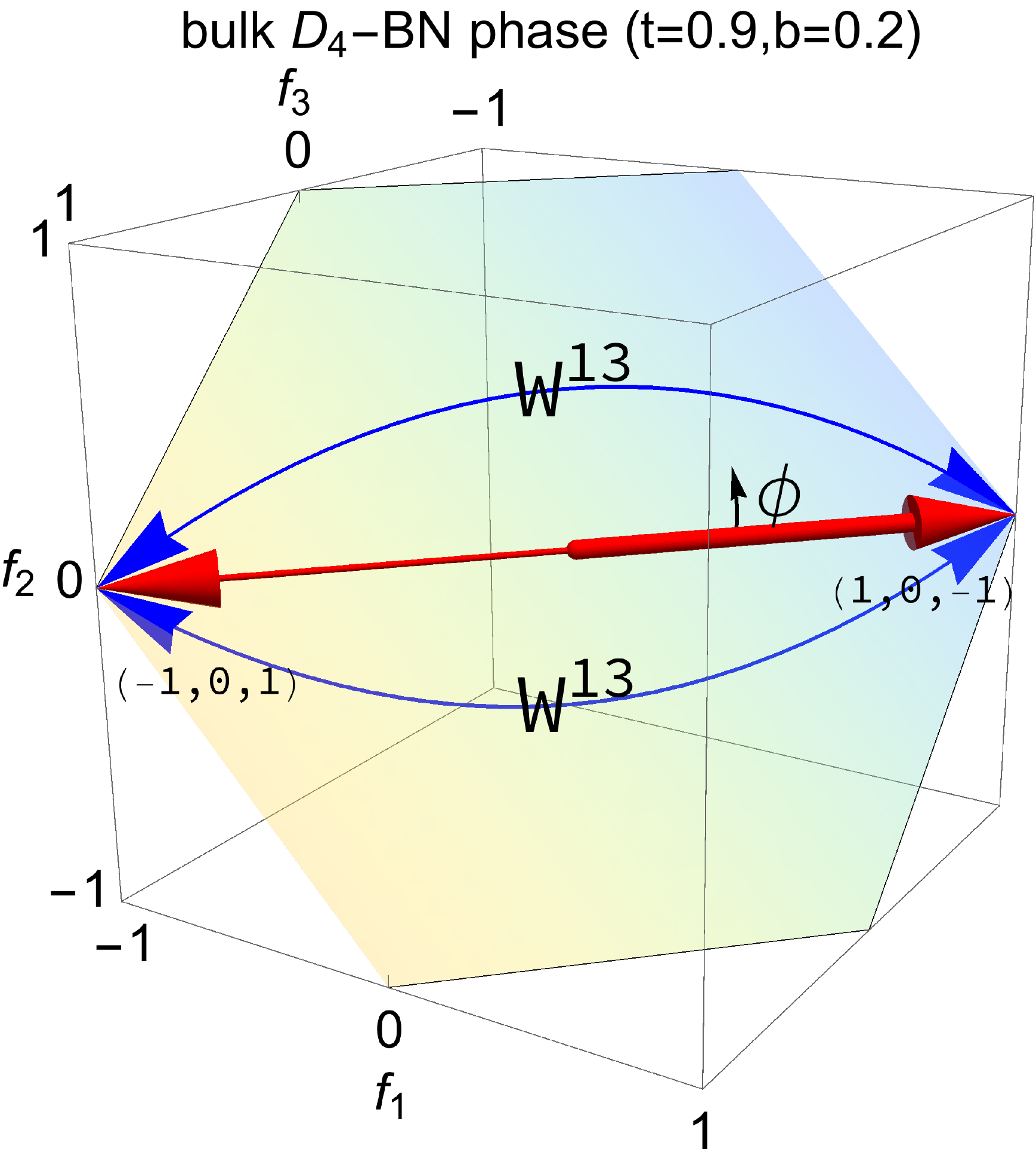}
\caption{Upper panels: We show the examples of the GL free energy on the plane with axes $f_{0} \cos\phi$ and $f_{0} \sin\phi$ for the bulk UN, D$_{2}$-BN, and D$_{4}$-BN phases. The degenerate vacua are shown by the colorful points in each bulk phase (orange for the bulk UN phase, magenta for the bulk D$_{2}$-BN phase, and blue for the bulk D$_{4}$-BN phase), and they are connected by the domain walls $W^{\alpha}$ ($\alpha=1$, $2$, $3$, $13$). Lower panels: The three-dimensional vectors $(f_{1},f_{2},f_{3})$ with $f_{1}+f_{2}+f_{3}=0$ are shown by the colorful arrows for the bulk UN, D$_{2}$-BN, and D$_{4}$-BN phases. The domain walls are indicated by the arrows, $W^{2}$, $W^{1}$, and $W^{3}$, for the bulk UN phase, $W^{2}$ and $W^{13}$ for the bulk D$_{2}$-BN phase, and $W^{13}$ for the bulk D$_{4}$-BN phase. The arrows lies on the plane for the traceless condition: $f_{1}+f_{2}+f_{3}=0$.
\label{fig:GL_potential}
}
\end{center}
\end{figure}

\begin{figure}%[tb]
\begin{center}
\includegraphics[scale=0.189]{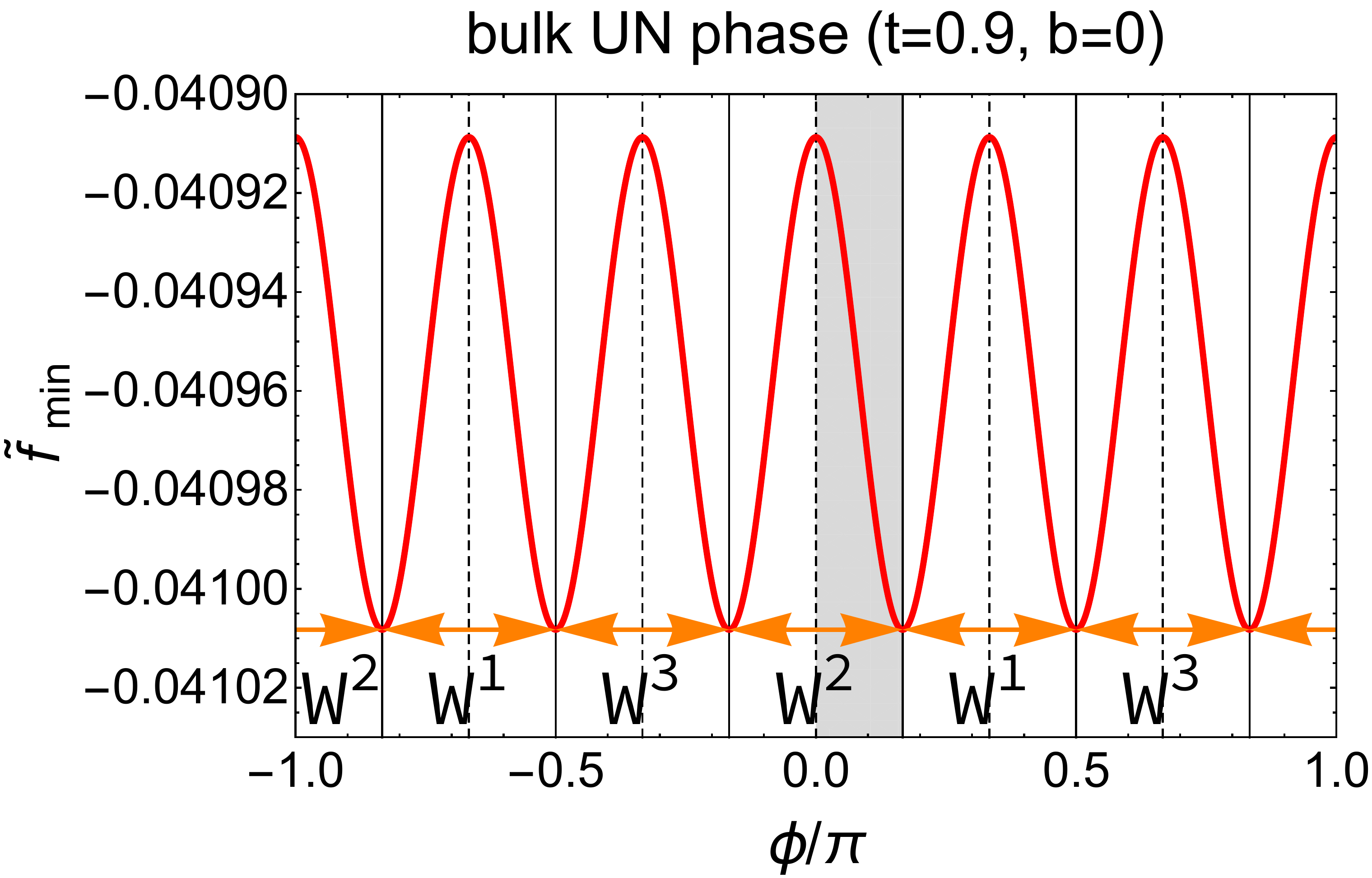}
\hspace{0.05em}
\includegraphics[scale=0.189]{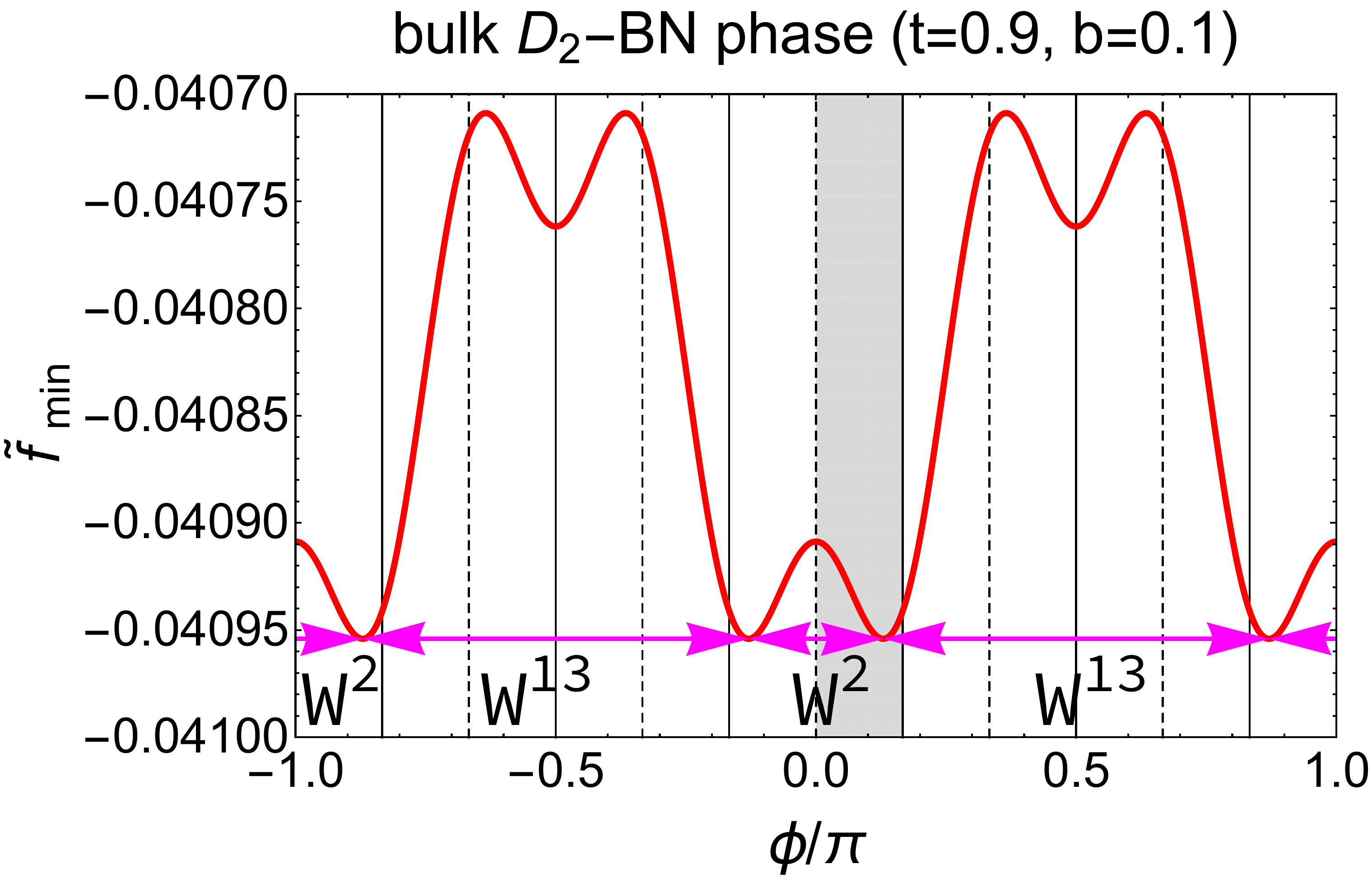}
\hspace{0.05em}
\includegraphics[scale=0.189]{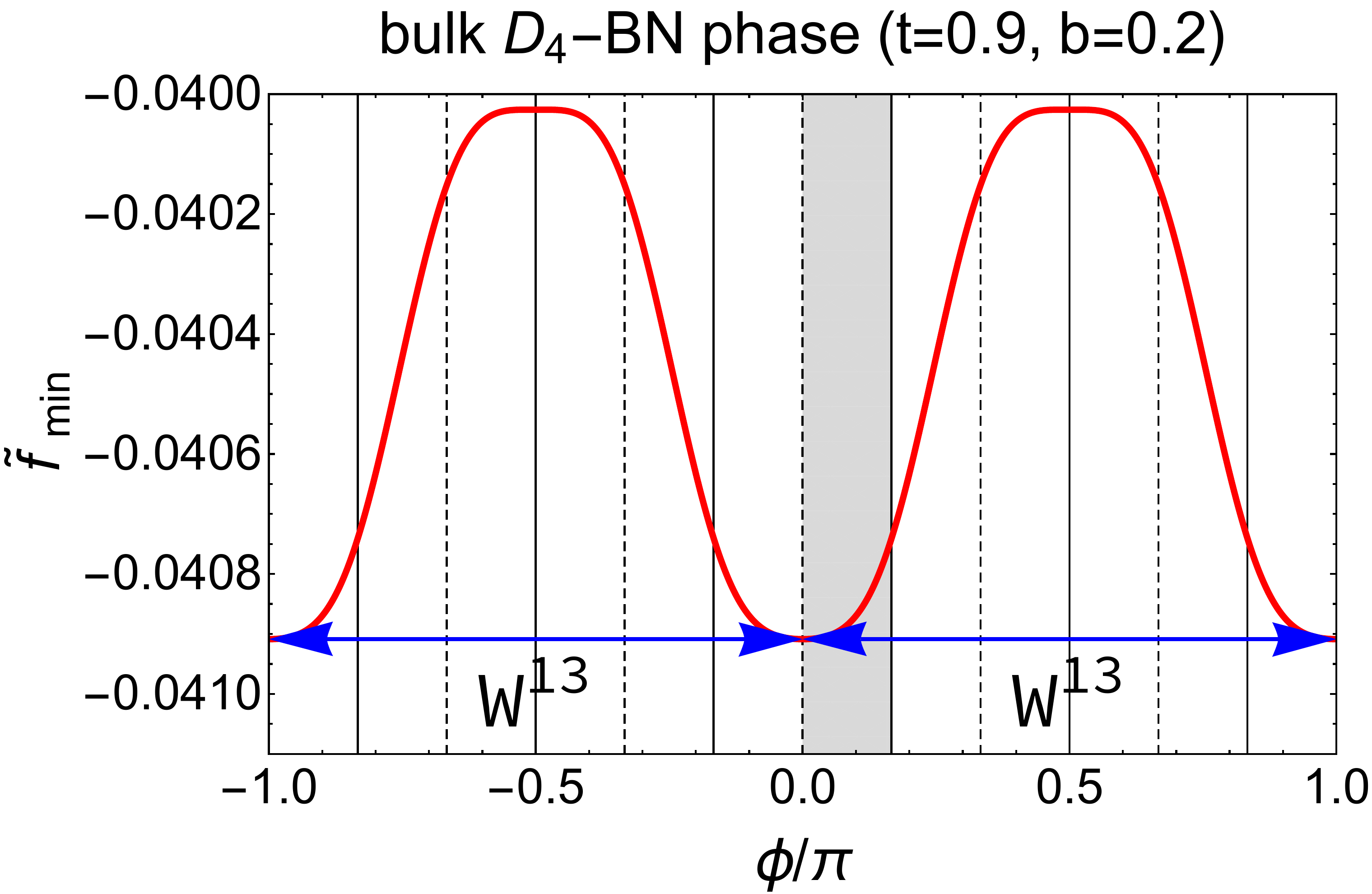}
\caption{The GL free energy $\tilde{f}_{\min}=\tilde{f}_{\min}(\phi)$ as functions of the angle $\phi$. The gray regions ($0\le\phi/\pi<1/6$) corresponds to the diagonal form parametrized in the right-hand side of Eq.~\eqref{eq:A_matrix_0}. The arrows indicate the domain walls connecting the different minima. The horizontal solid and dashed lines indicate the angle for the UN and D$_{4}$-BN phases, respectively, and the regions between them correspond to the D$_{2}$-BN phase.}
\label{fig:alpha_UN}
\end{center}
\end{figure}

For convenience in the analysis, instead of Eq.~\eqref{eq:A_matrix_0}, we may express the order parameter in terms of $f_{0}$ and $\phi$,
\begin{eqnarray}
   f_{1} = \biggl( \frac{\cos\phi}{\sqrt{2}} - \frac{\sin\phi}{\sqrt{6}} \biggr) f_{0}, \quad
   f_{2} = \sqrt{\frac{2}{3}} \bigl( \sin\phi \bigr) f_{0}, \quad
   f_{3} = \biggl( -\frac{\cos\phi}{\sqrt{2}} - \frac{\sin\phi}{\sqrt{6}} \biggr) f_{0},
   \label{eq:alpha_f_0}
\end{eqnarray}
by dropping $\tilde{d}$ and $\vec{n}$ in Eq.~\eqref{eq:alpha_f}.
This is the parametrization that the three-dimensional vector $\vec{f}\equiv(f_{1},f_{2},f_{3})$ is confined on the plane by $f_{1}+f_{2}+f_{3}=0$, and $\phi$ is the rotation angle on this plane.
We notice that the restriction of the range for $\phi$ to $0 \le \phi \le \pi/6$ recovers the parametrization in the diagonal form in Eq.~\eqref{eq:A_matrix_0}.
We show the GL potential 
 with the coordinate $(f_{0}\cos\phi,f_{0}\sin\phi)$ in Fig.~\ref{fig:GL_potential}, where we set the temperature $t=0.9$ and the magnetic field $b=0$ (the UN phase), $b=0.1$ (the D$_{2}$-BN phase), and $b=0.2$ (the D$_{4}$-BN phase).
In each phase, there are several degenerate states in the ground state: six degenerate states in the UN phase, four degenerate states in the D$_{2}$-BN phase, and two degenerate states in the D$_{4}$-BN phase.
Those degenerate states are more clearly seen by defining the GL free energy only with $\phi$,
\begin{eqnarray}
 \tilde{f}_{\min}(\phi) = \min_{f_{0} \ge 0} \tilde{f}(f_{0},\phi),
 \label{eq:fmin}
\end{eqnarray}
where the right-hand side means that a minimum value of $\tilde{f}(f_{0},\phi)$ is chosen in the variational calculation with respect to $f_{0}$.
We show the result of $\tilde{f}_{\min}(\phi)$ in Fig.~\ref{fig:alpha_UN}.
We confirm that the degenerate states: $\phi/\pi=\pm1/6$, $\pm1/2$, $\pm5/6$ in the UN phase;
$\phi/\pi=\pm0.129$, $\pm0.870$ in the D$_{2}$-BN phase;
and $\phi/\pi=0$, $1$ in the D$_{4}$-BN phase.

%%%%%%%%%%%%%%%%%%%%%%%%%%%%%%
%%%%%%%%%%%%%%%%%%%%%%%%%%%%%%

%%%%%%%%%%%%%%%%%%%%%%%%%%%%%%
\bigskip
%%%%%%%%%%%%%%%%%%%%%%%%%%%%%%

\subsection{Domain walls: $W^{\alpha}_{i}(\mathrm{UN})$, $W^{\alpha}_{i}(\mathrm{D}_{2}\mathrm{BN})$, and $W^{\alpha}_{i}(\mathrm{D}_{4}\mathrm{BN})$}

We consider the domain walls that connect the vacua in the GL free energy, {\it i.e.}, $\tilde{A}(+\infty;\vec{n})$ and $\tilde{A}(-\infty;\vec{n})$, where $\tilde{A}(\pm\infty;\vec{n})$ are the different (dimensionless) condensates in the degenerate ground states in the bulk space at $\tilde{d} \rightarrow \pm\infty$, respectively, along the line with the direction $\vec{n}$ [see Eq.~\eqref{eq:A_matrix_diagonal}, and also Figs.~\ref{fig:GL_potential} and~\ref{fig:alpha_UN}].
We denote the domain wall by $W^{\alpha}$ where $\alpha$ indicates the label to classify the domain walls, 
and explain our definitions of $W^{\alpha}$ in each bulk phase in the followings.

% UN
In the bulk UN phase (relevant for zero magnetic field),
we consider the domain walls $W^{\alpha}$ ($\alpha=1$, $2$, $3$) which connect the neighboring degenerate states in the angles $(\phi\,\mathrm{mod}\,2\pi)/\pi=-1/6$, $1/6$, $1/2$, $5/6$, $7/6$, $3/2$, and $11/2=-1/6$ and denote them in the following way:
\begin{eqnarray}
W^{1}(\mathrm{UN}):
\frac{f_{0}}{\sqrt{6}}
\left(
\begin{array}{ccc}
 1 & 0 & 0 \\
 0 & 1 & 0 \\
 0 & 0 & -2 
\end{array}
\right)
\longleftrightarrow
\frac{f_{0}}{\sqrt{6}}
\left(
\begin{array}{ccc}
 -1 & 0 & 0 \\
 0 & -1 & 0 \\
 0 & 0 & 2 
\end{array}
\right),
\quad %%%
\frac{f_{0}}{\sqrt{6}}
\left(
\begin{array}{ccc}
 -1 & 0 & 0 \\
 0 & -1 & 0 \\
 0 & 0 & 2 
\end{array}
\right)
\longleftrightarrow
\frac{f_{0}}{\sqrt{6}}
\left(
\begin{array}{ccc}
 1 & 0 & 0 \\
 0 & -2 & 0 \\
 0 & 0 & 1 
\end{array}
\right),
\label{eq:bc_WUN1}
\end{eqnarray}
between $(\phi\,\mathrm{mod}\,2\pi)/\pi=1/6$ and $1/2$ and between $(\phi\,\mathrm{mod}\,2\pi)/\pi=7/6$ and $3/2$, respectively;
%%%
\begin{eqnarray}
W^{2}(\mathrm{UN}):
\frac{f_{0}}{\sqrt{6}}
\left(
\begin{array}{ccc}
 2 & 0 & 0 \\
 0 & -1 & 0 \\
 0 & 0 & -1 
\end{array}
\right)
\longleftrightarrow
\frac{f_{0}}{\sqrt{6}}
\left(
\begin{array}{ccc}
 1 & 0 & 0 \\
 0 & 1 & 0 \\
 0 & 0 & -2 
\end{array}
\right),
\quad %%%
\frac{f_{0}}{\sqrt{6}}
\left(
\begin{array}{ccc}
 -2 & 0 & 0 \\
 0 & 1 & 0 \\
 0 & 0 & 1 
\end{array}
\right)
\longleftrightarrow
\frac{f_{0}}{\sqrt{6}}
\left(
\begin{array}{ccc}
 -1 & 0 & 0 \\
 0 & -1 & 0 \\
 0 & 0 & 2 
\end{array}
\right),
\label{eq:bc_WUN2}
\end{eqnarray}
between $(\phi\,\mathrm{mod}\,2\pi)/\pi=1/6$ and $1/2$, and between $(\phi\,\mathrm{mod}\,2\pi)/\pi=7/6$ and $3/2$, respectively;
and
%%%
\begin{eqnarray}
W^{3}(\mathrm{UN}):
\frac{f_{0}}{\sqrt{6}}
\left(
\begin{array}{ccc}
 -1 & 0 & 0 \\
 0 & 2 & 0 \\
 0 & 0 & -1 
\end{array}
\right)
\longleftrightarrow
\frac{f_{0}}{\sqrt{6}}
\left(
\begin{array}{ccc}
 -2 & 0 & 0 \\
 0 & 1 & 0 \\
 0 & 0 & 1 
\end{array}
\right),
\quad %%%
\frac{f_{0}}{\sqrt{6}}
\left(
\begin{array}{ccc}
 1 & 0 & 0 \\
 0 & -2 & 0 \\
 0 & 0 & 1 
\end{array}
\right)
\longleftrightarrow
\frac{f_{0}}{\sqrt{6}}
\left(
\begin{array}{ccc}
 2 & 0 & 0 \\
 0 & -1 & 0 \\
 0 & 0 & -1 
\end{array}
\right),
\label{eq:bc_WUN3}
\end{eqnarray}
between $(\phi\,\mathrm{mod}\,2\pi)/\pi=1/2$ and $5/6$ and between $(\phi\,\mathrm{mod}\,2\pi)/\pi=3/2$ and $11/6=-1/6$, respectively.
We notice that $W^{\alpha}$ ($\alpha=$1, $2$, $3$) leaves the $\alpha$th diagonal components unchanged and exchange the other two diagonal components up to the overall minus sign.

% D2BN
In the bulk D$_{2}$-BN phase in setting $b=0.1$,
we consider the domain walls connecting the neighboring degenerate states in the angles $(\phi\,\mathrm{mod}\,2\pi)/\pi=0.129$, $0.870$, $1.129$, and $1.870$ and denote them in the following way\footnote{We notice that different values for $\phi$ in the degenerate states will be realized for different strengths of the magnetic field in the D$_{2}$-BN phase. In contrast, the values of $\phi$ in the degenerate states in the UN and D$_{4}$-BN phases have no dependence on the magnetic field.}:
\begin{eqnarray}
W^{13}(\mathrm{D}_{2}\mathrm{BN}): \,
&&
f_{0}
\left(
\begin{array}{ccc}
 0.506 & 0 & 0 \\
 0 & 0.302 & 0 \\
 0 & 0 & -0.808
\end{array}
\right)
\longleftrightarrow
f_{0}
\left(
\begin{array}{ccc}
 -0.808 & 0 & 0 \\
 0 & 0.302 & 0 \\
 0 & 0 & 0.506 
\end{array}
\right), \nonumber \\
\quad %%%
&&
f_{0}
\left(
\begin{array}{ccc}
 -0.506 & 0 & 0 \\
 0 & -0.302 & 0 \\
 0 & 0 & 0.808
\end{array}
\right)
\longleftrightarrow
f_{0}
\left(
\begin{array}{ccc}
 0.808 & 0 & 0 \\
 0 & -0.302 & 0 \\
 0 & 0 & -0.506 
\end{array}
\right),
\label{eq:bc_WD213}
\end{eqnarray}
between $(\phi\,\mathrm{mod}\,2\pi)/\pi=0.120$ and $0.870$ and between $(\phi\,\mathrm{mod}\,2\pi)/\pi=1.120$ and $1.870$, respectively,
and
\begin{eqnarray}
W^{2}(\mathrm{D}_{2}\mathrm{BN}): \,
&&
f_{0}
\left(
\begin{array}{ccc}
 0.808 & 0 & 0 \\
 0 & -0.302 & 0 \\
 0 & 0 & -0.506
\end{array}
\right)
\longleftrightarrow
f_{0}
\left(
\begin{array}{ccc}
 0.506 & 0 & 0 \\
 0 & 0.302 & 0 \\
 0 & 0 & -0.808
\end{array}
\right), \nonumber \\
\quad %%%
&&
f_{0}
\left(
\begin{array}{ccc}
 -0.808 & 0 & 0 \\
 0 & 0.302 & 0 \\
 0 & 0 & 0.506 
\end{array}
\right)
\longleftrightarrow
f_{0}
\left(
\begin{array}{ccc}
 -0.506 & 0 & 0 \\
 0 & -0.302 & 0 \\
 0 & 0 & 0.808
\end{array}
\right),
\label{eq:bc_WD22}
\end{eqnarray}
between $(\phi\,\mathrm{mod}\,2\pi)/\pi=-0.120$ and $0.120$, and between $(\phi\,\mathrm{mod}\,2\pi)/\pi=0.870$ and $1.120$, respectively.
We notice that $W^{\alpha}(\mathrm{D}_{2}\mathrm{BN})$ ($\alpha=13$ and $2$) leaves the second diagonal components unchanged and exchange the other two diagonal components up to the overall minus sign.

% D4BN
In the bulk D$_{4}$-BN phase,
we consider the domain walls connecting the neighboring degenerate states in the angles $(\phi\,\mathrm{mod}\,2\pi)/\pi=0$ and $1$ and denote them in the following way:
\begin{eqnarray}
W^{13}(\mathrm{D}_{4}\mathrm{BN}): \,
\frac{f_{0}}{\sqrt{2}}
\left(
\begin{array}{ccc}
 1 & 0 & 0 \\
 0 & 0 & 0 \\
 0 & 0 & -1
\end{array}
\right)
\longleftrightarrow
\frac{f_{0}}{\sqrt{2}}
\left(
\begin{array}{ccc}
 -1 & 0 & 0 \\
 0 & 0 & 0 \\
 0 & 0 & 1 
\end{array}
\right),
\quad
&&
\frac{f_{0}}{\sqrt{2}}
\left(
\begin{array}{ccc}
 -1 & 0 & 0 \\
 0 & 0 & 0 \\
 0 & 0 & 1 
\end{array}
\right)
\longleftrightarrow
\frac{f_{0}}{\sqrt{2}}
\left(
\begin{array}{ccc}
 1 & 0 & 0 \\
 0 & 0 & 0 \\
 0 & 0 & -1
\end{array}
\right),
\label{eq:bc_WD413}
\end{eqnarray}
between $(\phi\,\mathrm{mod}\,2\pi)/\pi=0$ and $1$, and between $(\phi\,\mathrm{mod}\,2\pi)/\pi=1$ and $2$, respectively.
$W^{13}(\mathrm{D}_{4}\mathrm{BN})$ leaves the second diagonal components unchanged and exchanges the other two diagonal components.

Several comments are in order. In the above definitions, we remark that 
$W^{\alpha}(\mathrm{UN})$ ($\alpha=1$, $2$, $3$), $W^{\alpha}(\mathrm{D}_{2}\mathrm{BN})$ ($\alpha=13$, $2$), and $W^{\alpha}(\mathrm{D}_{4}\mathrm{BN})$ ($\alpha=13$) leave the (absolutely) minimum components unchanged up to the minus sign at the left and right infinities: 
$\pm f_{0}/\sqrt{6}$ in the case of $W^{\alpha}(\mathrm{UN})$ ($\alpha=1$, $2$, $3$), $\pm0.302 f_{0}$ in the case of $W^{\alpha}(\mathrm{D}_{2}\mathrm{BN})$ ($\alpha=13$, $2$), and $0$ in the case of $W^{\alpha}(\mathrm{D}_{4}\mathrm{BN})$ ($\alpha=13$).
We also notice that the domain wall configuration passes through various different phases. To see more details, in Fig.~\ref{fig:alpha_UN}, we show the angles $\phi/\pi=1/6+(3/2)n$ corresponding to the UN phase by the horizontal solid lines, and the angles $\phi/\pi=(3/2)n$ corresponding to the D$_{4}$-BN phase by the horizontal dashed lines ($n$ an integer).
The regions between the solid lines and the dashed lines correspond to the D$_{2}$-BN phase.
From them, we then find that the domain walls go across several phases as follows:
For example, the domain wall $W^{\alpha}$ ($\alpha=1$, $2$, $3$) connecting the two vacua in the bulk UN phase pass through the D$_{2}$-BN phase and the D$_{4}$-BN phase in the following order:
\begin{eqnarray}
 W^{\alpha}(\mathrm{UN}):
 \mathrm{UN} \!\cdot\! 
 \mathrm{D}_{2}\mathrm{BN} \!\cdot\! 
 \mathrm{D}_{4}\mathrm{BN} \!\cdot\! 
 \mathrm{D}_{2}\mathrm{BN} \!\cdot\! 
 \mathrm{UN},
\end{eqnarray}
where the boundary is fixed by the UN phase by definition.
In the bulk D$_{2}$-BN phase,
$W^{2}$(D$_{2}$BN) passes through only the D$_{4}$-BN phase, 
while $W^{13}$(D$_{2}$BN) passes through the UN and D$_{4}$-BN phases in the following order:
\begin{eqnarray}
 W^{2}(\mathrm{D}_{2}\mathrm{BN}):&&
  \mathrm{D}_{2}\mathrm{BN} \!\cdot\! 
  \mathrm{D}_{4}\mathrm{BN} \!\cdot\! 
  \mathrm{D}_{2}\mathrm{BN},\\  
 W^{13}(\mathrm{D}_{2}\mathrm{BN}):&&
  \mathrm{D}_{2}\mathrm{BN} \!\cdot\! 
  \mathrm{UN} \!\cdot\! 
  \mathrm{D}_{2}\mathrm{BN} \!\cdot\! 
  \mathrm{D}_{4}\mathrm{BN} \!\cdot\! 
  \mathrm{D}_{2}\mathrm{BN} \!\cdot\! 
  \mathrm{UN} \!\cdot\! 
  \mathrm{D}_{2}\mathrm{BN} \!\cdot\! 
  \mathrm{D}_{4}\mathrm{BN} \!\cdot\! 
  \mathrm{D}_{2}\mathrm{BN} \!\cdot\! 
  \mathrm{UN} \!\cdot\! 
  \mathrm{D}_{2}\mathrm{BN},
\end{eqnarray}
where the boundary is fixed by the D$_{2}$-BN phase by definition.
Finally, in the bulk D$_{4}$-BN phase,
$W^{13}$(D$_{4}$BN) passes through the UN and D$_{2}$-BN phases in the following order 
\begin{eqnarray}
  W^{13}(\mathrm{D}_{4}\mathrm{BN}):
  \mathrm{D}_{4}\mathrm{BN} \!\cdot\! 
  \mathrm{D}_{2}\mathrm{BN} \!\cdot\! 
  \mathrm{UN} \!\cdot\! 
  \mathrm{D}_{2}\mathrm{BN} \!\cdot\! 
  \mathrm{D}_{4}\mathrm{BN} \!\cdot\! 
  \mathrm{D}_{2}\mathrm{BN} \!\cdot\! 
  \mathrm{UN} \!\cdot\! 
  \mathrm{D}_{2}\mathrm{BN} \!\cdot\! 
  \mathrm{D}_{4}\mathrm{BN} \!\cdot\! 
  \mathrm{D}_{2}\mathrm{BN} \!\cdot\! 
  \mathrm{UN} \!\cdot\! 
  \mathrm{D}_{2}\mathrm{BN} \!\cdot\! 
  \mathrm{D}_{4}\mathrm{BN},
\end{eqnarray}
where the boundary is fixed by the D$_{4}$-BN phase by definition.
Accompanied by those internal phases, the number of the Nambu-Goldstone modes induced by the symmetry breaking can change locally at each internal phase inside the domain walls.\footnote{See e.g. Ref.~\cite{Masuda:2015jka} for the Nambu-Goldstone modes appearing in the UN, D$_{2}$-BN, and D$_{4}$-BN phases.}
In particular, the UN phase has less number of the NG modes. 
Therefore,  the domain wall  $W^{\alpha}(\mathrm{UN})$ in the UN phase contains one NG mode localized in its vicinity. 
On the other hand, 
 the continuous U(1) symmetry that the UN phase preserves restores
in the cores of the domain walls 
 $W^{13}(\mathrm{D}_{2}\mathrm{BN})$ and 
 $W^{13}(\mathrm{D}_{4}\mathrm{BN})$  in the D$_{2}$-BN and D$_{4}$-BN phases.

\begin{figure}[tb]
\begin{center}
\includegraphics[scale=0.3]{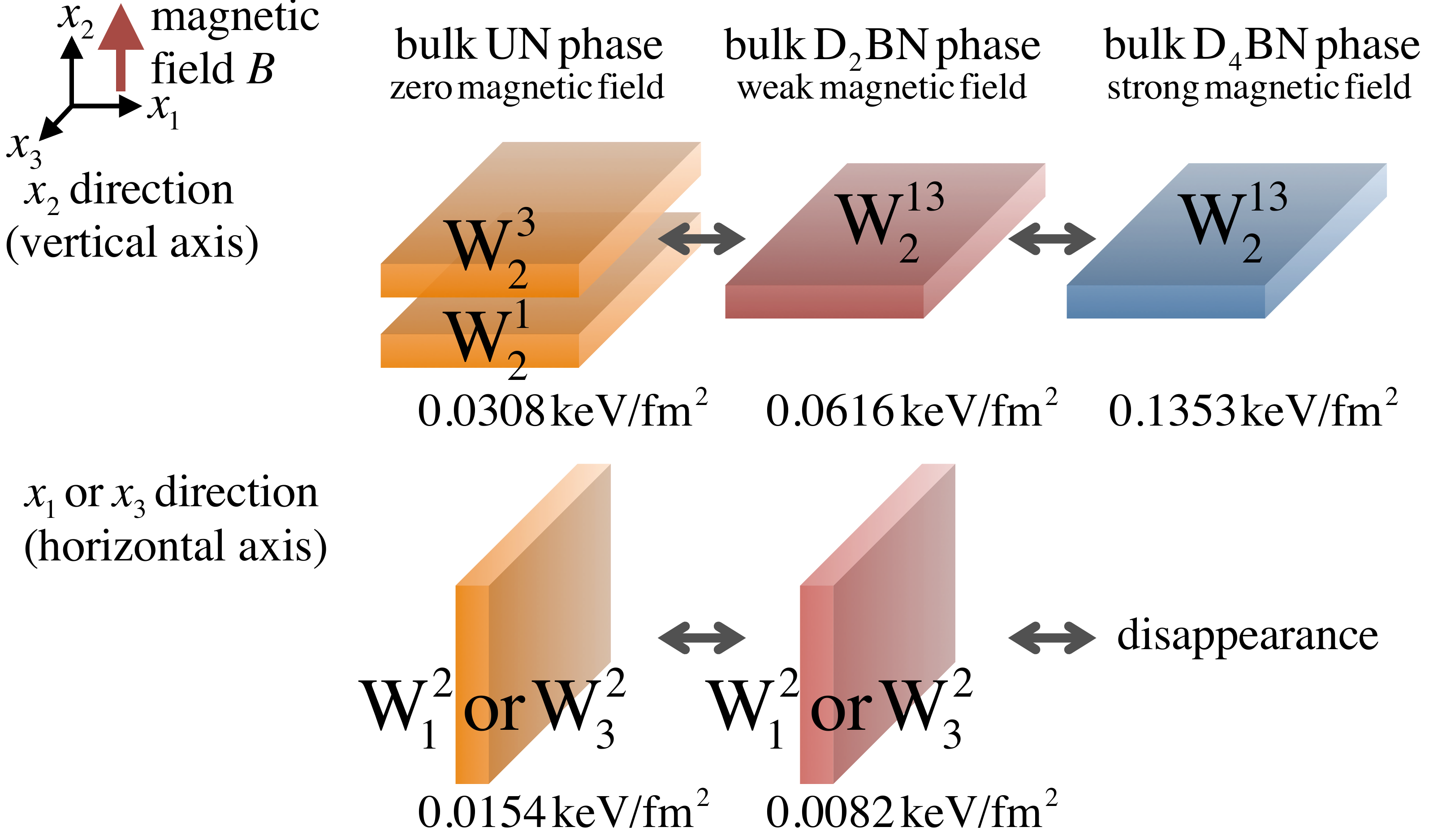}
\caption{The metamorphism of the domain walls with the fixed directions, with depending on the strength of the magnetic field. The magnetic fields in the D$_{2}$-BN and D$_{4}$-BN phases ($b=0.1$ and $0.2$) are applied to the $x_{2}$ (vertical) direction.}
\label{fig:Fig_190709}
\end{center}
\end{figure}

When a magnetic field $b$ is continuously changed, 
there happen phase transitions at some magnetic fields, 
among the UN phase, 
the D$_{2}$-BN phase, 
and the D$_{4}$-BN phase, 
 as shown in Fig.~\ref{fig:phase_diagram}.
Here, we discuss how the domain walls in each phase is connected to those of the other phases, see Fig.~\ref{fig:alpha_UN}.
When we gradually increase a magnetic field from zero to some finite value, the phase changes from the UN to D$_2$-BN phase.
In this process, the two domain walls $W^{1}(\mathrm{UN})$ and $W^{3}(\mathrm{UN})$ in the UN phase are bound together as a coalescence 
to become a single domain wall $W^{13}(\mathrm{D}_{2}\mathrm{BN})$ in the D$_{2}$-BN phase: $W^{1}+W^{3} \rightarrow W^{13}$.
This is because the potential minimum at $(\phi\,\mathrm{mod}\,2\pi)/\pi=\pm1/2$ existing between the two walls 
$W^{1}(\mathrm{UN})$ and $W^{3}(\mathrm{UN})$ is lifted in the D$_{2}$-BN phase 
and they are confined.
Therefore, the domain wall $W^{13}(\mathrm{D}_{2}\mathrm{BN})$ can be regarded as a composite domain wall.
If we further increase the magnetic field so that the phase becomes the D$_{4}$-BN phase, 
then the domain wall $W^{13}(\mathrm{D}_{2}\mathrm{BN})$ is changed to $W^{13}(\mathrm{D}_{4}\mathrm{BN})$, 
which can be regarded as a genuine elementary domain wall since the local energy minimum completely disappears finally in the D$_{4}$-BN phase.
On the other hand, returning back the zero magnetic field, 
the domain wall $W^{2}(\mathrm{UN})$ in the UN phase is transformed to the domain wall $W^{2}(\mathrm{D}_{2}\mathrm{BN})$ in the D$_{2}$-BN phase without causing coalescence or fragmentation of domain walls.
Further increasing the magnetic field to the D$_{4}$-BN phase, 
it disappears completely in the D$_{4}$-BN phase because the two degenerate ground states connected by $W^{2}(\mathrm{D}_{2}\mathrm{BN})$ merge to one ground state $(\phi\,\mathrm{mod}\,2\pi)/\pi=0$ or $1$ in the D$_{4}$-BN phase.
 In Fig.~\ref{fig:Fig_190709},
we summarize a part of the metamorphism of domain walls under the change of the strength of the magnetic field.

%%%%%%%%%%%%%%%%%%%%%%%%%%%%%%
%%%%%%%%%%%%%%%%%%%%%%%%%%%%%%

We consider that the direction of the domain walls is defined by the normal vector perpendicular to the surface.
In order to show the simple example, we consider that the domain walls are directed in the three special cases: (i) the $x_{1}$ direction, {\it i.e.}, $\vec{n}=(1,0,0)$, $(\theta,\varphi)=(\pi/2,0)$; (ii) the $x_{2}$ direction, {\it i.e.}, $\vec{n}=(0,1,0)$, $(\theta,\varphi)=(\pi/2,\pi/2)$; and  (iii) the $x_{3}$ direction {\it i.e.}, $\vec{n}=(0,0,1)$, $(\theta,\varphi)=(0,0)$.
At the end of the discussion, we will investigate the stability of the domain walls against small changes of the directions.
When we need to specify the directions of the domain wall, we use the notation $W^{\alpha}_{i}$ for $i=1$, $2$, and $3$ for the $x_{1}$, $x_{2}$, and $x_{3}$ directions: $W^{\alpha}_{i}$ ($\alpha=1$, $2$, $3$) in the UN phase, $W^{\alpha}_{i}$ ($\alpha=13$, $2$) in the D$_{2}$-BN phase, and $W^{\alpha}_{i}$ ($\alpha=13$) in the D$_{4}$-BN phase.

\subsection{Configurations of domain walls}

%%%%%%%%%%%%%%%%%%%%%%%%%%%%%%
\renewcommand{\arraystretch}{1.15}
\begin{table}[tb]
\begin{center}
\begin{tabular}{|c|c|c|c|c|c|c|c|c|c|}
\hline
bulk UN phase & \multicolumn{3}{|c|}{$W^{2}$(UN)} & \multicolumn{3}{|c|}{$W^{1}$(UN)} & \multicolumn{3}{|c|}{$W^{3}$(UN)} \\
\hline
   angle & \multicolumn{3}{|c|}{$-1/6\le(\phi\,\mathrm{mod}\,\pi)/\pi<1/6$} & \multicolumn{3}{|c|}{$1/6\le(\phi\,\mathrm{mod}\,\pi)/\pi<1/2$} & \multicolumn{3}{|c|}{$1/2\le(\phi\,\mathrm{mod}\,\pi)/\pi<5/6$} \\
\hline
 direction & $W^{2}_{1}$ & $W^{2}_{2}$ & $W^{2}_{3}$ & $W^{1}_{1}$ & $W^{1}_{2}$ & $W^{1}_{3}$ & $W^{3}_{1}$ & $W^{3}_{2}$ & $W^{3}_{3}$ \\
\hline
 $\sigma$ [keV/fm$^{2}$] & 0.0154 & 0.0199 & 0.0154 & 0.0199 & 0.0154 & 0.0154 & 0.0154 & 0.0154 & 0.0199 \\
\hline
\hline
bulk D$_{2}$-BN phase & \multicolumn{3}{|c|}{$W^{2}$(D$_{2}$BN)} & \multicolumn{6}{|c|}{$W^{13}$(D$_{2}$BN)} \\
\hline
   angle & \multicolumn{3}{|c|}{$-0.129\le(\phi\,\mathrm{mod}\,\pi)/\pi<0.129$} & \multicolumn{6}{|c|}{$0.129\le(\phi\,\mathrm{mod}\,\pi)/\pi<0.870$} \\
\hline
 direction & $W^{2}_{1}$ & $W^{2}_{2}$ & $W^{2}_{3}$ & \multicolumn{2}{|c|}{$W^{13}_{1}$} & \multicolumn{2}{|c|}{$W^{13}_{2}$} & \multicolumn{2}{|c|}{$W^{13}_{3}$} \\
\hline
 $\sigma$ [keV/fm$^{2}$] & 0.0082 & 0.0107 & 0.0082 & \multicolumn{2}{|c|}{0.0722} & \multicolumn{2}{|c|}{0.0616} & \multicolumn{2}{|c|}{0.0722} \\
\hline
\hline
bulk D$_{4}$-BN phase & \multicolumn{3}{|c|}{---} & \multicolumn{6}{|c|}{$W^{13}$(D$_{4}$BN)} \\
\hline
   angle & \multicolumn{3}{|c|}{---} & \multicolumn{6}{|c|}{$0\le(\phi\,\mathrm{mod}\,\pi)/\pi<1$} \\
\hline
 direction & \multicolumn{3}{|c|}{---} & \multicolumn{2}{|c|}{$W^{13}_{1}$} & \multicolumn{2}{|c|}{$W^{13}_{2}$} & \multicolumn{2}{|c|}{$W^{13}_{3}$} \\
\hline
 $\sigma$ [keV/fm$^{2}$] & \multicolumn{3}{|c|}{---} & \multicolumn{2}{|c|}{0.1533} & \multicolumn{2}{|c|}{0.1353} & \multicolumn{2}{|c|}{0.1533} \\
\hline
\end{tabular}
\end{center}
\caption{The surface energy densities by the domain walls with the directions along $x_{1}$, $x_{2}$, and $x_{3}$ directions. $\phi$ is defined in Eq.~\eqref{eq:alpha_f_0} (cf.~Fig. \ref{fig:alpha_UN}). The bulk UN, D$_{2}$-BN, and D$_{4}$-BN phases have $(t,b)=(0.9,0)$, $(0.9,0.1)$, and $(0.9,0.2)$, respectively.}
\label{table:surface_energy_dw_v2}
\end{table}%
\renewcommand{\arraystretch}{1}
%%%%%%%%%%%%%%%%%%%%%%%%%%%%%%

For the domain walls defined in the previous subsection, 
we consider their configurations in the three dimensional space, and calculate the energy per unit area (surface tension) on the surface of the domain walls.
We solve the EL equation \eqref{eq:EL_f} for the domain walls by introducing the boundary conditions in the bulk space.
The boundary conditions are given in Eqs.~\eqref{eq:bc_WUN1}, \eqref{eq:bc_WUN2}, and \eqref{eq:bc_WUN3} in the bulk UN phase, and Eqs.~\eqref{eq:bc_WD213}, \eqref{eq:bc_WD22} in the bulk D$_{2}$-BN phase, and Eq.~\eqref{eq:bc_WD413} in the bulk D$_{4}$-BN phase.
For the analysis, we consider the three-dimensional vector as functions of $\tilde{d}$ for a fixed $\vec{n}$,
\begin{eqnarray}
   \vec{f}(\tilde{d}) \equiv \big(f_{1}(\tilde{d};\vec{n}),f_{2}(\tilde{d};\vec{n}),f_{3}(\tilde{d};\vec{n})\bigr),
\end{eqnarray}
whose components have been introduced in Eq.~\eqref{eq:A_matrix_diagonal} in the three-dimensional space with the coordinate $(\tilde{x}_{1},\tilde{x}_{2},\tilde{x}_{3})$.
The schematic figures are presented in the bottom panels in Fig.~\ref{fig:GL_potential}.
The numerical results of $\vec{f}(\tilde{d})$ are shown in Figs.~\ref{fig:190521_soliton_3P2_f1f2_UN}, \ref{fig:190521_soliton_3P2_f1f2_D2BN}, and \ref{fig:190521_soliton_3P2_f1f2_D4BN} in Appendix~\ref{sec:plots_configuration_DW}.

With the solutions of $\vec{f}(\tilde{d})$, we consider the surface energy density per unit area of the domain walls surface directed along $\vec{n}$, which are expressed as
\begin{eqnarray}
   \sigma(\vec{n})
\equiv
   \int_{-\infty}^{\infty}
   \bigl( f(d;\vec{n}) - f_{\bulk} \bigr)
   \mathrm{d}d,
\label{eq:surface_energy_density}
\end{eqnarray}
for with the GL free energy densities $f(d;\vec{n})$ in the domain.
Here $f_{\bulk}$ is the GL free energy density in the bulk space ($|d| \rightarrow \infty$).\footnote{For the convenience of the calculation, we can express the surface energy density as
\begin{eqnarray}
   \sigma(\vec{n})
= \frac{p_{F}^{2}T_{c0}}{2\pi^{2}}
   \tilde{\sigma}(\vec{n}), 
\end{eqnarray}
where we have defined the dimensionless surface energy density by
\begin{eqnarray}
   \tilde{\sigma}(\vec{n})
\equiv
   \int_{-\infty}^{\infty}
   \bigl( \tilde{f}(\tilde{d};\vec{n}) - \tilde{f}_{\bulk} \bigr)
   \mathrm{d}\tilde{d},
\end{eqnarray}
for the dimensionless GL free energy densities, $\tilde{f}(\tilde{d};\vec{n})$ in the domain wall, and $\tilde{f}_{\bulk}$ in the bulk space ($|\tilde{d}| \rightarrow \infty$).}
We show the numerical results of the surface energy density for $W^{\alpha}_{i}(\mathrm{UN})$ with $\alpha=1$, $2$, $3$, $W^{\alpha}_{i}(\mathrm{D}_{2}\mathrm{BN})$ with $\alpha=13$, $2$, and $W^{13}_{i}(\mathrm{D}_{4}\mathrm{BN})$ ($i=1$, $2$, $3$) in Table~\ref{table:surface_energy_dw_v2}.
From the table, we observe the following properties of the domain walls, depending on the strengths of the magnetic field:
\begin{itemize}
% bulk UN
\item In the zero magnetic field, {\it i.e.}, in the bulk UN phase, the domain walls $W^{\alpha}_{i}(\mathrm{UN})$ with the condition $\alpha \neq i$ satisfied are the most stable ones, while $W^{\alpha}_{i}(\mathrm{UN})$ with $\alpha = i$ have higher energy. 
The directions $x_{1}$, $x_{2}$, and $x_{3}$ of the domain walls 
yield essentially the same results 
by cyclic transformations, because the rotational symmetry exists in the absence of a  magnetic field.
% bulk D2BN
\item In the weak magnetic fields, {\it i.e.}, in the bulk D$_{2}$-BN phase, 
the domain walls in the bulk UN phase are transformed as follows: 
the domain walls $W^{1}_{i}$ and $W^{3}_{i}$ in the bulk UN phase merge to the single domain wall $W^{13}_{i}$ in the bulk D$_{2}$-BN phase: $W^{1}_{i}+W^{3}_{i} \rightarrow W^{13}_{i}$, 
while the domain wall $W^{2}_{i}$ in the bulk UN phase becomes the one $W^{2}_{i}$ in the bulk D$_{2}$-BN phase.
We notice that, in the bulk D$_{2}$-BN phase, $W^{13}_{i}$ has higher energy than $W^{2}_{i}$, 
because of the composite nature of $W^{13}_{i}$.
% bulk D4BN
\item In the strong magnetic fields, {\it i.e.}, in the bulk D$_{4}$-BN phase, 
the domain wall $W^{2}_{i}$ in the D$_2$-BN phase disappears while the domain wall $W^{13}_{i}$ survives %
to become an elementary domain wall, which is the unique domain wall in the D$_4$-BN phase.  
\end{itemize}

%%%%%%%%%%%%%%%%%%%%%%%%%%%%%%
\begin{figure}[tb]
\begin{center}
\includegraphics[scale=0.2]{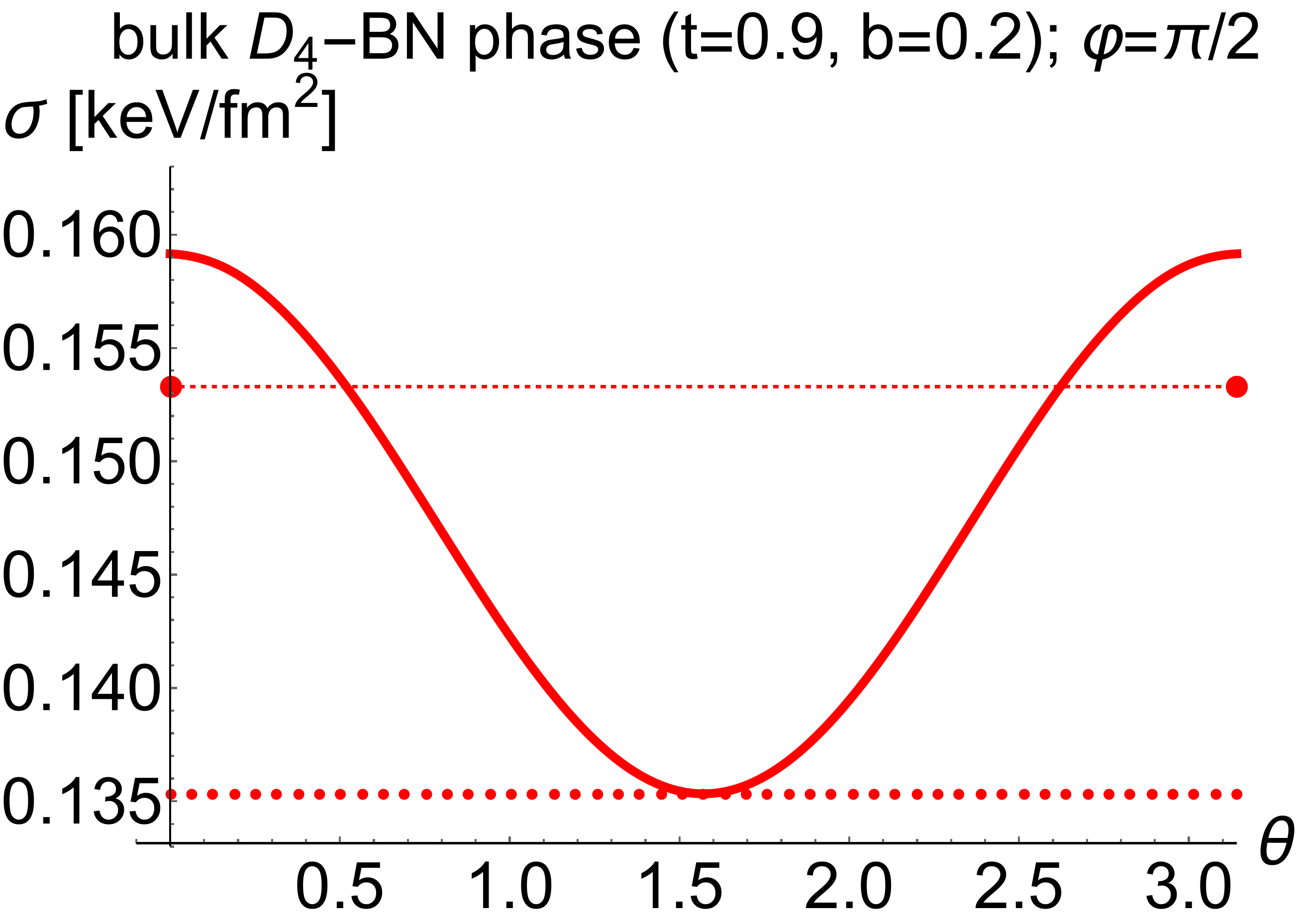}
\hspace{3em}
\includegraphics[scale=0.2]{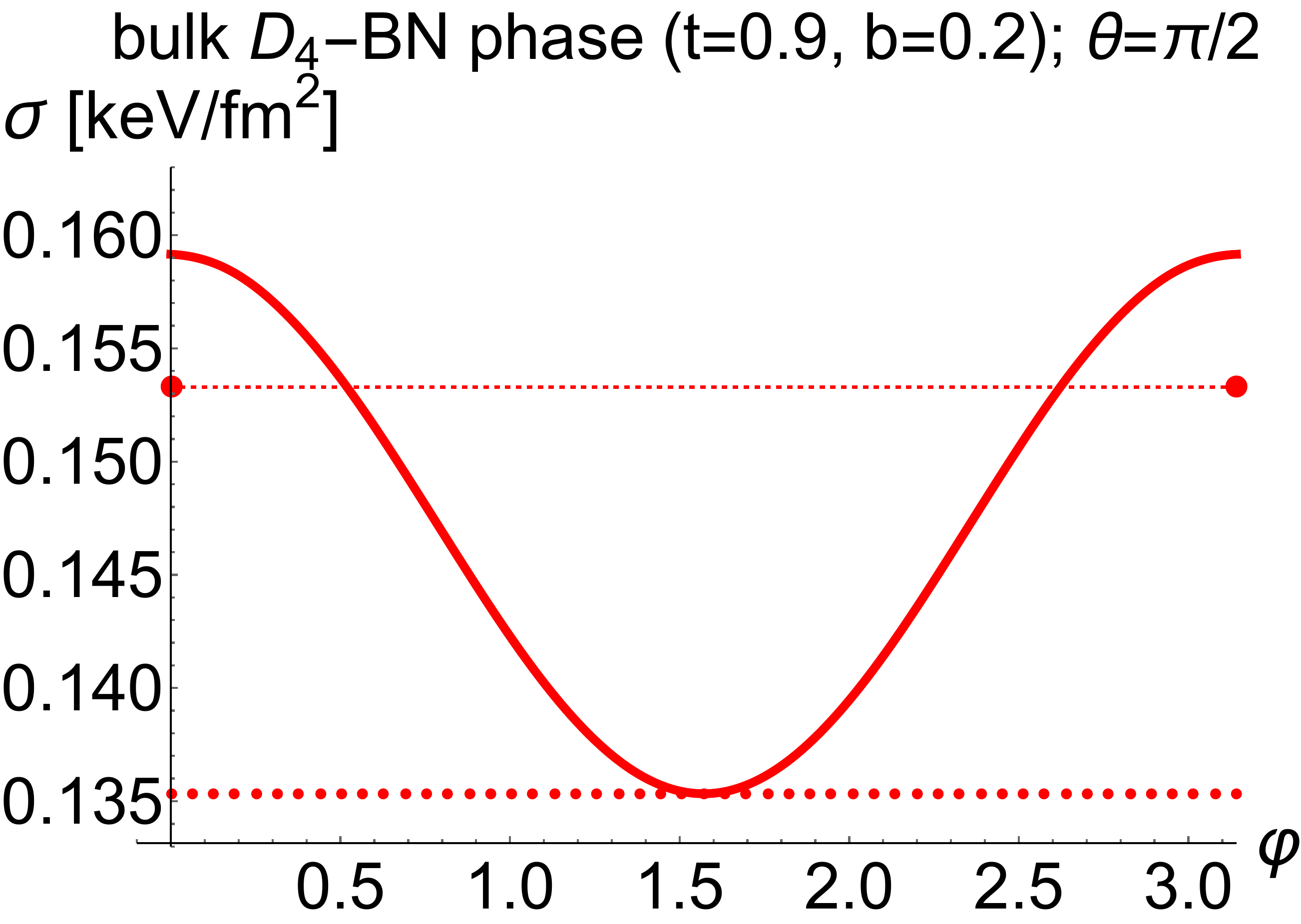}
\caption{The approximate values of the surface energy density $\sigma=\sigma(\vec{n})$ for the domain wall $W^{13}_{2}$ in the bulk D$_{4}$-BN phase. Left: The function of $\theta$ for $\varphi=\pi/2$ fixed. Right: The function of $\varphi$ for $\theta=\pi/2$ fixed. The horizontal thick-dotted lines indicate the value $\sigma=0.1353$ keV/fm$^{2}$, {\it i.e.}, the value at the $x_{2}$ direction ($\theta=\pi/2$ and $\varphi=\pi/2$). The horizontal thin-dotted lines indicate the value $\sigma=0.1533$ keV/fm$^{2}$, {\it i.e.}, the value at the $x_{1}$ direction ($\theta=\pi/2$ and $\varphi=0$) or at the $x_{3}$ direction ($\theta=0$). See the text for more details.}
\label{fig:sigma0_varphi_theta_D4BN_theta}
\end{center}
\end{figure}
%%%%%%%%%%%%%%%%%%%%%%%%%%%%%%

So far we have fixed the directions of the domain walls 
to be along the coordinate axes 
$x_{1}$, $x_{2}$, and $x_{3}$,
by setting the normal vector as $\vec{n}=(1,0,0)$, $(0,1,0)$, and $(0,0,1)$.
Now we discuss the stability of the domain walls against a {\it continuous} rotation of $\vec{n}$.
For example, we pick up $W^{13}_{2}$ as the most stable domain wall in the bulk D$_{4}$-BN phase, and investigate the angle dependence in the surface energy density of $W^{13}_{2}$.
For this purpose, we substitute the profile solution for $W^{13}_{2}$, which has been solved for $\vec{n}=(0,1,0)$ ($\theta=\pi/2$ and $\varphi=\pi/2$), 
into Eq.~\eqref{eq:surface_energy_density}, where the direction of $\vec{n}$ in Eq.~\eqref{eq:surface_energy_density} is set to be arbitrary in the vicinity of $\vec{n}=(0,1,0)$.
We then calculate approximately the surface energy density for the domain wall whose normal vector $\vec{n}$ is different from $\vec{n}=(0,1,0)$.
The result is shown in Fig.~\ref{fig:sigma0_varphi_theta_D4BN_theta}.
It is found that $\theta=\pi/2$ and $\varphi=\pi/2$, {\it i.e.}, the $x_{2}$ direction, still gives the minimum point.
It is also found that the values of $\tilde{\sigma}$ for the $x_{1}$ direction and for the $x_{3}$ direction are reproduced approximately, 
while the true solution (points in the figure) of course has less energy
 (see also~Table~\ref{table:surface_energy_dw_v2}).

\subsection{Piling domain walls}

%%%%%%%%%%%%%%%%%%%%%%%%%%%%%%
\begin{figure}[tb]
\begin{center}
\includegraphics[scale=0.25]{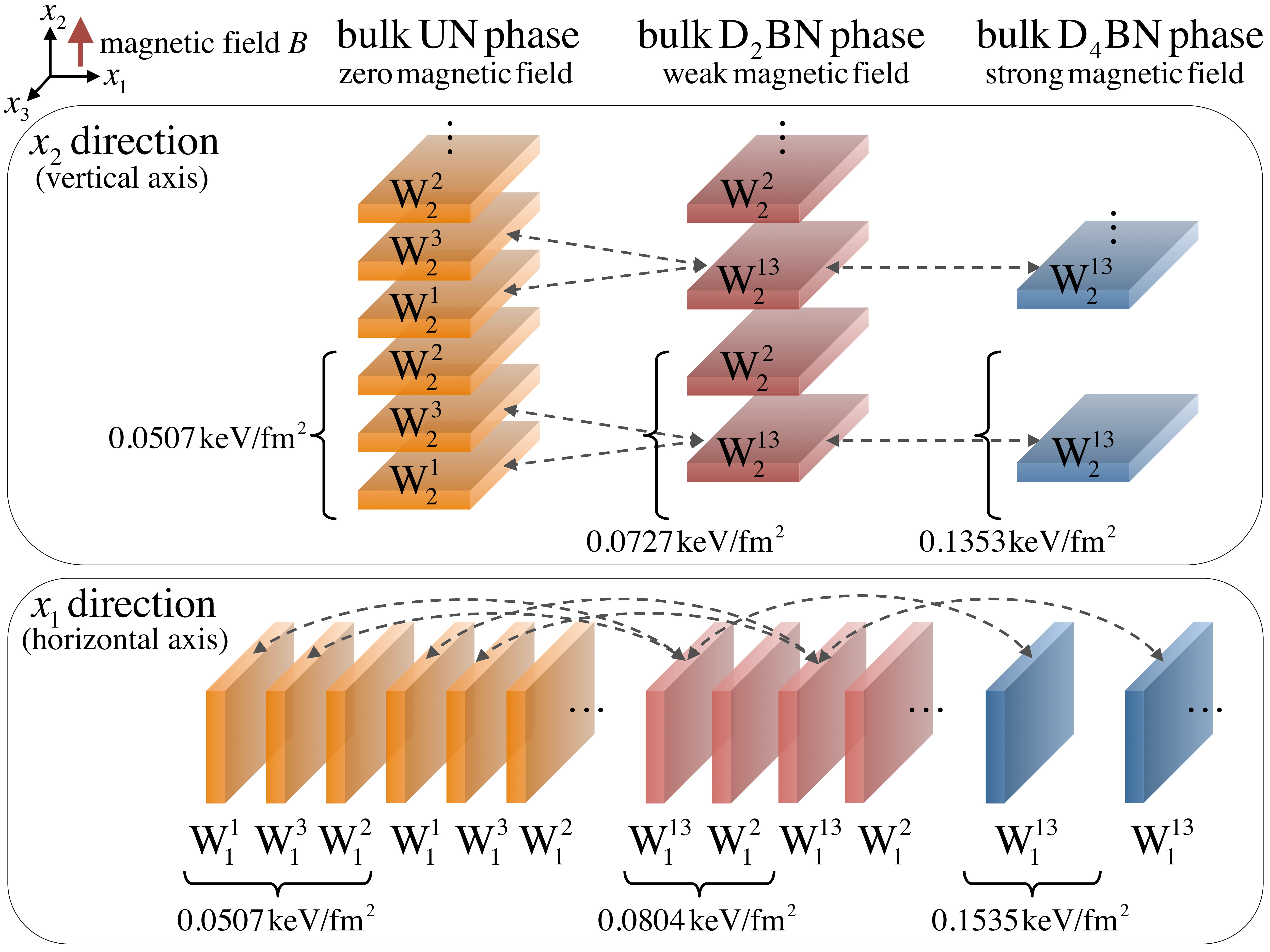}
\caption{The schematic figures for the metamorphism of the domain walls with the fixed directions ($i=1$, $2$), depending on the strength of the magnetic field. The magnetic fields in the D$_{2}$-BN and D$_{4}$-BN phases ($b=0.1$ and $0.2$) are applied to the $x_{2}$ (vertical) direction. The configurations $W^{1}_{2}$, $W^{3}_{2}$, and $W^{13}_{2}$ are the most stable states in the $x_{2}$ direction along the magnetic field, and the configurations $W^{2}_{1}$ or $W^{2}_{3}$ are the most stable states in the $x_{1}$ or $x_{3}$ direction perpendicular to the magnetic field. The surface energy densities in the unit cells are shown. See Table~\ref{table:surface_energy_dw_v2}.}
\label{fig:Fig_190711}
\end{center}
\end{figure}
%%%%%%%%%%%%%%%%%%%%%%%%%%%%%%

Here we discuss piling multiple domain walls.
We display the schematic image for a pile of domain walls which is the most stable for the given direction in Fig.~\ref{fig:Fig_190711}.
We notice that, when the domain walls are piled up, the phase function $\phi(\tilde{d};\vec{n})$ defined in Eq.~\eqref{eq:alpha_f} should change continuously throughout the piled domain walls.
From the continuity of $\phi(\tilde{d};\vec{n})$ modulo $2\pi$,
the ordering of the domain walls is determined uniquely 
(see also the upper panels of Fig.~\ref{fig:GL_potential}):
\begin{eqnarray}
  \cdots- (W^{1}_{i} - W^{3}_{i} - W^{2}_{i}) - (W^{1}_{i} - W^{3}_{i} - W^{2}_{i}) - \cdots,
\end{eqnarray}
in the bulk UN phase, 
\begin{eqnarray}
  \cdots - (W^{13}_{i} - W^{2}_{i}  )-( W^{13}_{i} - W^{2}_{i})- \cdots,
\end{eqnarray}
in the bulk D$_{2}$-BN phase, and
\begin{eqnarray}
  \cdots - (W^{13}_{i}) - (W^{13}_{i}) -\cdots,
\end{eqnarray}
in the bulk D$_{4}$-BN phase, 
along the directions of the $x_i$ axis 
($i=1$, $2$, $3$).\footnote{We can define the ``chirality" for right-winding and left-winding for the traveling direction of the domain walls in the bulk UN phase. We may assign the left-winding for a pile of domain walls $W^{1}_{i} - W^{3}_{i} - W^{2}_{i}$.  As the opposite chirality, we may also assign the right-winding for a pile of domain walls $W^{1}_{i} - W^{2}_{i} - W^{3}_{i}$.
}
We have introduced the unit cells by brackets as $(W^{1}_{i} - W^{3}_{i} - W^{2}_{i})$ in the bulk UN phase, $(W^{13}_{i} - W^{2}_{i})$ in the bulk D$_{2}$-BN phase, and $(W^{13}_{i})$ in the bulk D$_{4}$-BN phase,
and the piled domain walls are regarded as a repetition of the unit cell.

Let us discuss the surface energy density per a unit cell for the pile along the $x_{1}$ (or $x_{3}$) direction and along the $x_{2}$ direction. 
It is
\begin{eqnarray}
 \sigma (W^{1}_{i} - W^{3}_{i} - W^{2}_{i}) 
 = \sigma (W^{1}_{i}) + \sigma (W^{3}_{i}) + \sigma (W^{2}_{i}) + \sigma_{\rm int}
 = 0.0507  {\rm keV/fm}^{2}  \mbox{ for } i=1,2,3
\end{eqnarray}
for the UN phase,
\begin{eqnarray}
 \sigma (W^{13}_{i} - W^{2}_{i}) 
 = \sigma (W^{13}_{i}) + \sigma (W^{2}_{i}) + \sigma_{\rm int}
 = \left\{
 \begin{array}{c}
            0.0727  {\rm keV/fm}^{2}  \mbox{ for } i=2\\
            0.0804  {\rm keV/fm}^{2}  \mbox{ for } i=1,3 
    \end{array} \right.
\end{eqnarray}
for the D$_2$-BN phase, 
and 
\begin{eqnarray}
 \sigma (W^{13}_{i} - W^{2}_{i}) 
 = \sigma (W^{13}_{i}) + \sigma (W^{2}_{i}) + \sigma_{\rm int}
 = \left\{
 \begin{array}{c}
            0.1353  {\rm keV/fm}^{2}  \mbox{ for } i=2\\
            0.1535  {\rm keV/fm}^{2}  \mbox{ for } i=1,3 
    \end{array} \right.
\end{eqnarray}
for the D$_4$-BN phase, 
 where we have ignored the interaction energy $\sigma_{\rm int}$.
 We summarized the situation in Fig.~\ref{fig:Fig_190711}. 
Thus, at finite magnetic field, we conclude that a pile of the domain walls perpendicular to the magnetic field (the $x_{2}$ direction) is realized as the most stable states. 
As we did for a single domain wall, we can rotate the piled domain wall configurations with an arbitrary angle.

\section{Estimation of domain wall energy released from a neutron star}
\label{sec:discussion}

\begin{figure}[tb]
\begin{center}
\includegraphics[scale=0.22]{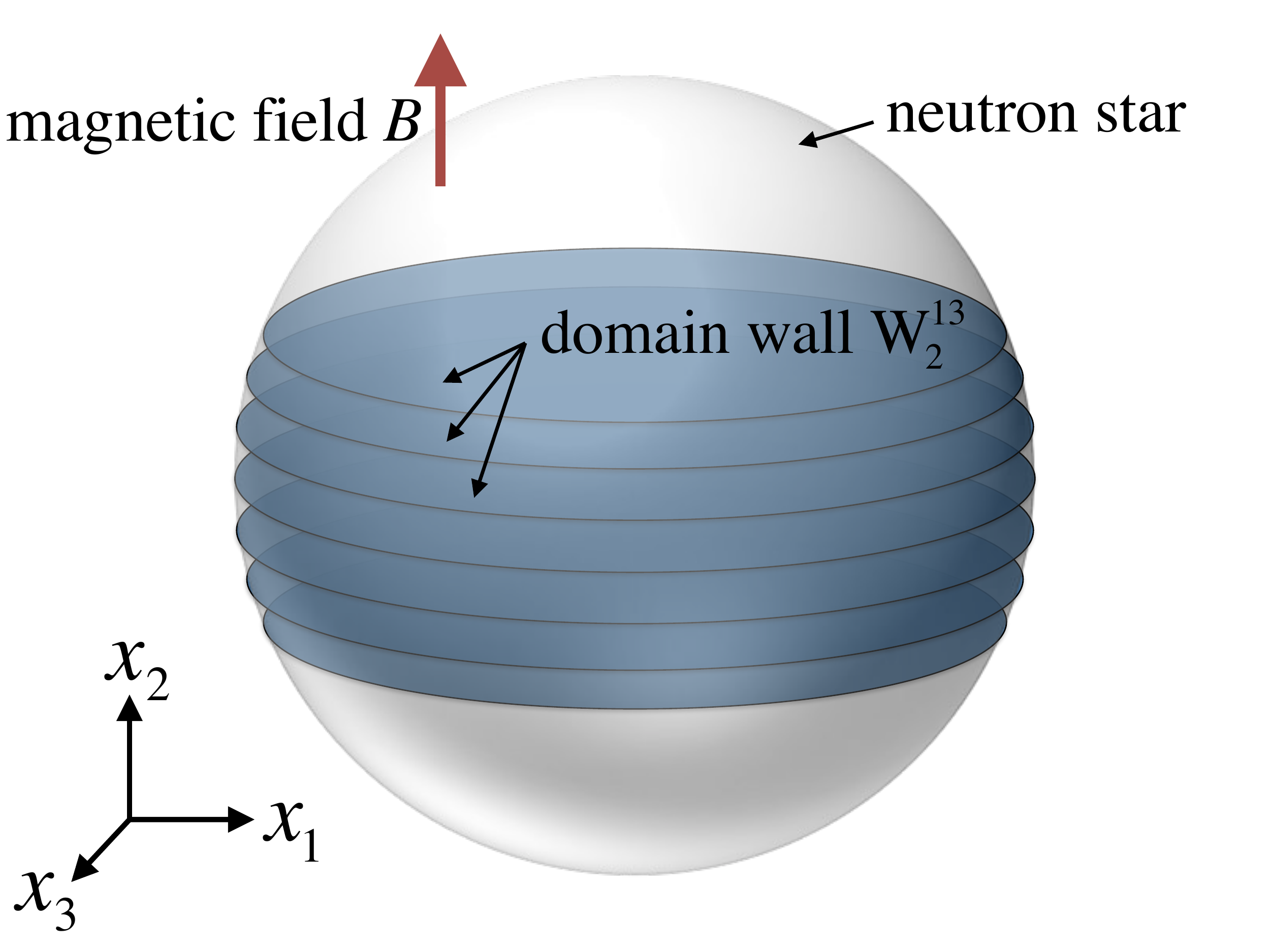}
\caption{The schematic figure of the bunch of the domain walls $W^{13}_{2}$ in the bulk D$_{4}$-BN phase in the neutron star.}
\label{fig:Fig_190619}
\end{center}
\end{figure}

We estimate the energy released from domain walls when they exist inside neutron stars.
As an example, we consider $W^{13}_{2}$ in the bulk D$_{4}$BN phase, which is the most stable state along the $x_{2}$ direction perpendicular to the applied magnetic field (cf.~Table~\ref{table:surface_energy_dw_v2}).
We suppose simply that domain walls are disks at constant $x_2$ inside a spherical condensation of a $^3P_2$ superfluid of the radius  $R \approx 10$ km 
of a neutron star.
Utilizing the surface energy density $\sigma=0.1353$ keV/fm$^{2}$ for $W^{13}_{2}$ in Table~\ref{table:surface_energy_dw_v2},
we obtain the total energy of $W^{13}_{2}$, $E_{\mathrm{DW}}=\sigma \times \pi R^{2}=0.85\times10^{29}$ erg.\footnote{We use the unit conversion $1\,\mathrm{keV} \approx 2.0\times10^{-9}\,\mathrm{erg}$.}
Here we suppose that the domain walls are densely piled in the neutron star as shown schematically in Fig.~\ref{fig:Fig_190619}.
In order to estimate the maximum number of the domain walls in the neutron stars, we suppose that the domain walls are packed maximally, where the interdistance between two domain walls will reach the healing distance 
as expected from 
defect formations at a phase transition \cite{Kibble:1976sj,Zurek:1985qw}: %}:
$\xi \approx p_{F}/mT_{c0}=355$ fm for $p_{F}=338$ MeV and $T=0.9T_{c0}$ ($t=0.9$).
This is obtained from $d = (p_{F}/(mT_{c0}))\tilde{d}$ by substituting $\tilde{d}\approx1$, where the value of $\tilde{d}$ can be read from Fig.~\ref{fig:190521_soliton_3P2_f1f2_D4BN} in Appendix~\ref{sec:plots_configuration_DW}.
Then, roughly speaking, the number of the domain walls will reach $N \approx R/\xi = 2.8\times10^{16}$.
Therefore, the energy stemming from the piled domain walls can be estimated as $E_{\mathrm{DW}} N \approx 2.4 \times 10^{45}$ erg.
In reality, the number $N$ would be smaller than the present estimation, and hence the energy $E_{\mathrm{DW}} N$ would be reduced also.
Nevertheless, we consider that a huge energy can be released from the domain walls,  leading to a possibility that the domain walls can be found in the astrophysical phenomena such as glitches as  sudden speed-up events of the rotation of neutron stars.

\begin{figure}[tb]
\begin{center}
\includegraphics[scale=0.22]{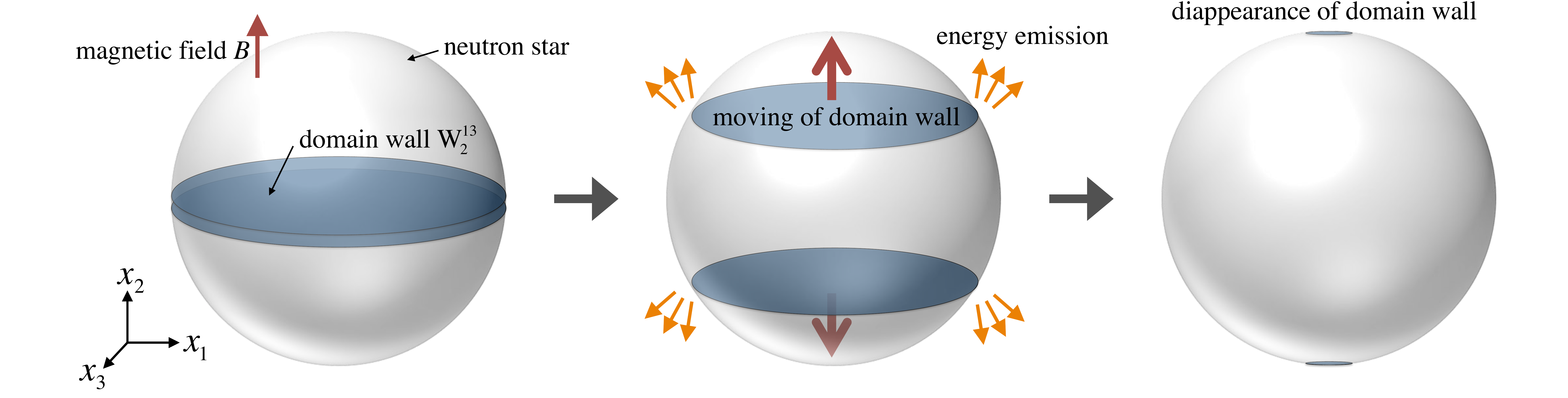}
\caption{The schematic figure for the domain walls $W^{13}_{2}$ in the bulk the D$_{4}$-BN phase, moving toward the north and south poles. The orientation of the domain walls is along the magnetic field (the $x_{2}$ direction). After arriving at the poles, the domain walls will disappear in the end.}
\label{fig:Fig_190620}
\end{center}
\end{figure}

We may consider the following situation that the domain wall trapped around the equator of the neutron star will not be the static objects, but they move to the north or south poles by releasing the energy of the domain wall.
This is because the smaller radius should be favored energetically, and therefore the domain walls will try to reduce the radius.
In this process, the domain walls would finally disappear when they reach the north or south poles, see Fig.~\ref{fig:Fig_190620}.
We may consider that the released energy by the domain walls, which will be emitted from the surface of the neutron stars, can be detected in astrophysical observation.

We comment that the healing distance $\xi$ is the temperature-dependent quantity as the order parameter. Because the healing distance should be very large near the critical temperature and it should become divergent just at the critical temperature, the number of domain walls in the neutron stars decreases accordingly. On the other hand, when the temperature becomes lower, the healing distance becomes smaller, and hence the number of domain walls in the neutron stars increases. Therefore, the released energies of the domain walls from the neutron stars are sensitive to the temperature inside the neutron stars.

%%%%%%%%%%%%%%%%%%%%%%%%%%%%%%
%%%%%%%%%%%%%%%%%%%%%%%%%%%%%%

%%%%%%%%%%%%%%%%%%%%%%%%%%%%%%
%%%%%%%%%%%%%%%%%%%%%%%%%%%%%%

\section{Summary and discussion}
\label{sec:conclusion}

We have worked out the domain walls in a neutron $^{3}P_{2}$ superfluid in neutron stars.
As an effective theory, we have adopted the GL equation for describing the domain walls as the nonuniform systems.
Considering the bulk UN, D$_{2}$-BN, and D$_{4}$-BN phases in the neutron $^{3}P_{2}$ superfluids,
we have constructed the configurations of the domain walls and have calculated the surface energy densities.
As examples, we have discussed the domain walls denoted by $W^{\alpha}_{i}$ ($\alpha=1$, $2$, $3$) in the bulk UN phase, $W^{\alpha}_{i}$ ($\alpha=13$, $2$) in the bulk D$_{2}$-BN phase, and $W^{\alpha}_{i}$ ($\alpha=13$) in the bulk D$_{4}$-BN phase for the fixed direction $x_{i}$ ($i=1$, $2$, $3$) in space.
As for symmetry structures, domain wall configurations pass thorough different phases 
in their vicinities. 
Consequently, unbroken symmetries around their cores are different from those in the bulk.
For instance, the phases of domain walls in the bulk UN phase are either D$_2$-BN or D$_4$-BN phase.
As a result, there appears one extra Nambu-Goldstone mode localized in the vicinity of the domain walls 
in the UN phase.
On the other hand,  in the D$_{2}$-BN phase, 
 a U(1) symmetry restores 
 in the vicinity of the domain wall $W^{13}(\mathrm{D}_{2}\mathrm{BN})$ 
 but not of the other case $W^2(\mathrm{D}_{2}\mathrm{BN})$.
 The same symmetry restoration occurs in the vicinities of all domain walls
 $W^{13}(\mathrm{D}_{4}\mathrm{BN})$ in the D$_{4}$-BN phase. 
Considering a pile of domain walls, we have found that the domain walls are lined up perpendicular to the magnetic field.
We have estimated the energy released from a pile of the domain walls existing inside neutron stars and have shown that the emitted energy can reach a huge amount 
which may be found in signals from the neutron stars, such as glitches, in the astrophysical observation.

In this paper, we have restricted ourselves to 
the diagonal components in $A(d;\vec{n})$. 
Taking into the off-diagonal components, the configuration may be unstable or metastable.
This is because a set two different ground states that we were considering in this paper 
can be connected by a global transformation of $\mathrm{U}(1) \times \mathrm{SO}(3)$, 
and therefore they are not disconnected in the full space. 
Physically, the configurations should decay to a ground state by emitting Nambu-Goldstone modes 
because symmetry transformation exchange diagonal components through off-diagonal components.
In spite of the limitation in the present research, 
we expect that the domain walls can give an impact on the astrophysical observation of the neutron stars as long as the domain walls survive within the time scale of the decay. 
It will be a
 next subject to study the lifetime of the domain walls, for which we have to consider the off-diagonal components in the order parameter.
It is also important to study the interaction between two domain walls which will be caused by the exchange of massive particles or sometimes Nambu-Goldstone modes.
Although the interaction is expected to be repulsive, it is still an open problem to discuss carefully the strength of the interaction, the dependence on the bulk phases, and so on.

Since the phases in the vicinities of domain walls are different from those in the bulk, 
unbroken symmetries are also different. 
This may allow topological defects living inside domain walls. 
For instance,  the U(1) symmetry unbroken in the bulk UN phase is further broken to 
D$_2$ or D$_4$ in the vicinity of the domain wall, admitting vortices inside domain walls.\footnote{  
This situation is similar to the boundary surface of superfluid studied recently \cite{Yasui:2019pgb}.}
What do such  composite configuration represent will be one of future directions.\footnote{See, for instance, Refs.\cite{Eto:2013hoa} and \cite{Eto:2006pg}, for composite soliton configurations in dense QCD and supersymmetric gauge theories, respectively.}

We also will need to study the edge effect of the domain walls, because the domain walls realistically should be affected by the surface of the neutron stars.
The surface of the neutron $^{3}P_{2}$ superfluids has the topological property~\cite{Yasui:2019pgb}.
The properties of the domain walls in the rotating system are also interesting subjects, in which the interaction between the domain walls and the quantum vortices may become important.\footnote{See, {\it e.g.}, Refs.~\cite{Masuda:2016vak,Masuda:2015jka,Chatterjee:2016gpm} for the recent works on the $^{3}P_{2}$ superfluid vortices in neutron matter.}
Beyond the description by the GL theory, it is an interesting question 
whether 
fermion (quasi-)gapless modes are present inside domain walls, by investigating  
fermionic degrees of freedom in terms of the BdG equation~\cite{Mizushima:2016fbn}, 
in which $^3P_2$ surface was shown to allow gapless Majorana fermions. 

There exist superfluid vortices in $^{3}P_{2}$ superfluids and they form a lattice under a rotation~\cite{Muzikar:1980as,Sauls:1982ie,Fujita1972,Richardson:1972xn,Masuda:2015jka,Chatterjee:2016gpm,Masuda:2016vak}. How these vortices interact with domain walls is an important question. Also, vortices may terminate on a domain wall with forming so-called a D-brane soliton, which is known to exist in two-component Bose-Einstein condensates~\cite{Kasamatsu:2010aq, Nitta:2012hy, Kasamatsu:2013lda, Kasamatsu:2013qia} and supersymmetric field theory~\cite{Gauntlett:2000de, Isozumi:2004vg}, where the endpoint of the vortex is called as a boojum. 
Another possibility is that a domain wall may terminate on a vortex, or in other words, a vortex is attached by a domain wall, as axion strings.

Domain walls also exist in color superconductivity in the quark matter considered to exist deeply inside
of neutron stars, see, {\it e.g.}, Ref.~\cite{Eto:2013hoa}.
Then, a similar situation occurs in this case. It is also an interesting question how domain walls interact with other topological defects.
It has been known that, in a rotating neutron star, Abelian quantum superfluid vortices are created along the rotation axis in the hadron matter as well as non-Abelian quantum vortices (color magnetic flux tubes) in the quark matter~\cite{Balachandran:2005ev,Nakano:2007dr,Eto:2009kg,Eto:2009bh,Eto:2009tr}.
Recently, the presence or absence of boojums have been discussed, which are defects at endpoints (or junction points) of these vortices at the interface~\cite{Cipriani:2012hr,Alford:2018mqj,Chatterjee:2018nxe,Chatterjee:2019tbz,Cherman:2018jir}.
The interactions between the domain walls and the boojum may influence the dynamics of neutron stars. Those problems are left for future works.

\section*{Acknowledgment}
This work is supported by the Ministry of Education, Culture, Sports, Science (MEXT)-Supported Program for the Strategic Research Foundation at Private Universities ``Topological Science" (Grant No. S1511006). 
This work is also supported in part by 
JSPS Grant-in-Aid for Scientific Research [KAKENHI Grants No.~17K05435 (S.Y.), 
No.~16H03984 (M.N.), and No.~18H01217 (M.N.)], 
and also by MEXT KAKENHI Grant-in-Aid for Scientific Research on Innovative Areas ``Topological Materials Science'' No.~15H05855 (M.N.).

\appendix

\section{Euler-Lagrange equations}
\label{sec:EL_appendix}

We show the concrete expressions of the terms in the left-hand sides in the EL equations \eqref{eq:EL_f} and \eqref{eq:EL_g}:
\begin{eqnarray}
&&
   - \nabla_{\!d} \frac{\delta {f}}{\delta (\nabla_{\!d}F_{1})}
    + \frac{\delta {f}}{\delta F_{1}}
\nonumber \\ %%
&=&
   -\frac{{K}^{(0)}}{4}
   \Bigl(
      2\bigl( 2 - \sin^{2}\theta \, \sin^{2}\varphi \bigr) \nabla_{\!d}^{2}
      F_{1}
   + \bigl( 1 +2 \cos^{2}\theta \bigr)\nabla_{\!d}^{2}
      F_{2}
   + 2
      \cos\theta\sin\theta\sin\varphi
      \nabla_{\!d}^{2}
      G_{1}
    - 2
      \sin^{2}\theta\cos\varphi\sin\varphi
      \nabla_{\!d}^{2}
      G_{3}
   \Bigr)
\nonumber \\ && %%
+
   {\alpha}^{(0)} 2 (2 F_{1}+F_{2})
   \nonumber \\ && %%
+ {\beta}^{(0)} 4
   (2 F_{1}+F_{2})
   \bigl(
         F_{1}^{2}+F_{1}F_{2}+F_{2}^{2}+G_{1}^{2}+G_{2}^{2}+G_{3}^{2}
   \bigr)
   \nonumber \\ && %%
+ {\gamma}^{(0)}24
   \Bigl(
         12 F_{1}^{5}
      + 30 F_{2} F_{1}^{4}
      + 52 F_{2}^{2} F_{1}^{3}
      + 24 G_{1}^{2} F_{1}^{3}
      + 24 G_{2}^{2} F_{1}^{3}
      + 24 G_{3}^{2} F_{1}^{3}
      + 48 F_{2}^{3} F_{1}^{2}
      + 42 F_{2} G_{1}^{2} F_{1}^{2}
      + 36 F_{2} G_{2}^{2} F_{1}^{2}
         \nonumber \\ && \hspace{4em} %%
      + 30 F_{2} G_{3}^{2} F_{1}^{2}
      + 26 F_{2}^{4} F_{1}
      + 14 G_{1}^{4} F_{1}
      + 12 G_{2}^{4} F_{1}
      + 14 G_{3}^{4} F_{1}
      + 40 F_{2}^{2} G_{1}^{2} F_{1}
      + 40 F_{2}^{2} G_{2}^{2} F_{1}
      + 24 G_{1}^{2} G_{2}^{2} F_{1}
      + 28 F_{2}^{2} G_{3}^{2} F_{1}
         \nonumber \\ && \hspace{4em} %%
      + 20 G_{1}^{2} G_{3}^{2} F_{1}
      + 24 G_{2}^{2} G_{3}^{2} F_{1}
      + 8 F_{2} G_{1} G_{2} G_{3} F_{1}
      + 6 F_{2}^{5}
      + 6 F_{2} G_{1}^{4}
      + 6 F_{2} G_{2}^{4}
      + 8 F_{2} G_{3}^{4}
       - 4 G_{1} G_{2} G_{3}^{3}
      + 12 F_{2}^{3} G_{1}^{2}
         \nonumber \\ && \hspace{4em} %%
      + 14 F_{2}^{3} G_{2}^{2}
      + 14 F_{2} G_{1}^{2} G_{2}^{2}
      + 10 F_{2}^{3} G_{3}^{2}
      + 10 F_{2} G_{1}^{2} G_{3}^{2}
      + 10 F_{2} G_{2}^{2} G_{3}^{2}
      + 4 G_{1}^{3} G_{2} G_{3}
      + 4 F_{2}^{2} G_{1} G_{2} G_{3}
      \Bigr)
   \nonumber \\ && %%
+ {\delta}^{(0)} 192
   \Bigl(
         16 F_{1}^{7}
      + 56 F_{2} F_{1}^{6}
      + 138 F_{2}^{2} F_{1}^{5}
      + 48 G_{1}^{2} F_{1}^{5}
      + 48 G_{2}^{2} F_{1}^{5}
      + 48 G_{3}^{2} F_{1}^{5}
      + 205 F_{2}^{3} F_{1}^{4}
      + 150 F_{2} G_{1}^{2} F_{1}^{4}
      + 120 F_{2} G_{2}^{2} F_{1}^{4}
         \nonumber \\ && \hspace{4em} %%
      + 90 F_{2} G_{3}^{2} F_{1}^{4}
      + 200 F_{2}^{4} F_{1}^{3}
      + 60 G_{1}^{4} F_{1}^{3}
      + 48 G_{2}^{4} F_{1}^{3}
      + 60 G_{3}^{4} F_{1}^{3}
      + 252 F_{2}^{2} G_{1}^{2} F_{1}^{3}
      + 228 F_{2}^{2} G_{2}^{2} F_{1}^{3}
      + 96 G_{1}^{2} G_{2}^{2} F_{1}^{3}
         \nonumber \\ && \hspace{4em} %%
      + 132 F_{2}^{2} G_{3}^{2} F_{1}^{3}
      + 72 G_{1}^{2} G_{3}^{2} F_{1}^{3}
      + 96 G_{2}^{2} G_{3}^{2} F_{1}^{3}
      + 48 F_{2} G_{1} G_{2} G_{3} F_{1}^{3}
      + 123 F_{2}^{5} F_{1}^{2}
      + 99 F_{2} G_{1}^{4} F_{1}^{2}
      + 72 F_{2} G_{2}^{4} F_{1}^{2}
         \nonumber \\ && \hspace{4em} %%
      + 81 F_{2} G_{3}^{4} F_{1}^{2}
       - 36 G_{1} G_{2} G_{3}^{3} F_{1}^{2}
      + 222 F_{2}^{3} G_{1}^{2} F_{1}^{2}
       + 222 F_{2}^{3} G_{2}^{2} F_{1}^{2}
       + 180 F_{2} G_{1}^{2} G_{2}^{2} F_{1}^{2}
       + 114 F_{2}^{3} G_{3}^{2} F_{1}^{2}
       + 108 F_{2} G_{1}^{2} G_{3}^{2} F_{1}^{2}
         \nonumber \\ && \hspace{4em} %%
       + 108 F_{2} G_{2}^{2} G_{3}^{2} F_{1}^{2}
       + 36 G_{1}^{3} G_{2} G_{3} F_{1}^{2}
       + 72 F_{2}^{2} G_{1} G_{2} G_{3} F_{1}^{2}
       + 46 F_{2}^{6} F_{1}
       + 22 G_{1}^{6} F_{1}
       + 16 G_{2}^{6} F_{1}
       + 22 G_{3}^{6} F_{1}
       + 90 F_{2}^{2} G_{1}^{4} F_{1}
         \nonumber \\ && \hspace{4em} %%
       + 90 F_{2}^{2} G_{2}^{4} F_{1}
       + 48 G_{1}^{2} G_{2}^{4} F_{1}
       + 72 F_{2}^{2} G_{3}^{4} F_{1}
       + 42 G_{1}^{2} G_{3}^{4} F_{1}
       + 54 G_{2}^{2} G_{3}^{4} F_{1}
        - 24 F_{2} G_{1} G_{2} G_{3}^{3} F_{1}
       + 114 F_{2}^{4} G_{1}^{2} F_{1}
         \nonumber \\ && \hspace{4em} %%
       + 126 F_{2}^{4} G_{2}^{2} F_{1}
       + 54 G_{1}^{4} G_{2}^{2} F_{1}
       + 180 F_{2}^{2} G_{1}^{2} G_{2}^{2} F_{1}
       + 66 F_{2}^{4} G_{3}^{2} F_{1}
       + 42 G_{1}^{4} G_{3}^{2} F_{1}
       + 48 G_{2}^{4} G_{3}^{2} F_{1}
       + 108 F_{2}^{2} G_{1}^{2} G_{3}^{2} F_{1}
         \nonumber \\ && \hspace{4em} %%
       + 108 F_{2}^{2} G_{2}^{2} G_{3}^{2} F_{1}
       + 108 G_{1}^{2} G_{2}^{2} G_{3}^{2} F_{1}
       + 48 F_{2} G_{1} G_{2}^{3} G_{3} F_{1}
       + 48 F_{2} G_{1}^{3} G_{2} G_{3} F_{1}
       + 48 F_{2}^{3} G_{1} G_{2} G_{3} F_{1}
       + 8 F_{2}^{7}
         \nonumber \\ && \hspace{4em} %%
       + 8 F_{2} G_{1}^{6}
       + 8 F_{2} G_{2}^{6}
       + 14 F_{2} G_{3}^{6}
        - 12 G_{1} G_{2} G_{3}^{5}
       + 24 F_{2}^{3} G_{1}^{4}
       + 33 F_{2}^{3} G_{2}^{4}
       + 30 F_{2} G_{1}^{2} G_{2}^{4}
       + 27 F_{2}^{3} G_{3}^{4}
       + 24 F_{2} G_{1}^{2} G_{3}^{4}
         \nonumber \\ && \hspace{4em} %%
       + 24 F_{2} G_{2}^{2} G_{3}^{4}
        - 12 G_{1} G_{2}^{3} G_{3}^{3}
        - 12 F_{2}^{2} G_{1} G_{2} G_{3}^{3}
       + 24 F_{2}^{5} G_{1}^{2}
       + 30 F_{2}^{5} G_{2}^{2}
       + 30 F_{2} G_{1}^{4} G_{2}^{2}
       + 60 F_{2}^{3} G_{1}^{2} G_{2}^{2}
       + 18 F_{2}^{5} G_{3}^{2}
         \nonumber \\ && \hspace{4em} %%
       + 18 F_{2} G_{1}^{4} G_{3}^{2}
       + 18 F_{2} G_{2}^{4} G_{3}^{2}
       + 36 F_{2}^{3} G_{1}^{2} G_{3}^{2}
       + 36 F_{2}^{3} G_{2}^{2} G_{3}^{2}
       + 54 F_{2} G_{1}^{2} G_{2}^{2} G_{3}^{2}
       + 12 G_{1}^{3} G_{2}^{3} G_{3}
       + 24 F_{2}^{2} G_{1} G_{2}^{3} G_{3}
         \nonumber \\ && \hspace{4em} %%
       + 12 G_{1}^{5} G_{2} G_{3}
       + 24 F_{2}^{2} G_{1}^{3} G_{2} G_{3}
       +12 F_{2}^{4} G_{1} G_{2} G_{3}
   \Bigr)
   \nonumber \\ && %%
+ {\beta}^{(2)}
   \Bigl(
         2 F_{1} b_{x}^{2}+2 b_{z}^{2} (F_{1}+F_{2})
   \Bigr)
   \nonumber \\ && %%
+ {\beta}^{(4)} 
   \Bigl(
         b_{x}^{2}+b_{y}^{2}+b_{z}^{2}
   \Bigr)
   \Bigl(
         2 F_{1} b_{x}^{2}+2 b_{z}^{2} (F_{1}+F_{2})
   \Bigr)
   \nonumber \\ && %%
+ {\gamma}^{(2)}
   \Bigl(
       - 12b_{x}^{2}
         \Bigl(
                 8F_{1}^{3}
              + 3F_{1}^{2} F_{2}
              + 2F_{1} F_{2}^{2}
              + 2F_{1} G_{1}^{2}
              + 8F_{1} G_{2}^{2}
              + F_{2} G_{2}^{2}
               - 2G_{1} G_{2} G_{3}
              + 10F_{1} G_{3}^{2}
              + 3F_{2} G_{3}^{2}
         \Bigr)
         \nonumber \\ && \hspace{3em} %%
       - 12b_{y}^{2}
         \Bigl(
                 2F_{1} F_{2}^{2}
              + F_{2}^{3}
              + 4F_{1} G_{1}^{2}
              + F_{2} G_{1}^{2}
              + 4F_{1} G_{2}^{2}
              + 3F_{2} G_{3}^{2}
         \Bigr)
         \nonumber \\ && \hspace{3em} %%
       - 12b_{z}^{2}
         \Bigl(
                 8F_{1}^{3}
              + 21F_{1}^{2} F_{2}
              + 20F_{1}F_{2}^{2}
              + 7F_{2}^{3}
              + 10F_{1}G_{1}^{2}
              + 7F_{2} G_{1}^{2}
              + 8F_{1} G_{2}^{2}
              + 7F_{2} G_{2}^{2}
              + 2G_{1} G_{2} G_{3}
              + 2F_{1} G_{3}^{2}
              + 2F_{2} G_{3}^{2}
         \Bigr)
   \Bigr),
\nonumber \\
\end{eqnarray}
for $F_{1}$,
\begin{eqnarray}
&&
   - \nabla_{\!d} \frac{\delta {f}}{\delta (\nabla_{\!d}F_{2})}
   + \frac{\delta {f}}{\delta F_{2}}
\nonumber \\ %%
&=&
   -\frac{{K}^{(0)}}{4}
   \Bigl(
        \bigl( 1 +2 \cos^{2}\theta \bigr)\nabla_{\!d}^{2}
        F_{1}
     + 2\bigl( 2 - \sin^{2}\theta \, \cos^{2}\varphi \bigr) \nabla_{\!d}^{2}
        F_{2}
      - 2
        \sin^{2}\theta\cos\varphi\sin\varphi
        \nabla_{\!d}^{2}
        G_{3}
     + 2
        \cos\theta\sin\theta\cos\varphi
        \nabla_{\!d}^{2}
        G_{2}
   \Bigr)
\nonumber \\ && %%
+
{\alpha}^{(0)}
   2 (F_{1}+2 F_{2}) 
   \nonumber \\ && %%
+{\beta}^{(0)}
   4 (F_{1}+2 F_{2})
   \bigl(F_{1}^{2}+F_{2}
   F_{1}+F_{2}^{2}+G_{1}^{2}+G_{2}^{2}+G_{3}^{2}\bigr)
   \nonumber \\ && %%
+{\gamma}^{(0)}
   24
   \Bigl(
           6 F_{1}^{5}
         +26 F_{2} F_{1}^{4}
         +48 F_{2}^{2} F_{1}^{3}
         +14 G_{1}^{2} F_{1}^{3}
         +12 G_{2}^{2} F_{1}^{3}
         +10 G_{3}^{2} F_{1}^{3}
         +52 F_{2}^{3} F_{1}^{2}
         +40 F_{2} G_{1}^{2} F_{1}^{2}
         +40 F_{2} G_{2}^{2} F_{1}^{2}
         +28 F_{2} G_{3}^{2} F_{1}^{2}
            \nonumber \\ && \hspace{4em} %%
         +4 G_{1} G_{2} G_{3} F_{1}^{2}
         +30 F_{2}^{4} F_{1}
         +6 G_{1}^{4} F_{1}
         +6 G_{2}^{4} F_{1}
         +8 G_{3}^{4} F_{1}
         +36 F_{2}^{2} G_{1}^{2} F_{1}
         +42 F_{2}^{2} G_{2}^{2} F_{1}
         +14 G_{1}^{2} G_{2}^{2} F_{1}
         +30 F_{2}^{2} G_{3}^{2} F_{1}
            \nonumber \\ && \hspace{4em} %%
         +10 G_{1}^{2} G_{3}^{2} F_{1}
         +10 G_{2}^{2} G_{3}^{2} F_{1}
         +8 F_{2} G_{1} G_{2} G_{3} F_{1}
         +12 F_{2}^{5}
         +12 F_{2} G_{1}^{4}
         +14 F_{2} G_{2}^{4}
         +14 F_{2} G_{3}^{4}
         -4 G_{1} G_{2} G_{3}^{3}
            \nonumber \\ && \hspace{4em} %%
         +24 F_{2}^{3} G_{1}^{2}
         +24 F_{2}^{3} G_{2}^{2}
         +24 F_{2} G_{1}^{2} G_{2}^{2}
         +24 F_{2}^{3} G_{3}^{2}
         +24 F_{2} G_{1}^{2} G_{3}^{2}
         +20 F_{2} G_{2}^{2} G_{3}^{2}
         +4 G_{1} G_{2}^{3} G_{3}
   \Bigr)
   \nonumber \\ && %%
+{\delta}^{(0)}
   192
   \Bigl(
           8 F_{1}^{7}
         +46 F_{2} F_{1}^{6}
         +123 F_{2}^{2} F_{1}^{5}
         +30 G_{1}^{2} F_{1}^{5}
         +24 G_{2}^{2} F_{1}^{5}
         +18 G_{3}^{2} F_{1}^{5}
         +200 F_{2}^{3} F_{1}^{4}
         +126 F_{2} G_{1}^{2} F_{1}^{4}
         +114 F_{2} G_{2}^{2} F_{1}^{4}
           \nonumber \\ && \hspace{4em} %%
         +66 F_{2} G_{3}^{2} F_{1}^{4}
         +12 G_{1} G_{2} G_{3} F_{1}^{4}
         +205 F_{2}^{4} F_{1}^{3}
         +33 G_{1}^{4} F_{1}^{3}
         +24 G_{2}^{4} F_{1}^{3}
         +27 G_{3}^{4} F_{1}^{3}
         +222 F_{2}^{2} G_{1}^{2} F_{1}^{3}
         +222 F_{2}^{2} G_{2}^{2} F_{1}^{3}
           \nonumber \\ && \hspace{4em} %%
         +60 G_{1}^{2} G_{2}^{2} F_{1}^{3}
         +114 F_{2}^{2} G_{3}^{2} F_{1}^{3}
         +36 G_{1}^{2} G_{3}^{2} F_{1}^{3}
         +36 G_{2}^{2} G_{3}^{2} F_{1}^{3}
         +48 F_{2} G_{1} G_{2} G_{3} F_{1}^{3}
         +138 F_{2}^{5} F_{1}^{2}
         +90 F_{2} G_{1}^{4} F_{1}^{2}
           \nonumber \\ && \hspace{4em} %%
         +90 F_{2} G_{2}^{4} F_{1}^{2}
         +72 F_{2} G_{3}^{4} F_{1}^{2}
         -12 G_{1} G_{2} G_{3}^{3} F_{1}^{2}
         +228 F_{2}^{3} G_{1}^{2} F_{1}^{2}
         +252 F_{2}^{3} G_{2}^{2} F_{1}^{2}
         +180 F_{2} G_{1}^{2} G_{2}^{2} F_{1}^{2}
         +132 F_{2}^{3} G_{3}^{2} F_{1}^{2}
           \nonumber \\ && \hspace{4em} %%
         +108 F_{2} G_{1}^{2} G_{3}^{2} F_{1}^{2}
         +108 F_{2} G_{2}^{2} G_{3}^{2} F_{1}^{2}
         +24 G_{1} G_{2}^{3} G_{3} F_{1}^{2}
         +24 G_{1}^{3} G_{2} G_{3} F_{1}^{2}
         +72 F_{2}^{2} G_{1} G_{2} G_{3} F_{1}^{2}
         +56 F_{2}^{6} F_{1}
         +8 G_{1}^{6} F_{1}
           \nonumber \\ && \hspace{4em} %%
         +8 G_{2}^{6} F_{1}
         +14 G_{3}^{6} F_{1}
         +72 F_{2}^{2} G_{1}^{4} F_{1}
         +99 F_{2}^{2} G_{2}^{4} F_{1}
         +30 G_{1}^{2} G_{2}^{4} F_{1}
         +81 F_{2}^{2} G_{3}^{4} F_{1}
         +24 G_{1}^{2} G_{3}^{4} F_{1}
         +24 G_{2}^{2} G_{3}^{4} F_{1}
           \nonumber \\ && \hspace{4em} %%
         -24 F_{2} G_{1} G_{2} G_{3}^{3} F_{1}
         +120 F_{2}^{4} G_{1}^{2} F_{1}
         +150 F_{2}^{4} G_{2}^{2} F_{1}
         +30 G_{1}^{4} G_{2}^{2} F_{1}
         +180 F_{2}^{2} G_{1}^{2} G_{2}^{2} F_{1}
         +90 F_{2}^{4} G_{3}^{2} F_{1}
         +18 G_{1}^{4} G_{3}^{2} F_{1}
           \nonumber \\ && \hspace{4em} %%
         +18 G_{2}^{4} G_{3}^{2} F_{1}
         +108 F_{2}^{2} G_{1}^{2} G_{3}^{2} F_{1}
         +108 F_{2}^{2} G_{2}^{2} G_{3}^{2} F_{1}
         +54 G_{1}^{2} G_{2}^{2} G_{3}^{2} F_{1}
         +48 F_{2} G_{1} G_{2}^{3} G_{3} F_{1}
         +48 F_{2} G_{1}^{3} G_{2} G_{3} F_{1}
           \nonumber \\ && \hspace{4em} %%
         +48 F_{2}^{3} G_{1} G_{2} G_{3} F_{1}
         +16 F_{2}^{7}+16 F_{2} G_{1}^{6}
         +22 F_{2} G_{2}^{6}
         +22 F_{2} G_{3}^{6}
         -12 G_{1} G_{2} G_{3}^{5}
         +48 F_{2}^{3} G_{1}^{4}
         +60 F_{2}^{3} G_{2}^{4}
         +54 F_{2} G_{1}^{2} G_{2}^{4}
           \nonumber \\ && \hspace{4em} %%
         +60 F_{2}^{3} G_{3}^{4}
         +54 F_{2} G_{1}^{2} G_{3}^{4}
         +42 F_{2} G_{2}^{2} G_{3}^{4}
         -12 G_{1}^{3} G_{2} G_{3}^{3}
         -36 F_{2}^{2} G_{1} G_{2} G_{3}^{3}
         +48 F_{2}^{5} G_{1}^{2}
         +48 F_{2}^{5} G_{2}^{2}
         +48 F_{2} G_{1}^{4} G_{2}^{2}
           \nonumber \\ && \hspace{4em} %%
         +96 F_{2}^{3} G_{1}^{2} G_{2}^{2}
         +48 F_{2}^{5} G_{3}^{2}
         +48 F_{2} G_{1}^{4} G_{3}^{2}
         +42 F_{2} G_{2}^{4} G_{3}^{2}
         +96 F_{2}^{3} G_{1}^{2} G_{3}^{2}
         +72 F_{2}^{3} G_{2}^{2} G_{3}^{2}
         +108 F_{2} G_{1}^{2} G_{2}^{2} G_{3}^{2}
           \nonumber \\ && \hspace{4em} %%
         +12 G_{1} G_{2}^{5} G_{3}
         +12 G_{1}^{3} G_{2}^{3} G_{3}
         +36 F_{2}^{2} G_{1} G_{2}^{3} G_{3}
   \Bigr)
   \nonumber \\ && %%
+{\beta}^{(2)}
   \bigl(2 F_{2} b_{y}^{2}+2 b_{z}^{2} (F_{1}+F_{2})\bigr)   
   \nonumber \\ && %%
+{\beta}^{(4)}
   \bigl(b_{x}^{2}+b_{y}^{2}+b_{z}^{2}\bigr)
   \bigl(2 F_{2} b_{y}^{2}+2 b_{z}^{2} (F_{1}+F_{2})\bigr) 
   \nonumber \\ && %%
+{\gamma}^{(2)}
   \Bigl(
         - 12b_{x}^{2}
           \bigl(
                    F_{1}^{3}
                 + 2F_{1}^{2} F_{2}
                 + F_{1} G_{2}^{2}
                 + 4F_{2} G_{2}^{2}
                 + 3F_{1} G_{3}^{2}
                 + 4F_{2} G_{3}^{2}
           \bigr) 
           \nonumber \\ && \hspace{3em} %%
         - 12b_{y}^{2}
           \bigl(
                   2F_{1}^{2} F_{2}
                + 3F_{1} F_{2}^{2}
                + 8F_{2}^{3}
                + F_{1} G_{1}^{2}
                + 8F_{2} G_{1}^{2}
                + 2F_{2} G_{2}^{2}
                 - 2G_{1} G_{2} G_{3}
                + 3F_{1} G_{3}^{2}
                + 10F_{2} G_{3}^{2}
           \bigr)
           \nonumber \\ && \hspace{3em} %%
         - 12b_{z}^{2}
           \bigl(
                    7F_{1}^{3}
                 + 20F_{1}^{2} F_{2}
                 + 21F_{1} F_{2}^{2}
                 + 8F_{2}^{3}
                 + 7F_{1} G_{1}^{2}
                 + 8F_{2} G_{1}^{2}
                 + 7F_{1} G_{2}^{2}
                 + 10 F_{2} G_{2}^{2}
                 + 2G_{1} G_{2} G_{3}
                 + 2F_{1} G_{3}^{2}
                 + 2F_{2} G_{3}^{2}
           \bigr)
   \Bigr),
\nonumber \\
\end{eqnarray}
for $F_{2}$,
\begin{eqnarray}
&&
   - \nabla_{\!d} \frac{\delta {f}}{\delta (\nabla_{\!d}G_{1})}
   + \frac{\delta {f}}{\delta G_{1}}
\nonumber \\ %%
&=&
   -\frac{{K}^{(0)}}{4}
   \Bigl(
          2
          \cos\theta\sin\theta\sin\varphi
          \nabla_{\!d}^{2}
          F_{1}
       + 2
          \bigl( 2 - \sin^{2}\theta \cos^{2}\varphi \bigr) \nabla_{\!d}^{2}
          G_{1}
       + 2
          \sin^{2}\theta\cos\varphi\sin\varphi
          \nabla_{\!d}^{2}
          G_{2}
       + 2
          \cos\theta\sin\theta\cos\varphi
          \nabla_{\!d}^{2}
          G_{3}
   \Bigr)
\nonumber \\ && %%
+
  {\alpha}^{(0)} 4 G_{1}
   \nonumber \\ && %%
+{\beta}^{(0)} 8 G_{1}
   \bigl(
            F_{1}^{2}+F_{2} F_{1}
            +F_{2}^{2}+G_{1}^{2}
            +G_{2}^{2}+G_{3}^{2}
    \bigr) 
   \nonumber \\ && %%
+{\gamma}^{(0)}
   24
   \Bigl(
           12 G_{1}^{5}
         +28 F_{1}^{2} G_{1}^{3}
         +24 F_{2}^{2} G_{1}^{3}
         +24 G_{2}^{2} G_{1}^{3}
         +24 G_{3}^{2} G_{1}^{3}
         +24 F_{1} F_{2} G_{1}^{3}
         +12 F_{1} G_{2} G_{3} G_{1}^{2}
         +12 F_{1}^{4} G_{1}
         +12 F_{2}^{4} G_{1}
            \nonumber \\ && \hspace{4em} %%
         +12 G_{2}^{4} G_{1}
         +12 G_{3}^{4} G_{1}
         +24 F_{1} F_{2}^{3} G_{1}
         +40 F_{1}^{2} F_{2}^{2} G_{1}
         +24 F_{1}^{2} G_{2}^{2} G_{1}
         +24 F_{2}^{2} G_{2}^{2} G_{1}
         +28 F_{1} F_{2} G_{2}^{2} G_{1}
            \nonumber \\ && \hspace{4em} %%
         +20 F_{1}^{2} G_{3}^{2} G_{1}
         +24 F_{2}^{2} G_{3}^{2} G_{1}
         +32 G_{2}^{2} G_{3}^{2} G_{1}
         +20 F_{1} F_{2} G_{3}^{2} G_{1}
         +28 F_{1}^{3} F_{2} G_{1}
         -4 F_{1} G_{2} G_{3}^{3}
         -4 F_{2} G_{2} G_{3}^{3}
         +4 F_{2} G_{2}^{3} G_{3}
            \nonumber \\ && \hspace{4em} %%
         +4 F_{1} F_{2}^{2} G_{2} G_{3}
         +4 F_{1}^{2} F_{2} G_{2} G_{3}
   \Bigr)
   \nonumber \\ && %%
+{\delta}^{(0)}
   384
   \Bigl(
           8 G_{1}^{7}
         +33 F_{1}^{2} G_{1}^{5}
         +24 F_{2}^{2} G_{1}^{5}
         +24 G_{2}^{2} G_{1}^{5}
         +24 G_{3}^{2} G_{1}^{5}
         +24 F_{1} F_{2} G_{1}^{5}
         +30 F_{1} G_{2} G_{3} G_{1}^{4}
         +30 F_{1}^{4} G_{1}^{3}
         +24 F_{2}^{4} G_{1}^{3}
            \nonumber \\ && \hspace{4em} %%
         +24 G_{2}^{4} G_{1}^{3}
         +24 G_{3}^{4} G_{1}^{3}
         +48 F_{1} F_{2}^{3} G_{1}^{3}
         +90 F_{1}^{2} F_{2}^{2} G_{1}^{3}
         +54 F_{1}^{2} G_{2}^{2} G_{1}^{3}
         +48 F_{2}^{2} G_{2}^{2} G_{1}^{3}
         +60 F_{1} F_{2} G_{2}^{2} G_{1}^{3}
         +42 F_{1}^{2} G_{3}^{2} G_{1}^{3}
            \nonumber \\ && \hspace{4em} %%
         +48 F_{2}^{2} G_{3}^{2} G_{1}^{3}
         +72 G_{2}^{2} G_{3}^{2} G_{1}^{3}
         +36 F_{1} F_{2} G_{3}^{2} G_{1}^{3}
         +66 F_{1}^{3} F_{2} G_{1}^{3}
         -18 F_{2} G_{2} G_{3}^{3} G_{1}^{2}
         +18 F_{1} G_{2}^{3} G_{3} G_{1}^{2}
         +18 F_{2} G_{2}^{3} G_{3} G_{1}^{2}
            \nonumber \\ && \hspace{4em} %%
         +18 F_{1}^{3} G_{2} G_{3} G_{1}^{2}
         +36 F_{1} F_{2}^{2} G_{2} G_{3} G_{1}^{2}
         +36 F_{1}^{2} F_{2} G_{2} G_{3} G_{1}^{2}
         +8 F_{1}^{6} G_{1}
         +8 F_{2}^{6} G_{1}
         +8 G_{2}^{6} G_{1}
         +8 G_{3}^{6} G_{1}
         +24 F_{1} F_{2}^{5} G_{1}
            \nonumber \\ && \hspace{4em} %%
         +57 F_{1}^{2} F_{2}^{4} G_{1}
         +24 F_{1}^{2} G_{2}^{4} G_{1}
         +27 F_{2}^{2} G_{2}^{4} G_{1}
         +30 F_{1} F_{2} G_{2}^{4} G_{1}
         +21 F_{1}^{2} G_{3}^{4} G_{1}
         +27 F_{2}^{2} G_{3}^{4} G_{1}
         +36 G_{2}^{2} G_{3}^{4} G_{1}
            \nonumber \\ && \hspace{4em} %%
         +24 F_{1} F_{2} G_{3}^{4} G_{1}
         +74 F_{1}^{3} F_{2}^{3} G_{1}
         +63 F_{1}^{4} F_{2}^{2} G_{1}
         +24 F_{1}^{4} G_{2}^{2} G_{1}
         +24 F_{2}^{4} G_{2}^{2} G_{1}
         +60 F_{1} F_{2}^{3} G_{2}^{2} G_{1}
         +90 F_{1}^{2} F_{2}^{2} G_{2}^{2} G_{1}
            \nonumber \\ && \hspace{4em} %%
         +60 F_{1}^{3} F_{2} G_{2}^{2} G_{1}
         +18 F_{1}^{4} G_{3}^{2} G_{1}
         +24 F_{2}^{4} G_{3}^{2} G_{1}
         +36 G_{2}^{4} G_{3}^{2} G_{1}
         +36 F_{1} F_{2}^{3} G_{3}^{2} G_{1}
         +54 F_{1}^{2} F_{2}^{2} G_{3}^{2} G_{1}
         +54 F_{1}^{2} G_{2}^{2} G_{3}^{2} G_{1}
            \nonumber \\ && \hspace{4em} %%
         +54 F_{2}^{2} G_{2}^{2} G_{3}^{2} G_{1}
         +54 F_{1} F_{2} G_{2}^{2} G_{3}^{2} G_{1}
         +36 F_{1}^{3} F_{2} G_{3}^{2} G_{1}
         +30 F_{1}^{5} F_{2} G_{1}
         -6 F_{1} G_{2} G_{3}^{5}
         -6 F_{2} G_{2} G_{3}^{5}
         -6 F_{1} G_{2}^{3} G_{3}^{3}
            \nonumber \\ && \hspace{4em} %%
         -6 F_{1}^{3} G_{2} G_{3}^{3}
         -6 F_{2}^{3} G_{2} G_{3}^{3}
         -6 F_{1} F_{2}^{2} G_{2} G_{3}^{3}
         -6 F_{1}^{2} F_{2} G_{2} G_{3}^{3}
         +6 F_{2} G_{2}^{5} G_{3}
         +6 F_{2}^{3} G_{2}^{3} G_{3}
         +12 F_{1} F_{2}^{2} G_{2}^{3} G_{3}
            \nonumber \\ && \hspace{4em} %%
         +12 F_{1}^{2} F_{2} G_{2}^{3} G_{3}
         +6 F_{1} F_{2}^{4} G_{2} G_{3}
         +12 F_{1}^{2} F_{2}^{3} G_{2} G_{3}
         +12 F_{1}^{3} F_{2}^{2} G_{2} G_{3}
         +6 F_{1}^{4} F_{2} G_{2} G_{3}
   \Bigr)
   \nonumber \\ && %%
+{\beta}^{(2)} 2
   \bigl( b_{y}^{2}+b_{z}^{2} \bigr) G_{1}
   \nonumber \\ && %%
+{\beta}^{(4)} 2
   \bigl(b_{x}^{2}+b_{y}^{2}+b_{z}^{2}\bigr)
   \bigl( b_{y}^{2}+b_{z}^{2} \bigr) G_{1}
   \nonumber \\ && %%
+{\gamma}^{(2)}
   \Bigl(
       - 12b_{x}^{2}
         \bigl(
                2F_{1}^{2} G_{1}
             + 4G_{1} G_{2}^{2}
              - 2F_{1} G_{2} G_{3}
             + 4G_{1} G_{3}^{2}
         \bigr)
         \nonumber \\ && \hspace{3em} %%
      - 12b_{y}^{2}
         \bigl(
                4F_{1}^{2} G_{1}
             + 2F_{1} F_{2} G_{1}
             + 8F_{2}^{2} G_{1}
             + 8G_{1}^{3}
             + 4G_{1} G_{2}^{2}
              - 2F_{2} G_{2} G_{3}
             + 8G_{1} G_{3}^{2}
         \bigr)
         \nonumber \\ && \hspace{3em} %%
      - 12b_{z}^{2}
         \bigl(
                 10F_{1}^{2} G_{1}
              + 14F_{1} F_{2} G_{1}
              + 8F_{2}^{2} G_{1}
              + 8G_{1}^{3}
              + 8G_{1} G_{2}^{2}
              + 2F_{1} G_{2} G_{3}
              + 2F_{2} G_{2} G_{3}
              + 4G_{1} G_{3}^{2}
         \bigr)
   \Bigr),
\end{eqnarray}
for $G_{1}$,
\begin{eqnarray}
&&
   - \nabla_{\!d} \frac{\delta {f}}{\delta (\nabla_{\!d}G_{2})}
   + \frac{\delta {f}}{\delta G_{2}}
\nonumber \\ %%
&=&
   -\frac{{K}^{(0)}}{4}
   \Bigl(
          2
          \cos\theta\sin\theta\cos\varphi
          \nabla_{\!d}^{2}
          F_{2}
       + 2
          \sin^{2}\theta\cos\varphi\sin\varphi
          \nabla_{\!d}^{2}
          G_{1}
       + 2
          \bigl( 2 - \sin^{2}\theta \sin^{2}\varphi \bigr) \nabla_{\!d}^{2}
          G_{2}
       + 2
          \cos\theta\sin\theta\sin\varphi
          \nabla_{\!d}^{2}
          G_{3}
   \Bigr)
\nonumber \\ && %%
+{\alpha}^{(0)}
  4 G_{2}
   \nonumber \\ && %%
+{\beta}^{(0)}
  8 G_{2}
  \bigl(F_{1}^{2}+F_{2} F_{1}+F_{2}^{2}+G_{1}^{2}+G_{2}^{2}+G_{3}^{2}\bigr)
   \nonumber \\ && %%
+{\gamma}^{(0)}
   24
   \Bigl(
         12 G_{2}^{5}
        +24 F_{1}^{2} G_{2}^{3}
        +28 F_{2}^{2} G_{2}^{3}
        +24 G_{1}^{2} G_{2}^{3}
        +24 G_{3}^{2} G_{2}^{3}
        +24 F_{1} F_{2} G_{2}^{3}
        +12 F_{2} G_{1} G_{3} G_{2}^{2}
        +12 F_{1}^{4} G_{2}
        +12 F_{2}^{4} G_{2}
          \nonumber \\ && \hspace{4em} %%
        +12 G_{1}^{4} G_{2}
        +12 G_{3}^{4} G_{2}
        +28 F_{1} F_{2}^{3} G_{2}
        +40 F_{1}^{2} F_{2}^{2} G_{2}
        +24 F_{1}^{2} G_{1}^{2} G_{2}
        +24 F_{2}^{2} G_{1}^{2} G_{2}
        +28 F_{1} F_{2} G_{1}^{2} G_{2}
        +24 F_{1}^{2} G_{3}^{2} G_{2}
          \nonumber \\ && \hspace{4em} %%
        +20 F_{2}^{2} G_{3}^{2} G_{2}
        +32 G_{1}^{2} G_{3}^{2} G_{2}
        +20 F_{1} F_{2} G_{3}^{2} G_{2}
        +24 F_{1}^{3} F_{2} G_{2}
        -4 F_{1} G_{1} G_{3}^{3}
        -4 F_{2} G_{1} G_{3}^{3}
        +4 F_{1} G_{1}^{3} G_{3}
        +4 F_{1} F_{2}^{2} G_{1} G_{3}
          \nonumber \\ && \hspace{4em} %%
        +4 F_{1}^{2} F_{2} G_{1} G_{3}
   \Bigr)
   \nonumber \\ && %%
+{\delta}^{(0)}
   384
   \bigl(
         8 G_{2}^{7}
       +24 F_{1}^{2} G_{2}^{5}
       +33 F_{2}^{2} G_{2}^{5}
       +24 G_{1}^{2} G_{2}^{5}
       +24 G_{3}^{2} G_{2}^{5}
       +24 F_{1} F_{2} G_{2}^{5}
       +30 F_{2} G_{1} G_{3} G_{2}^{4}
       +24 F_{1}^{4} G_{2}^{3}
       +30 F_{2}^{4} G_{2}^{3}
         \nonumber \\ && \hspace{4em} %%
       +24 G_{1}^{4} G_{2}^{3}
       +24 G_{3}^{4} G_{2}^{3}
       +66 F_{1} F_{2}^{3} G_{2}^{3}
       +90 F_{1}^{2} F_{2}^{2} G_{2}^{3}
       +48 F_{1}^{2} G_{1}^{2} G_{2}^{3}
       +54 F_{2}^{2} G_{1}^{2} G_{2}^{3}
       +60 F_{1} F_{2} G_{1}^{2} G_{2}^{3}
       +48 F_{1}^{2} G_{3}^{2} G_{2}^{3}
         \nonumber \\ && \hspace{4em} %%
       +42 F_{2}^{2} G_{3}^{2} G_{2}^{3}
       +72 G_{1}^{2} G_{3}^{2} G_{2}^{3}
       +36 F_{1} F_{2} G_{3}^{2} G_{2}^{3}
       +48 F_{1}^{3} F_{2} G_{2}^{3}
       -18 F_{1} G_{1} G_{3}^{3} G_{2}^{2}
       +18 F_{1} G_{1}^{3} G_{3} G_{2}^{2}
       +18 F_{2} G_{1}^{3} G_{3} G_{2}^{2}
         \nonumber \\ && \hspace{4em} %%
       +18 F_{2}^{3} G_{1} G_{3} G_{2}^{2}
       +36 F_{1} F_{2}^{2} G_{1} G_{3} G_{2}^{2}
       +36 F_{1}^{2} F_{2} G_{1} G_{3} G_{2}^{2}
       +8 F_{1}^{6} G_{2}
       +8 F_{2}^{6} G_{2}
       +8 G_{1}^{6} G_{2}
       +8 G_{3}^{6} G_{2}
       +30 F_{1} F_{2}^{5} G_{2}
         \nonumber \\ && \hspace{4em} %%
       +63 F_{1}^{2} F_{2}^{4} G_{2}
       +27 F_{1}^{2} G_{1}^{4} G_{2}
       +24 F_{2}^{2} G_{1}^{4} G_{2}
       +30 F_{1} F_{2} G_{1}^{4} G_{2}
       +27 F_{1}^{2} G_{3}^{4} G_{2}
       +21 F_{2}^{2} G_{3}^{4} G_{2}
       +36 G_{1}^{2} G_{3}^{4} G_{2}
         \nonumber \\ && \hspace{4em} %%
       +24 F_{1} F_{2} G_{3}^{4} G_{2}
       +74 F_{1}^{3} F_{2}^{3} G_{2}
       +57 F_{1}^{4} F_{2}^{2} G_{2}
       +24 F_{1}^{4} G_{1}^{2} G_{2}
       +24 F_{2}^{4} G_{1}^{2} G_{2}
       +60 F_{1} F_{2}^{3} G_{1}^{2} G_{2}
       +90 F_{1}^{2} F_{2}^{2} G_{1}^{2} G_{2}
         \nonumber \\ && \hspace{4em} %%
       +60 F_{1}^{3} F_{2} G_{1}^{2} G_{2}
       +24 F_{1}^{4} G_{3}^{2} G_{2}
       +18 F_{2}^{4} G_{3}^{2} G_{2}
       +36 G_{1}^{4} G_{3}^{2} G_{2}
       +36 F_{1} F_{2}^{3} G_{3}^{2} G_{2}
       +54 F_{1}^{2} F_{2}^{2} G_{3}^{2} G_{2}
       +54 F_{1}^{2} G_{1}^{2} G_{3}^{2} G_{2}
         \nonumber \\ && \hspace{4em} %%
       +54 F_{2}^{2} G_{1}^{2} G_{3}^{2} G_{2}
       +54 F_{1} F_{2} G_{1}^{2} G_{3}^{2} G_{2}
       +36 F_{1}^{3} F_{2} G_{3}^{2} G_{2}
       +24 F_{1}^{5} F_{2} G_{2}
       -6 F_{1} G_{1} G_{3}^{5}
       -6 F_{2} G_{1} G_{3}^{5}
       -6 F_{2} G_{1}^{3} G_{3}^{3}
         \nonumber \\ && \hspace{4em} %%
       -6 F_{1}^{3} G_{1} G_{3}^{3}
       -6 F_{2}^{3} G_{1} G_{3}^{3}
       -6 F_{1} F_{2}^{2} G_{1} G_{3}^{3}
       -6 F_{1}^{2} F_{2} G_{1} G_{3}^{3}
       +6 F_{1} G_{1}^{5} G_{3}
       +6 F_{1}^{3} G_{1}^{3} G_{3}
         \nonumber \\ && \hspace{4em} %%
       +12 F_{1} F_{2}^{2} G_{1}^{3} G_{3}
       +12 F_{1}^{2} F_{2} G_{1}^{3} G_{3}
       +6 F_{1} F_{2}^{4} G_{1} G_{3}
       +12 F_{1}^{2} F_{2}^{3} G_{1} G_{3}
       +12 F_{1}^{3} F_{2}^{2} G_{1} G_{3}
       +6 F_{1}^{4} F_{2} G_{1} G_{3}
   \Bigr)
   \nonumber \\ && %%
+{\beta}^{(2)} 2
  \bigl( b_{x}^{2}+b_{z}^{2} \bigr) G_{2}
   \nonumber \\ && %%
+{\beta}^{(4)} 2
  \bigl(b_{x}^{2}+b_{y}^{2}+b_{z}^{2}\bigr)
  \bigl( b_{x}^{2}+b_{z}^{2} \bigr) G_{2}
   \nonumber \\ && %%
+{\gamma}^{(2)}
   \Bigl(
        - 12b_{x}^{2}
           \bigl(
                    8F_{1}^{2} G_{2}
                 + 2F_{1} F_{2} G_{2}
                 + 4F_{2}^{2} G_{2}
                 + 4G_{1}^{2} G_{2}
                 + 8G_{2}^{3}
                  - 2F_{1}G_{1}G_{3}
                 + 8G_{2} G_{3}^{2} 
            \bigr)
            \nonumber \\ && \hspace{3em} %%
         - 12b_{y}^{2}
            \bigl(
                    2F_{2}^{2} G_{2}
                 + 4G_{1}^{2} G_{2}
                  - 2F_{2} G_{1} G_{3}
                + 4G_{2} G_{3}^{2}
            \bigr)
            \nonumber \\ && \hspace{3em} %%
         - 12b_{z}^{2}
            \bigl(
                    8F_{1}^{2} G_{2}
                 + 14F_{1} F_{2} G_{2}
                 + 10F_{2}^{2} G_{2}
                 + 8G_{1}^{2} G_{2}
                 + 8G_{2}^{3}
                 + 2F_{1} G_{1} G_{3}
                 + 2F_{2} G_{1} G_{3}
                 + 4G_{2} G_{3}^{2}
            \bigr)
   \Bigr),
\end{eqnarray}
for $G_{2}$, and
\begin{eqnarray}
&&
   - \nabla_{\!d} \frac{\delta {f}}{\delta (\nabla_{\!d}G_{3})}
   + \frac{\delta {f}}{\delta G_{3}}
\nonumber \\ %%
&=&
   -\frac{{K}^{(0)}}{4}
   \Bigl(
       - 2
         \sin^{2}\theta\cos\varphi\sin\varphi
         \nabla_{\!d}^{2}
         F_{1}
       - 2
         \sin^{2}\theta\cos\varphi\sin\varphi
         \nabla_{\!d}^{2}
         F_{2}
         \nonumber \\ && \hspace{2em} %%
      + 2
         \cos\theta\sin\theta\cos\varphi
         \nabla_{\!d}^{2}
         G_{1}
      + 2
         \cos\theta\sin\theta\sin\varphi
         \nabla_{\!d}^{2}
         G_{2}
      + 2\bigl(1+\sin^{2}\theta\bigr)
         \nabla_{\!d}
         G_{3}
   \Bigr)
\nonumber \\ && %%
+{\alpha}^{(0)} 4 G_{3}
   \nonumber \\ && %%
+{\beta}^{(0)}
   8 G_{3}
   \bigl(F_{1}^{2}+F_{2} F_{1}+F_{2}^{2}+G_{1}^{2}+G_{2}^{2}+G_{3}^{2}\bigr)
   \nonumber \\ && %%
+{\gamma}^{(0)}
   24
   \Bigl(
           12 G_{3}^{5}
          +28 F_{1}^{2} G_{3}^{3}
          +28 F_{2}^{2} G_{3}^{3}
          +24 G_{1}^{2} G_{3}^{3}
          +24 G_{2}^{2} G_{3}^{3}
          +32 F_{1} F_{2} G_{3}^{3}
          -12 F_{1} G_{1} G_{2} G_{3}^{2}
          -12 F_{2} G_{1} G_{2} G_{3}^{2}
          +12 F_{1}^{4} G_{3}
            \nonumber \\ && \hspace{4em} %%
          +12 F_{2}^{4} G_{3}
          +12 G_{1}^{4} G_{3}
          +12 G_{2}^{4} G_{3}
          +20 F_{1} F_{2}^{3} G_{3}
          +28 F_{1}^{2} F_{2}^{2} G_{3}
          +20 F_{1}^{2} G_{1}^{2} G_{3}
          +24 F_{2}^{2} G_{1}^{2} G_{3}
          +20 F_{1} F_{2} G_{1}^{2} G_{3}
            \nonumber \\ && \hspace{4em} %%
          +24 F_{1}^{2} G_{2}^{2} G_{3}
          +20 F_{2}^{2} G_{2}^{2} G_{3}
          +32 G_{1}^{2} G_{2}^{2} G_{3}
          +20 F_{1} F_{2} G_{2}^{2} G_{3}
          +20 F_{1}^{3} F_{2} G_{3}
          +4 F_{2} G_{1} G_{2}^{3}
          +4 F_{1} G_{1}^{3} G_{2}
            \nonumber \\ && \hspace{4em} %%
          +4 F_{1} F_{2}^{2} G_{1} G_{2}
          +4 F_{1}^{2} F_{2} G_{1} G_{2}
   \Bigr)
   \nonumber \\ && %%
+{\delta}^{(0)}
   384
   \Bigl(
           8 G_{3}^{7}
         +33 F_{1}^{2} G_{3}^{5}
         +33 F_{2}^{2} G_{3}^{5}
         +24 G_{1}^{2} G_{3}^{5}
         +24 G_{2}^{2} G_{3}^{5}
         +42 F_{1} F_{2} G_{3}^{5}
         -30 F_{1} G_{1} G_{2} G_{3}^{4}
         -30 F_{2} G_{1} G_{2} G_{3}^{4}
         +30 F_{1}^{4} G_{3}^{3}
           \nonumber \\ && \hspace{4em} %%
         +30 F_{2}^{4} G_{3}^{3}
         +24 G_{1}^{4} G_{3}^{3}
         +24 G_{2}^{4} G_{3}^{3}
         +54 F_{1} F_{2}^{3} G_{3}^{3}
         +72 F_{1}^{2} F_{2}^{2} G_{3}^{3}
         +42 F_{1}^{2} G_{1}^{2} G_{3}^{3}
         +54 F_{2}^{2} G_{1}^{2} G_{3}^{3}
         +48 F_{1} F_{2} G_{1}^{2} G_{3}^{3}
           \nonumber \\ && \hspace{4em} %%
         +54 F_{1}^{2} G_{2}^{2} G_{3}^{3}
         +42 F_{2}^{2} G_{2}^{2} G_{3}^{3}
         +72 G_{1}^{2} G_{2}^{2} G_{3}^{3}
         +48 F_{1} F_{2} G_{2}^{2} G_{3}^{3}
         +54 F_{1}^{3} F_{2} G_{3}^{3}
         -18 F_{1} G_{1} G_{2}^{3} G_{3}^{2}
         -18 F_{2} G_{1}^{3} G_{2} G_{3}^{2}
           \nonumber \\ && \hspace{4em} %%
         -18 F_{1}^{3} G_{1} G_{2} G_{3}^{2}
         -18 F_{2}^{3} G_{1} G_{2} G_{3}^{2}
         -18 F_{1} F_{2}^{2} G_{1} G_{2} G_{3}^{2}
         -18 F_{1}^{2} F_{2} G_{1} G_{2} G_{3}^{2}
         +8 F_{1}^{6} G_{3}
         +8 F_{2}^{6} G_{3}
         +8 G_{1}^{6} G_{3}
         +8 G_{2}^{6} G_{3}
           \nonumber \\ && \hspace{4em} %%
         +18 F_{1} F_{2}^{5} G_{3}
         +33 F_{1}^{2} F_{2}^{4} G_{3}
         +21 F_{1}^{2} G_{1}^{4} G_{3}
         +24 F_{2}^{2} G_{1}^{4} G_{3}
         +18 F_{1} F_{2} G_{1}^{4} G_{3}
         +24 F_{1}^{2} G_{2}^{4} G_{3}
         +21 F_{2}^{2} G_{2}^{4} G_{3}
           \nonumber \\ && \hspace{4em} %%
         +36 G_{1}^{2} G_{2}^{4} G_{3}
         +18 F_{1} F_{2} G_{2}^{4} G_{3}
         +38 F_{1}^{3} F_{2}^{3} G_{3}
         +33 F_{1}^{4} F_{2}^{2} G_{3}
         +18 F_{1}^{4} G_{1}^{2} G_{3}
         +24 F_{2}^{4} G_{1}^{2} G_{3}
         +36 F_{1} F_{2}^{3} G_{1}^{2} G_{3}
           \nonumber \\ && \hspace{4em} %%
         +54 F_{1}^{2} F_{2}^{2} G_{1}^{2} G_{3}
         +36 F_{1}^{3} F_{2} G_{1}^{2} G_{3}
         +24 F_{1}^{4} G_{2}^{2} G_{3}
         +18 F_{2}^{4} G_{2}^{2} G_{3}
         +36 G_{1}^{4} G_{2}^{2} G_{3}
         +36 F_{1} F_{2}^{3} G_{2}^{2} G_{3}
         +54 F_{1}^{2} F_{2}^{2} G_{2}^{2} G_{3}
           \nonumber \\ && \hspace{4em} %%
         +54 F_{1}^{2} G_{1}^{2} G_{2}^{2} G_{3}
         +54 F_{2}^{2} G_{1}^{2} G_{2}^{2} G_{3}
         +54 F_{1} F_{2} G_{1}^{2} G_{2}^{2} G_{3}
         +36 F_{1}^{3} F_{2} G_{2}^{2} G_{3}
         +18 F_{1}^{5} F_{2} G_{3}
         +6 F_{2} G_{1} G_{2}^{5}
         +6 F_{1} G_{1}^{3} G_{2}^{3}
           \nonumber \\ && \hspace{4em} %%
         +6 F_{2} G_{1}^{3} G_{2}^{3}
         +6 F_{2}^{3} G_{1} G_{2}^{3}
         +12 F_{1} F_{2}^{2} G_{1} G_{2}^{3}
         +12 F_{1}^{2} F_{2} G_{1} G_{2}^{3}
         +6 F_{1} G_{1}^{5} G_{2}
         +6 F_{1}^{3} G_{1}^{3} G_{2}
         +12 F_{1} F_{2}^{2} G_{1}^{3} G_{2}
           \nonumber \\ && \hspace{4em} %%
         +12 F_{1}^{2} F_{2} G_{1}^{3} G_{2}
         +6 F_{1} F_{2}^{4} G_{1} G_{2}
         +12 F_{1}^{2} F_{2}^{3} G_{1} G_{2}
         +12 F_{1}^{3} F_{2}^{2} G_{1} G_{2}
         +6 F_{1}^{4} F_{2} G_{1} G_{2}
   \Bigr)
   \nonumber \\ && %%
+{\beta}^{(2)} 2\bigl( b_{x}^{2}+b_{y}^{2} \bigr) G_{3}
   \nonumber \\ && %%
+{\beta}^{(4)} 2
   \bigl(b_{x}^{2}+b_{y}^{2}+b_{z}^{2}\bigr)
   \bigl( b_{x}^{2}+b_{y}^{2} \bigr) G_{3}
   \nonumber \\ && %%
+{\gamma}^{(2)}
   \Bigl(
          - 12b_{x}^{2}
            \bigl(
                    - 2F_{1} G_{1} G_{2}
                   + 10F_{1}^{2} G_{3}
                   + 6F_{1} F_{2} G_{3}
                   + 4F_{2}^{2} G_{3}
                   + 4G_{1}^{2} G_{3}
                   + 8G_{2}^{2} G_{3}
                   + 8G_{3}^{3}
             \bigr)
             \nonumber \\ && \hspace{3em} %%
          - 12b_{y}^{2}
             \bigl(
                    - 2F_{2} G_{1} G_{2}
                   + 4F_{1}^{2} G_{3}
                   + 6F_{1} F_{2} G_{3}
                   + 10F_{2}^{2} G_{3}
                   + 8G_{1}^{2} G_{3}
                   + 4G_{2}^{2} G_{3}
                   + 8G_{3}^{3}
             \bigr)
             \nonumber \\ && \hspace{3em} %%
          - 12b_{z}^{2}
             \bigl(
                      2F_{1} G_{1} G_{2}
                   + 2F_{2} G_{1} G_{2}
                   + 2F_{1}^{2} G_{3}
                   + 4F_{1} F_{2} G_{3}
                   + 2F_{2}^{2} G_{3}
                   + 4G_{1}^{2} G_{3}
                   + 4G_{2}^{2} G_{3}
             \bigr)
   \Bigr),
\end{eqnarray}
for $G_{3}$.

\section{Plots of the three-dimensional vector $\vec{f}(\tilde{d})$ of the domain walls}
\label{sec:plots_configuration_DW}

\begin{figure}[ptb]
\begin{center}
\includegraphics[scale=0.18]{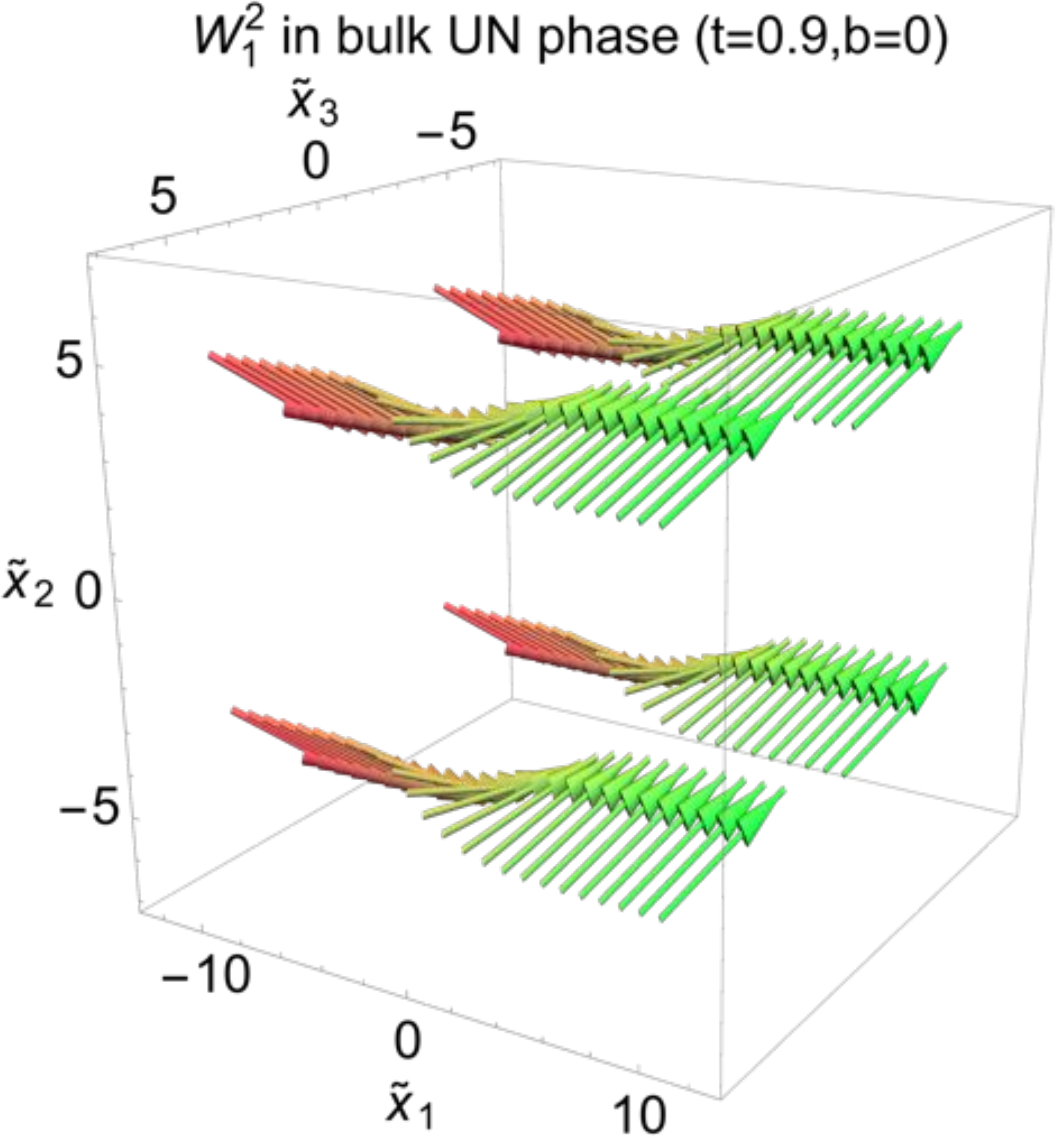}
\hspace{1em}
\includegraphics[scale=0.18]{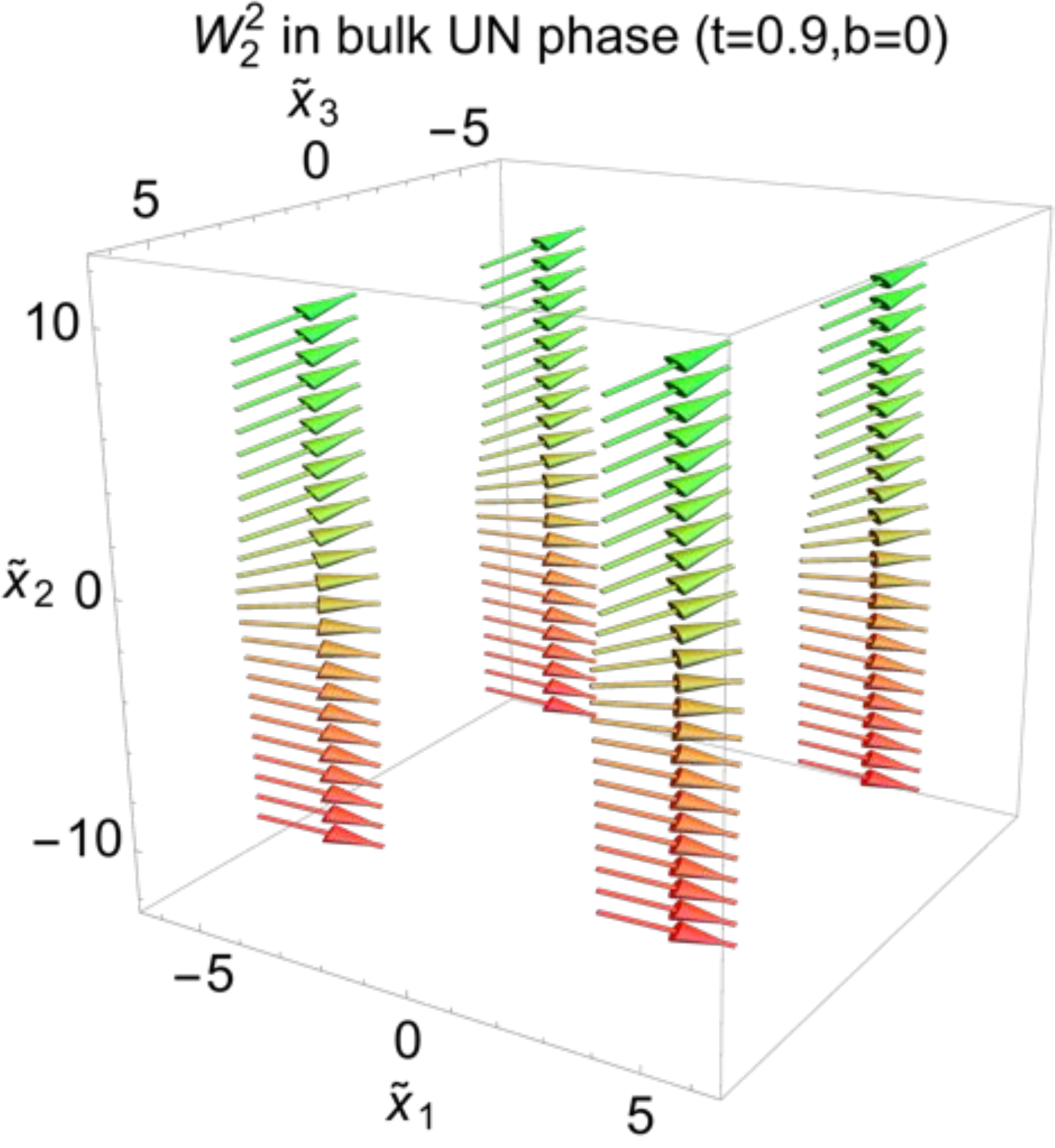}
\hspace{1em}
\includegraphics[scale=0.18]{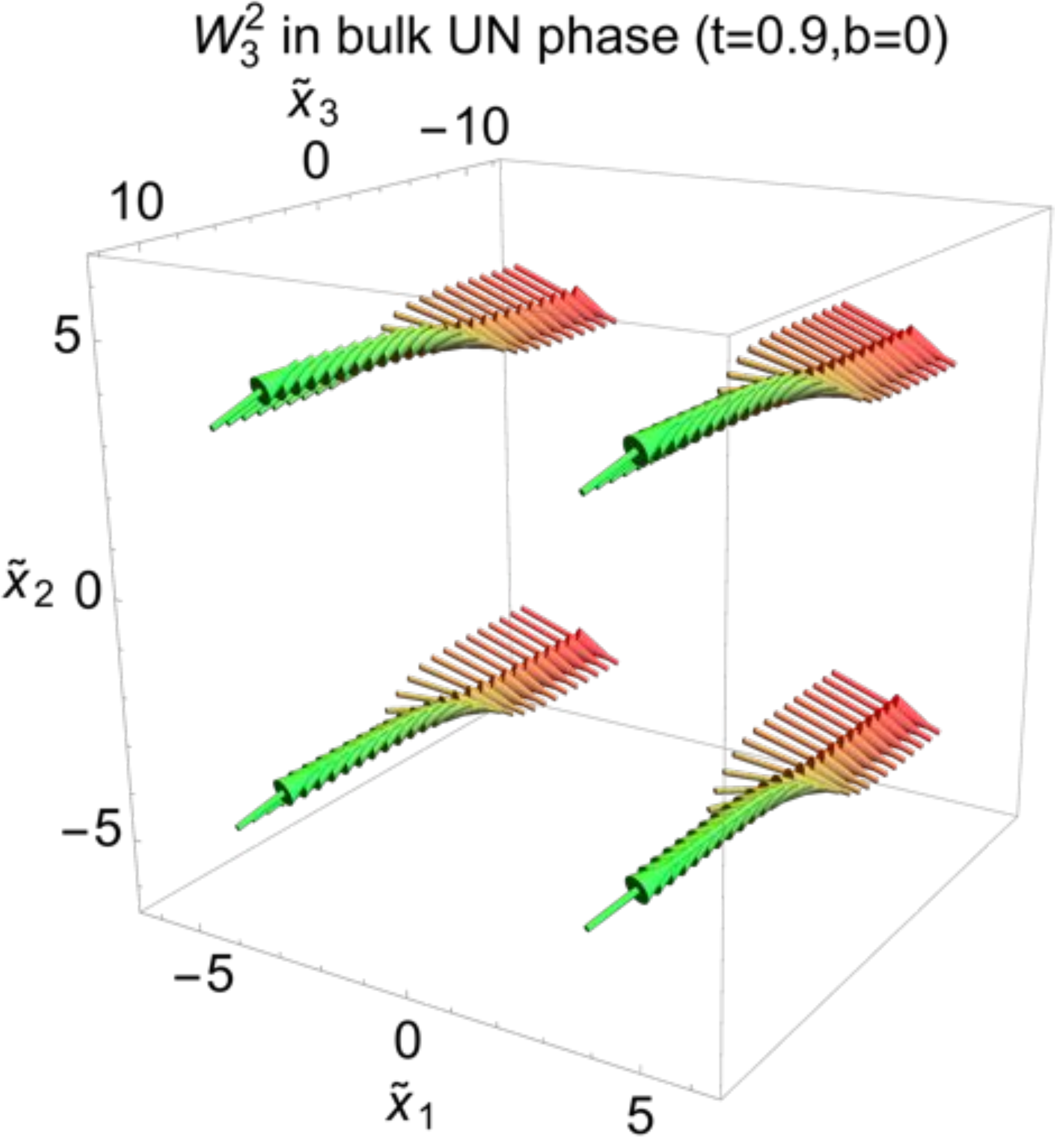}
\\
\vspace{1em}
\includegraphics[scale=0.18]{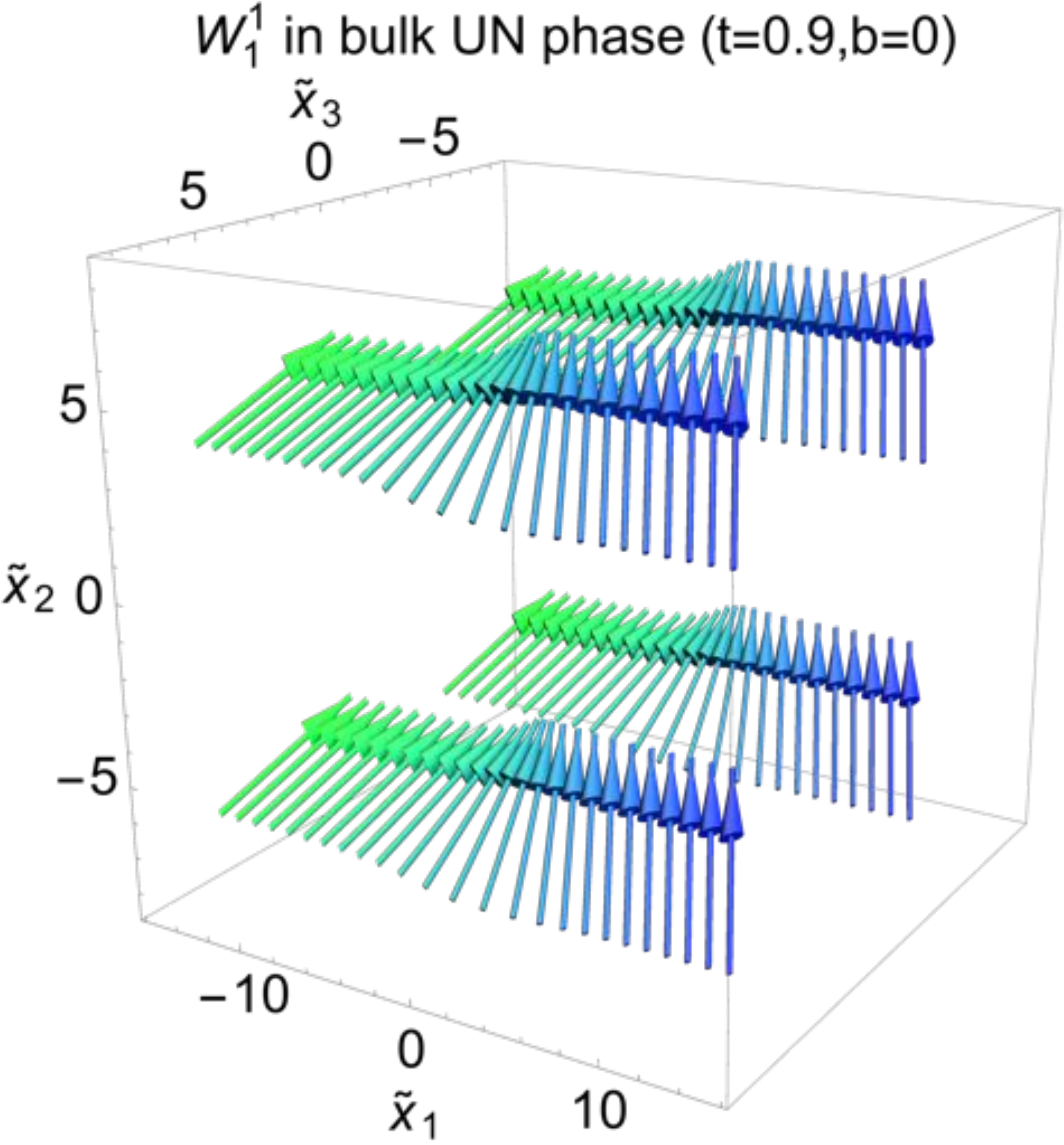}
\hspace{1em}
\includegraphics[scale=0.18]{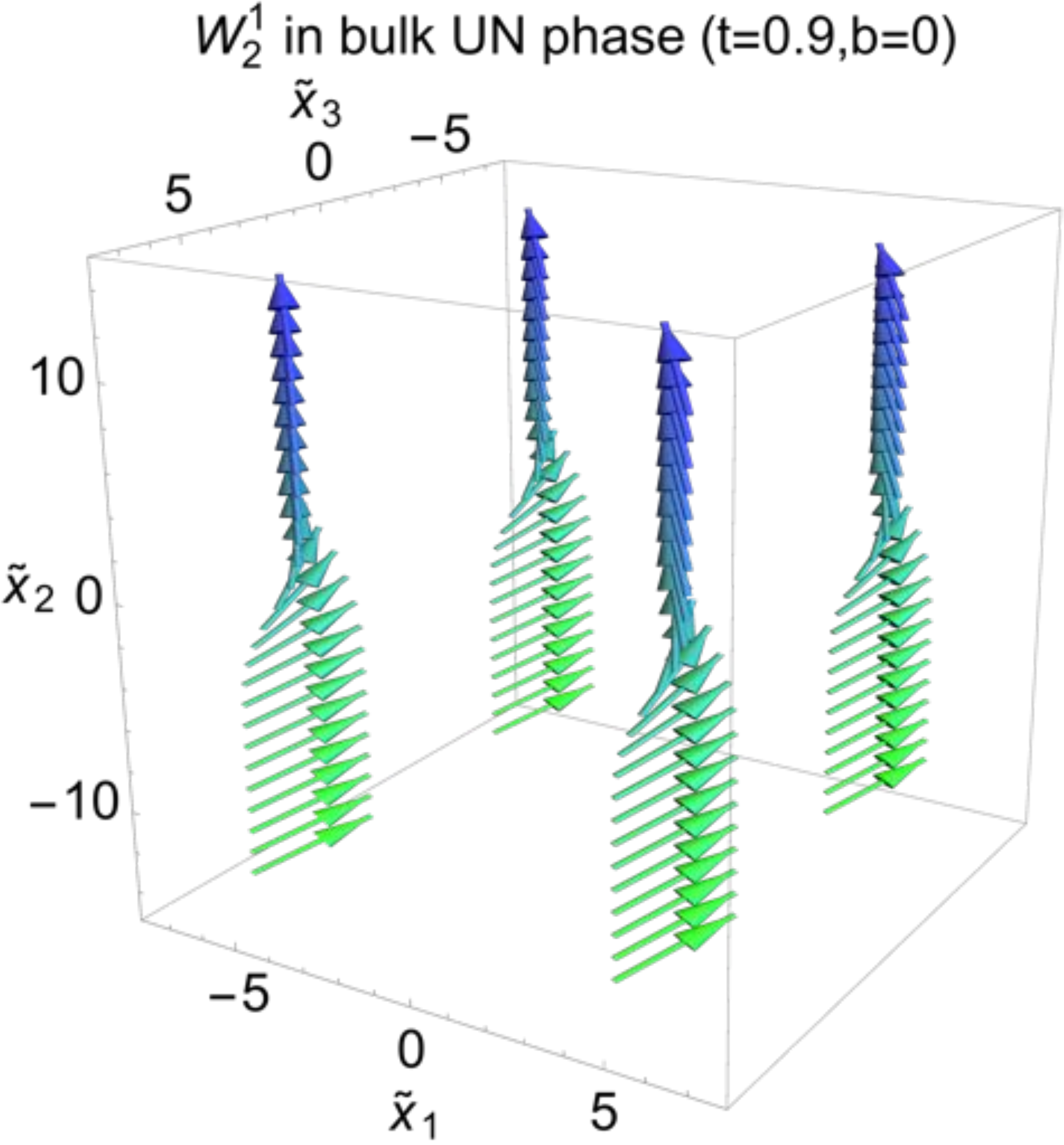}
\hspace{1em}
\includegraphics[scale=0.18]{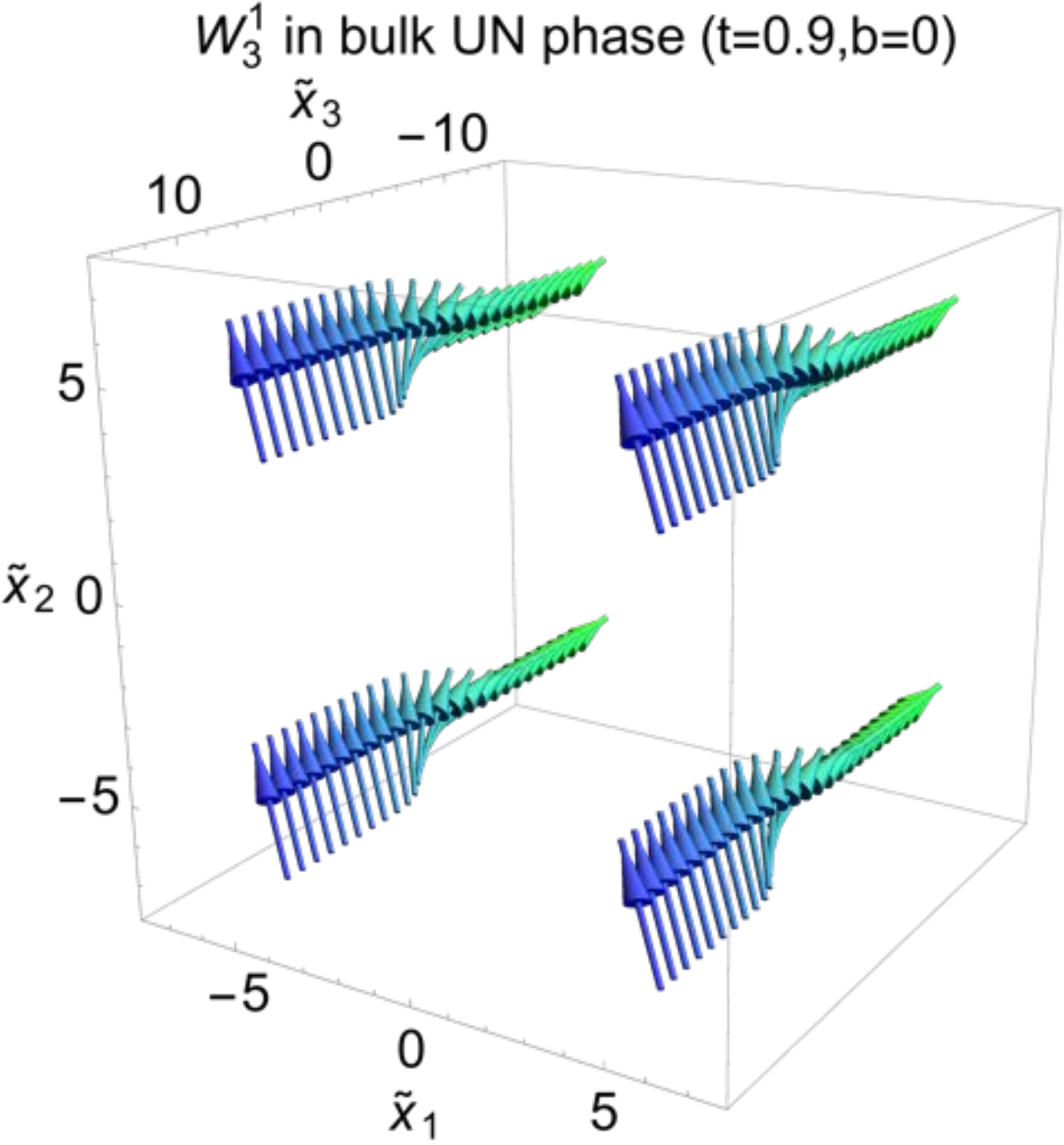}
\\
\vspace{1em}
\includegraphics[scale=0.18]{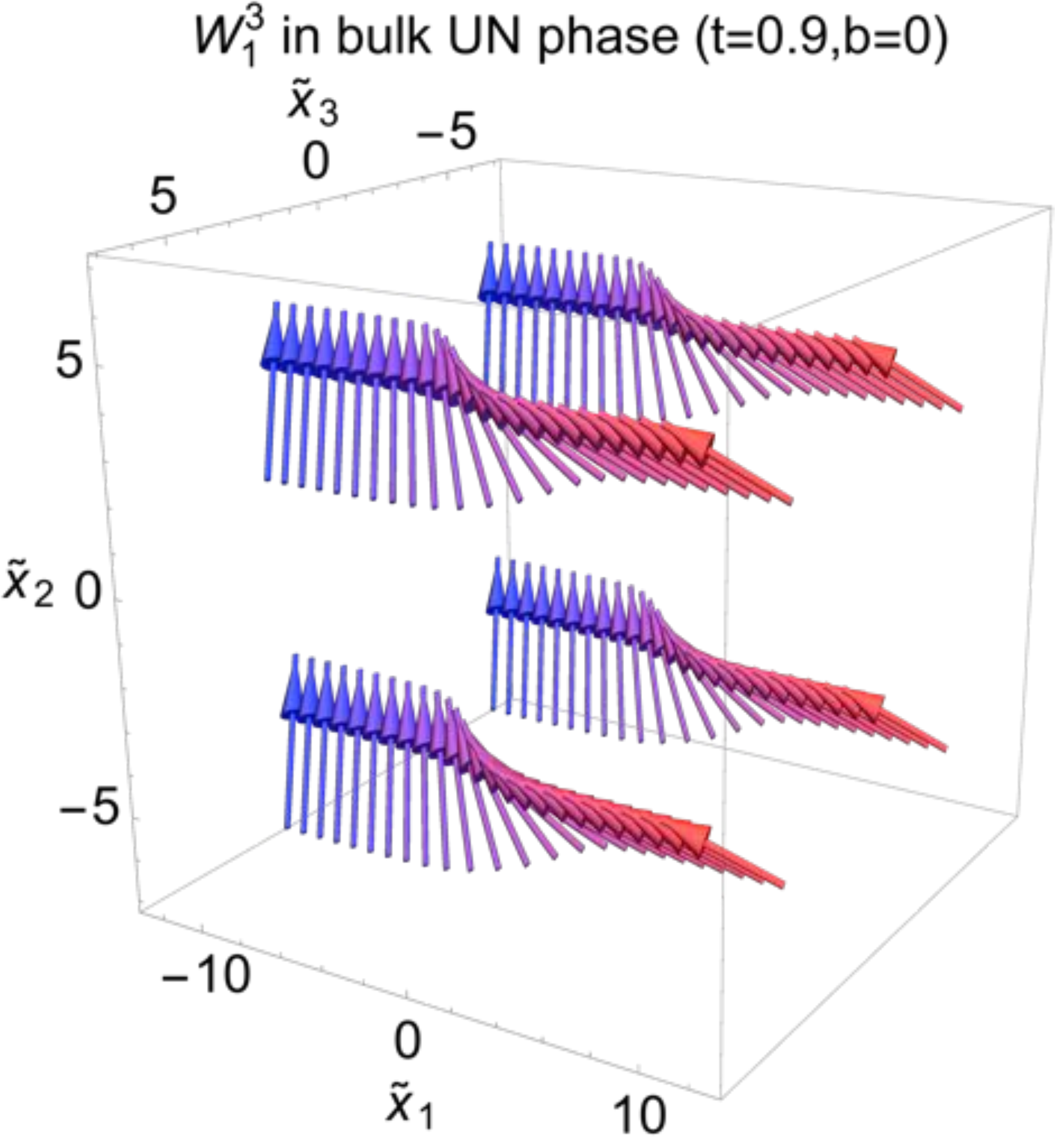}
\hspace{1em}
\includegraphics[scale=0.18]{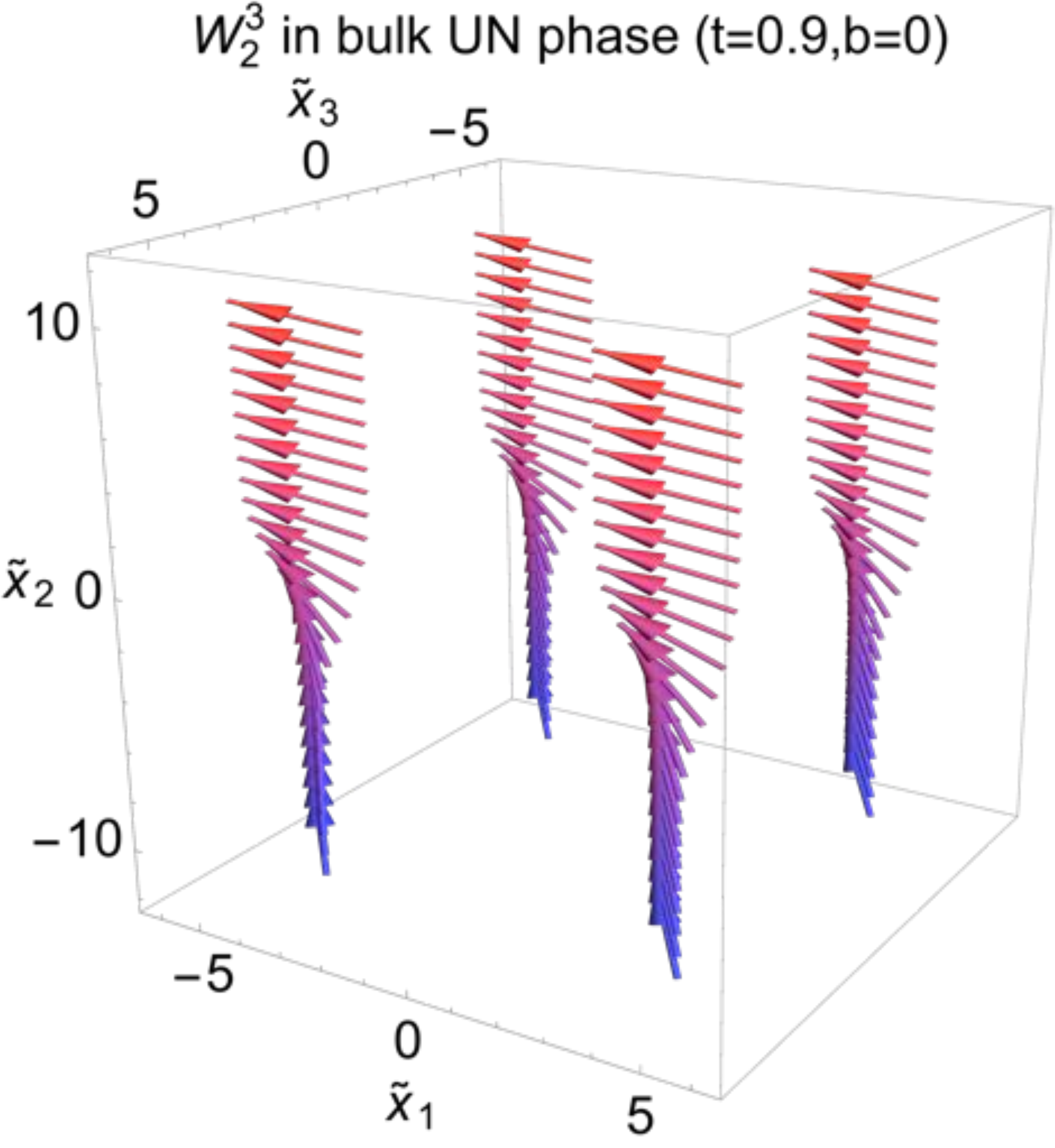}
\hspace{1em}
\includegraphics[scale=0.18]{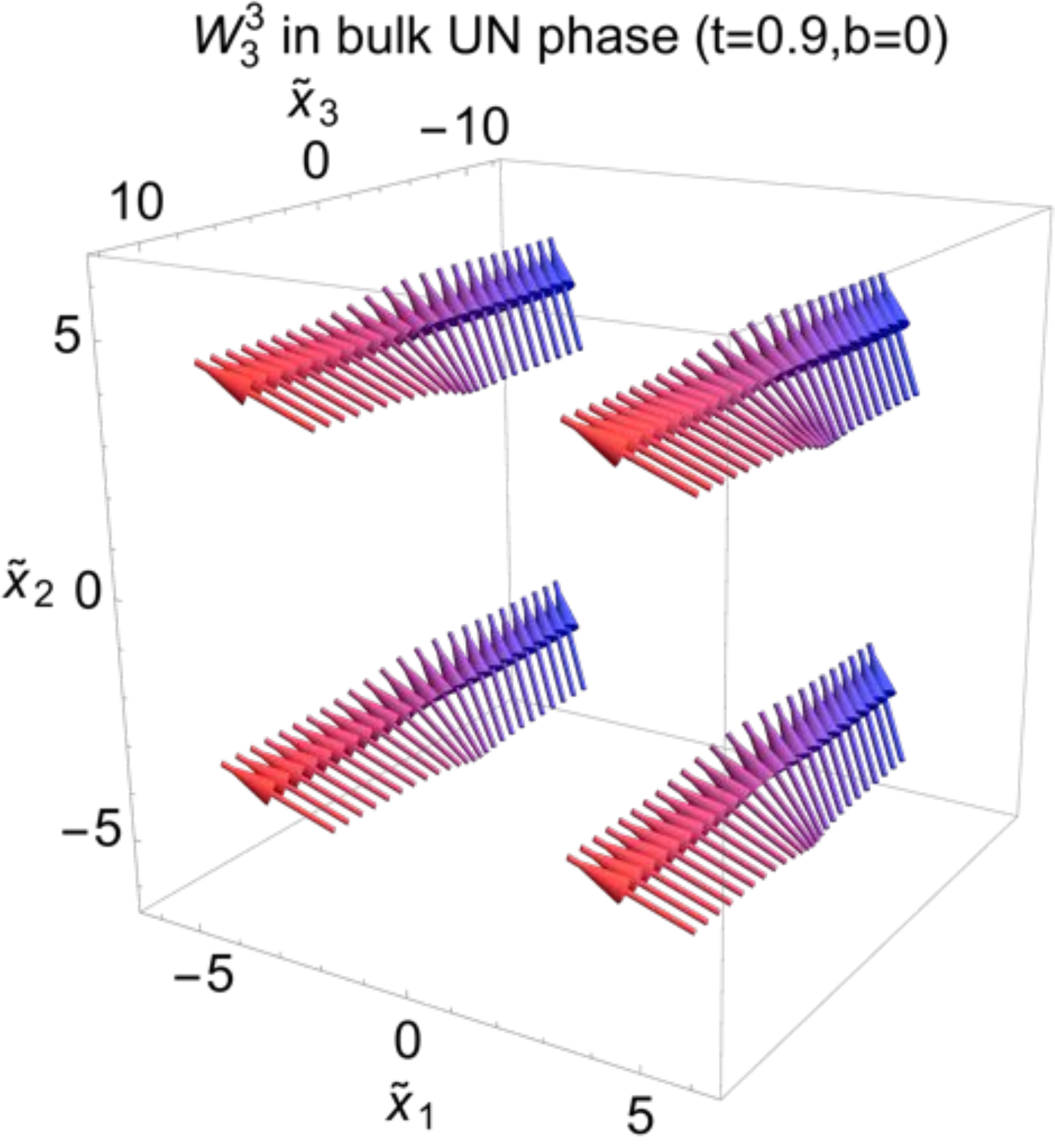}
\caption{The plots of the profile functions as the three-dimensional vectors $\vec{f}(\tilde{d}) = \bigl(f_{1}(\tilde{d};\vec{n}),f_{2}(\tilde{d};\vec{n}),f_{3}(\tilde{d};\vec{n})\bigr)$ with $f_{1}(\tilde{d};\vec{n})+f_{2}(\tilde{d};\vec{n})+f_{3}(\tilde{d};\vec{n})=0$ for the domain walls $W^{2}_{i}$, $W^{1}_{i}$, and $W^{3}_{i}$ ($i=1$, $2$, $3$ the directions along $\tilde{x}_{1}$, $\tilde{x}_{2}$, and $\tilde{x}_{3}$ directions) in the bulk UN phase.}
\label{fig:190521_soliton_3P2_f1f2_UN}
\end{center}
\end{figure}

\begin{figure}[tb]
\begin{center}
\includegraphics[scale=0.18]{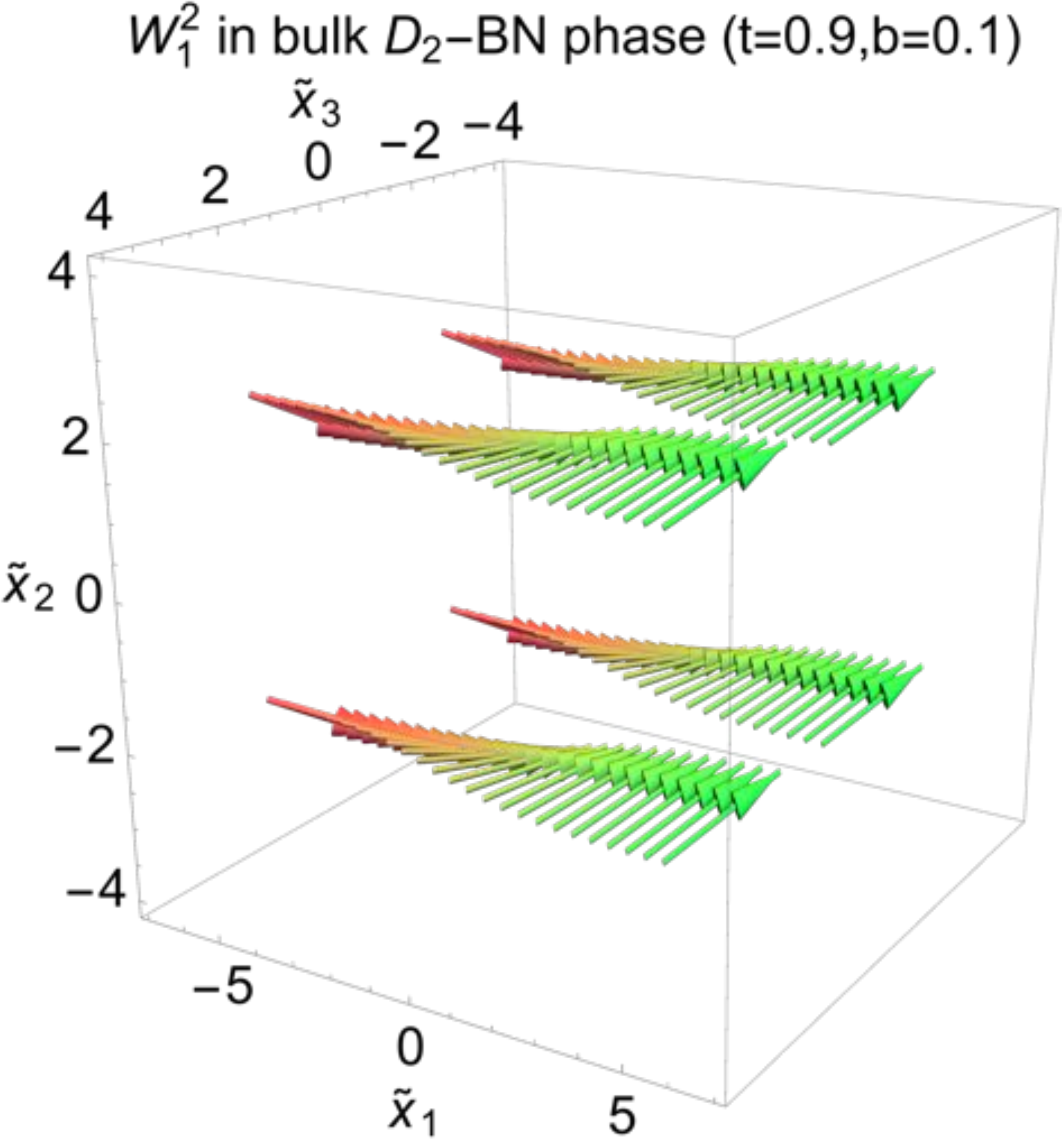}
\hspace{1em}
\includegraphics[scale=0.18]{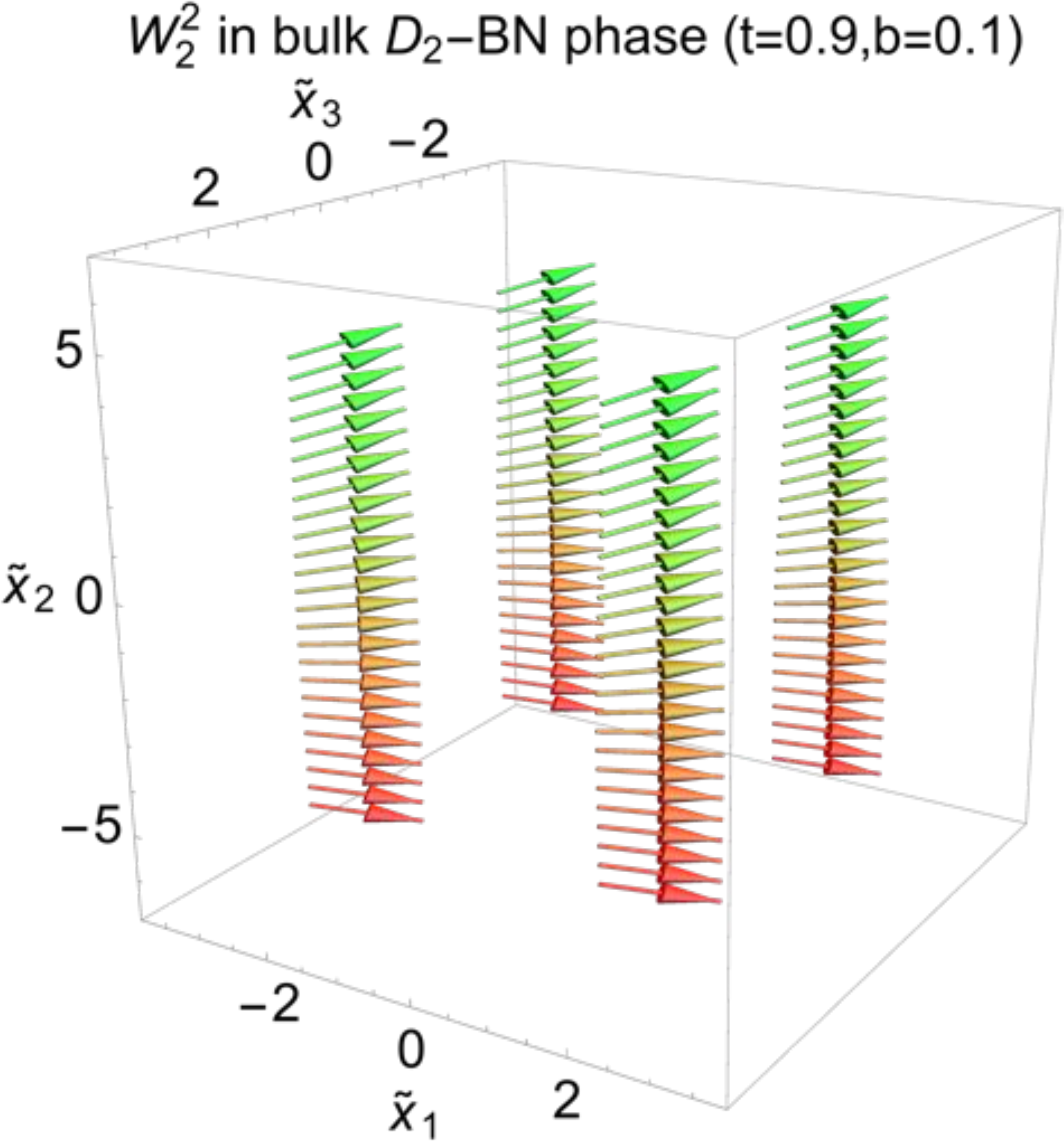}
\hspace{1em}
\includegraphics[scale=0.18]{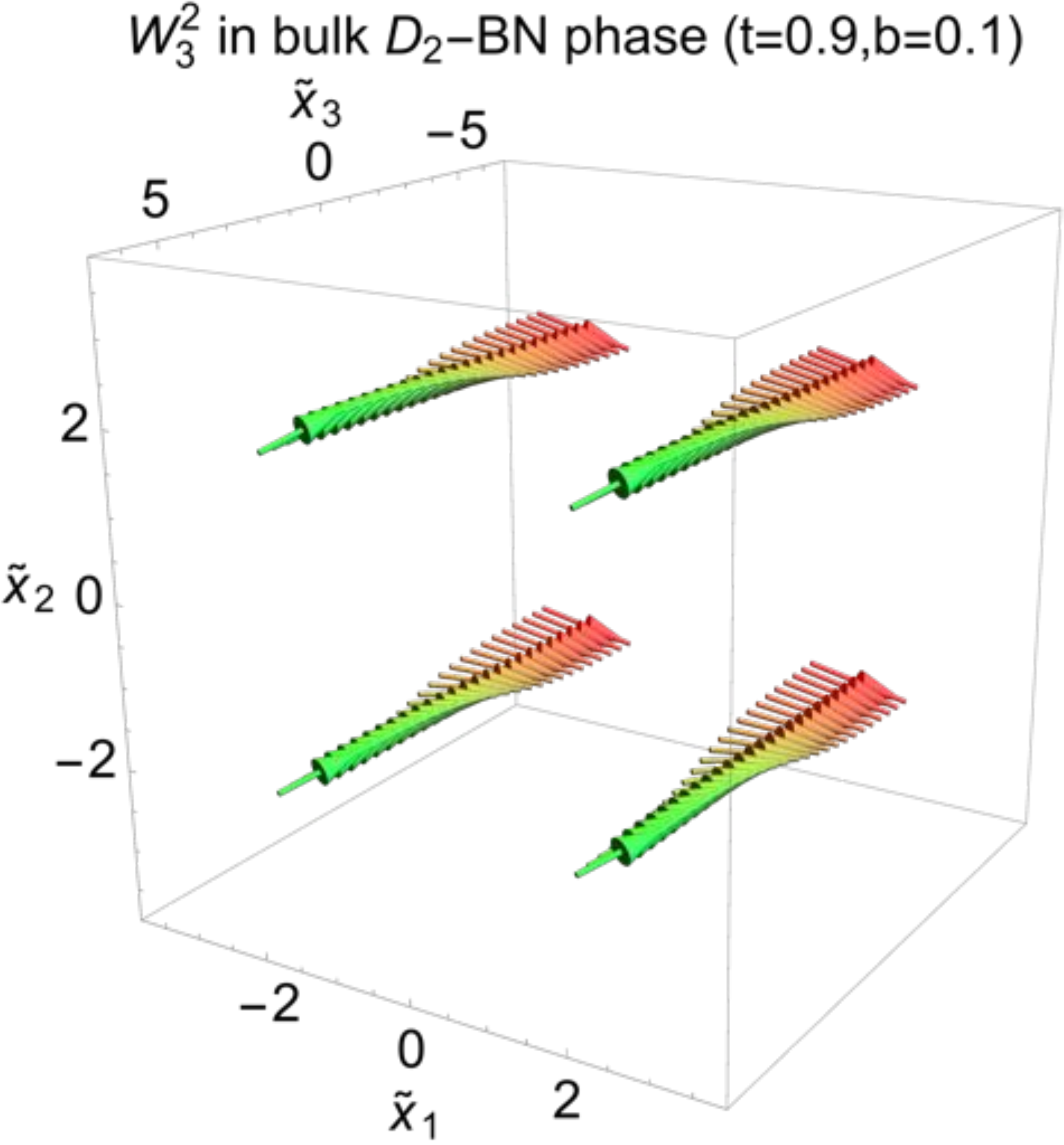}
\\
\vspace{1em}
\includegraphics[scale=0.18]{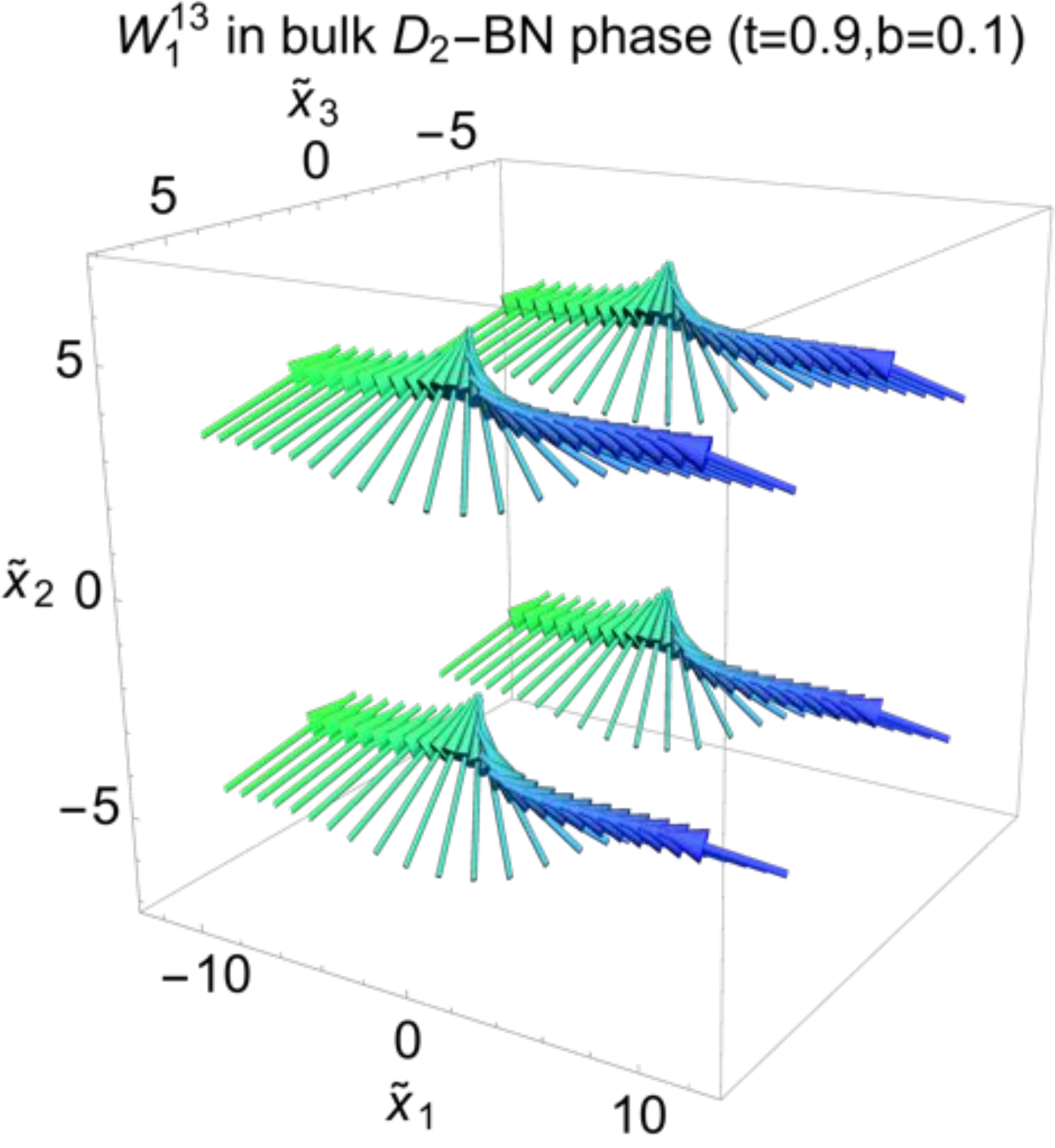}
\hspace{1em}
\includegraphics[scale=0.18]{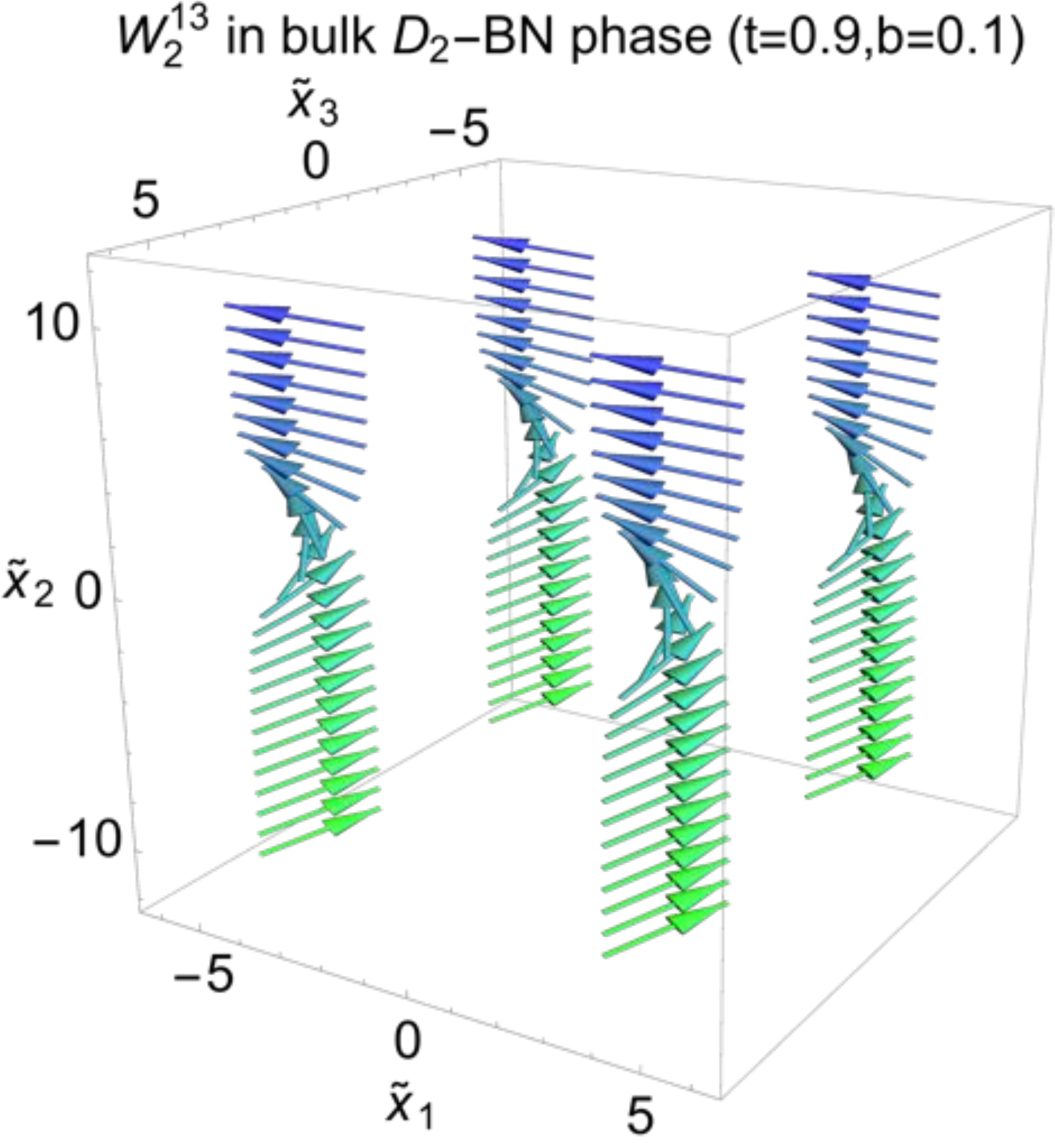}
\hspace{1em}
\includegraphics[scale=0.18]{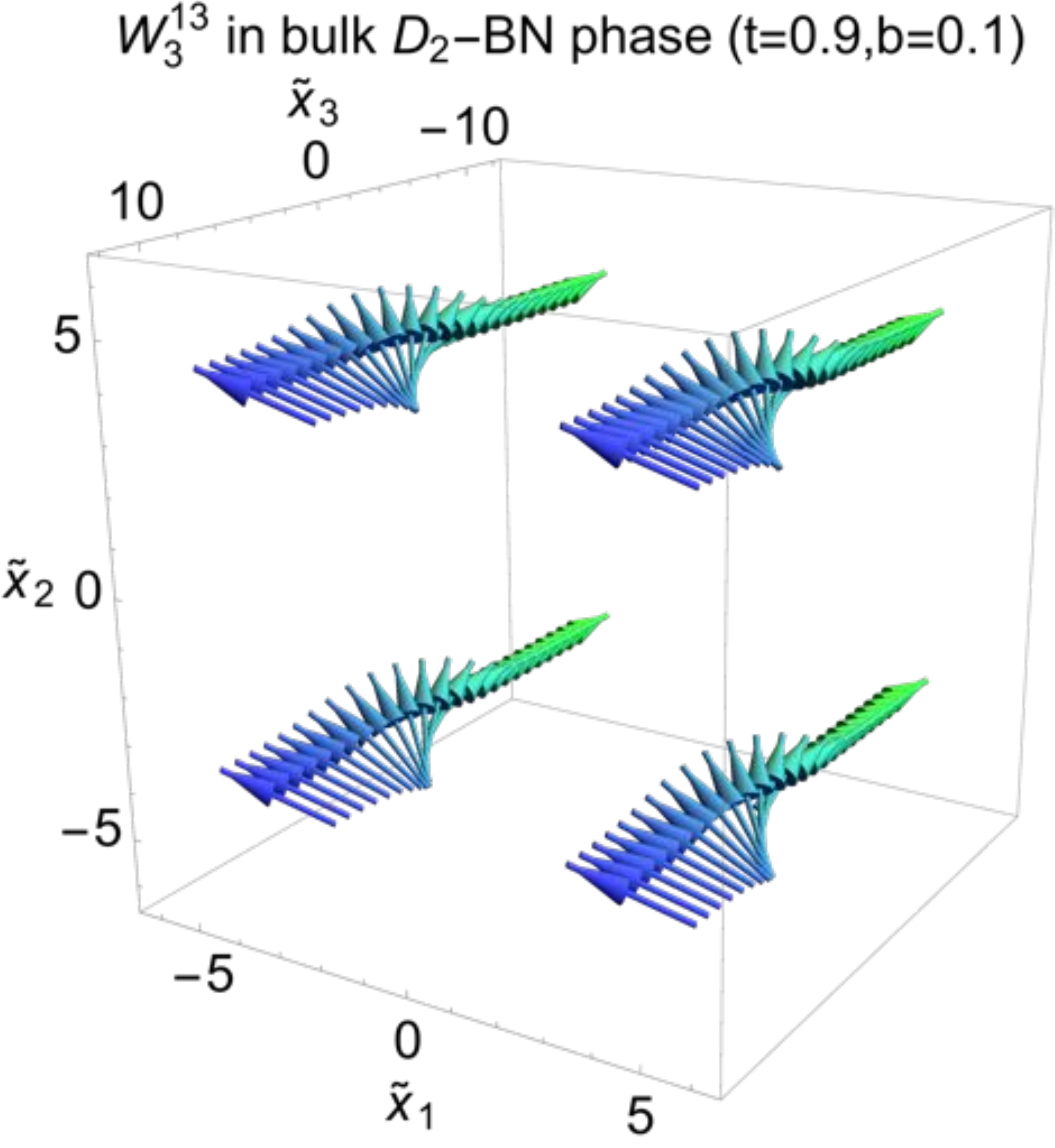}
\caption{The plots of the profile functions as the three-dimensional vectors $\vec{f}(\tilde{d}) = \bigl(f_{1}(\tilde{d};\vec{n}),f_{2}(\tilde{d};\vec{n}),f_{3}(\tilde{d};\vec{n})\bigr)$ with $f_{1}(\tilde{d};\vec{n})+f_{2}(\tilde{d};\vec{n})+f_{3}(\tilde{d};\vec{n})=0$ for the domain walls $W^{2}_{i}$ and $W^{13}_{i}$ ($i=1$, $2$, $3$ the directions along $\tilde{x}_{1}$, $\tilde{x}_{2}$, and $\tilde{x}_{3}$ directions) in the bulk D$_{2}$-BN phase.}
\label{fig:190521_soliton_3P2_f1f2_D2BN}
\end{center}
\end{figure}

\begin{figure}[tb]
\begin{center}
\includegraphics[scale=0.18]{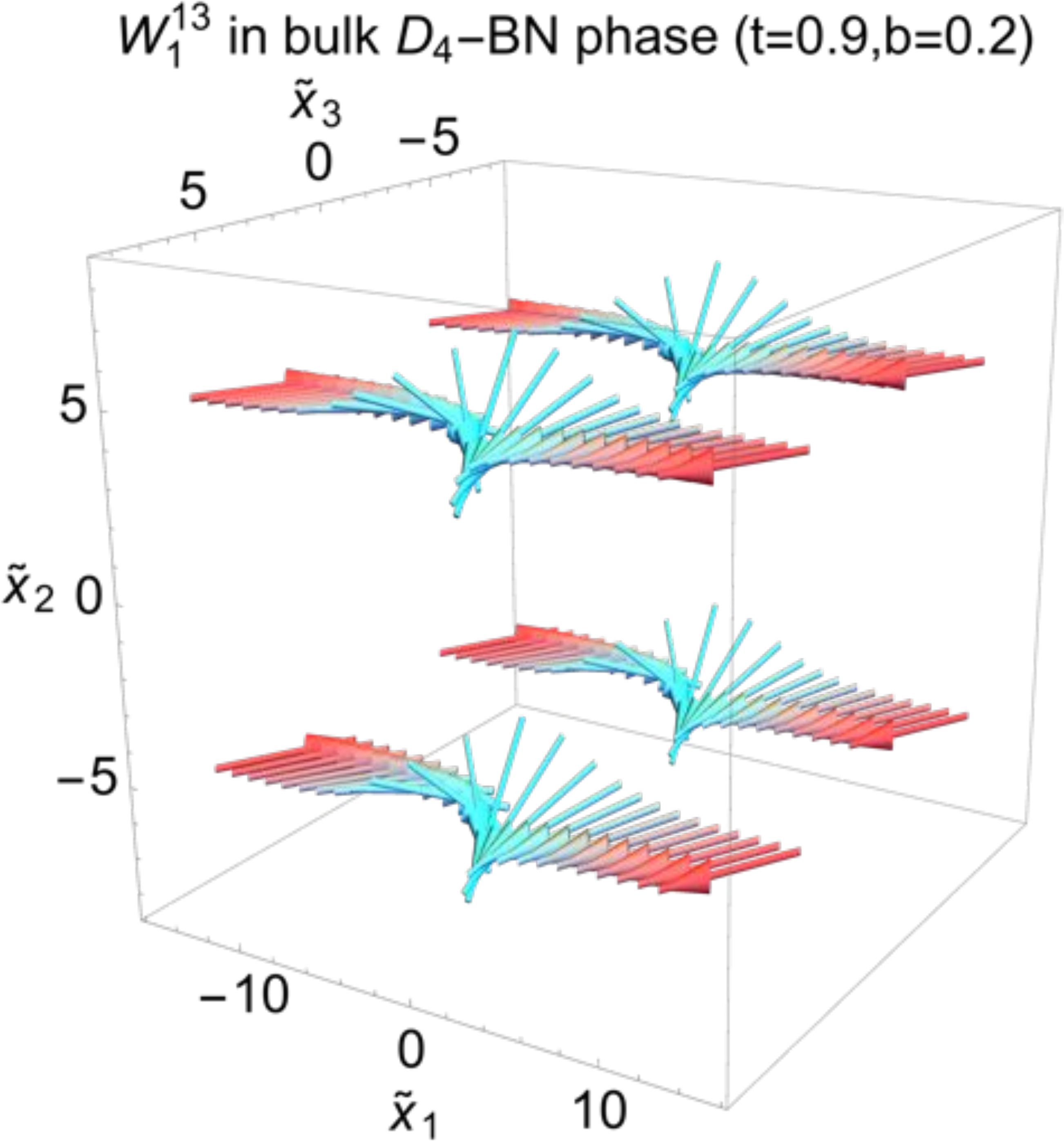}
\hspace{1em}
\includegraphics[scale=0.18]{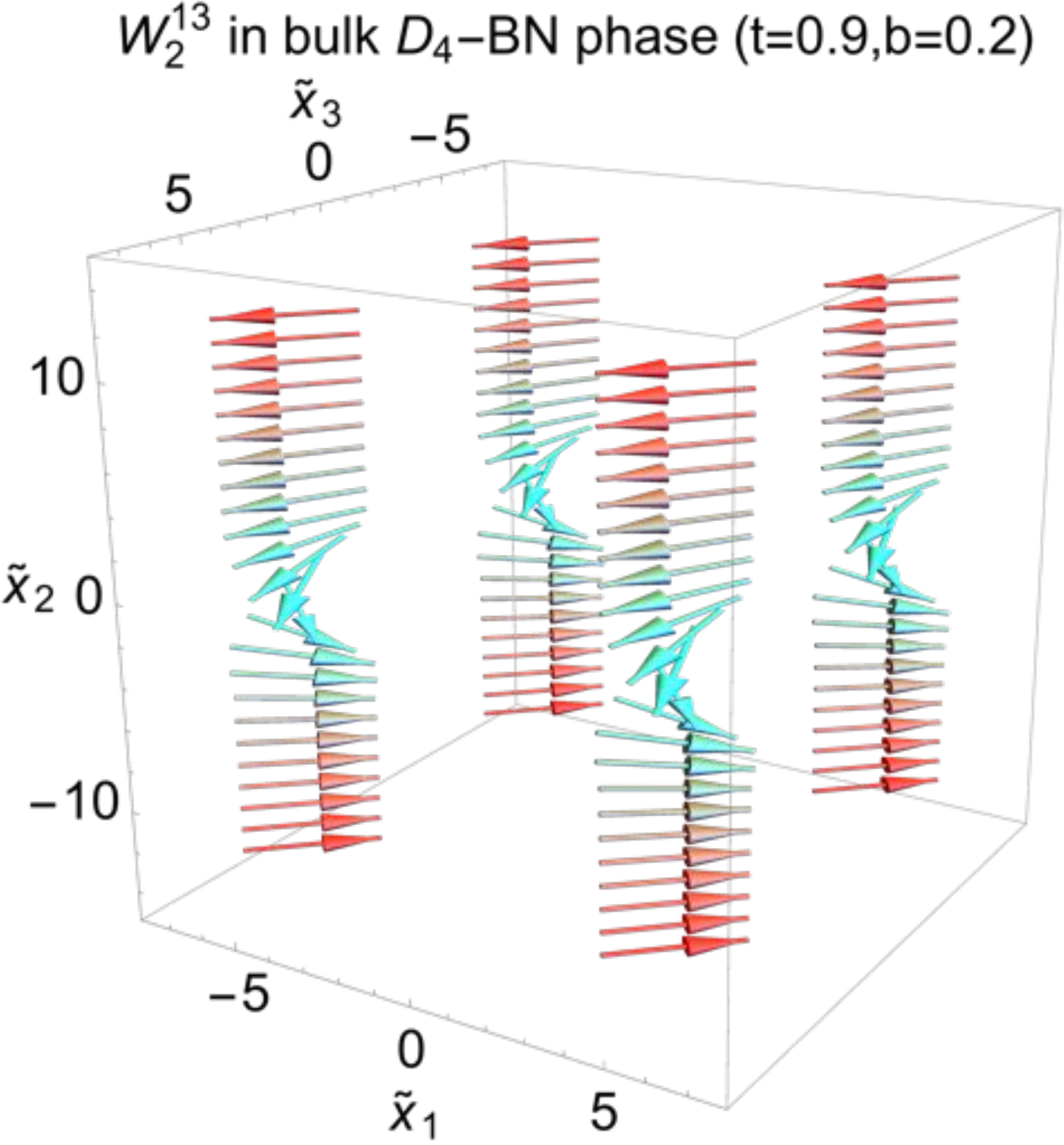}
\hspace{1em}
\includegraphics[scale=0.18]{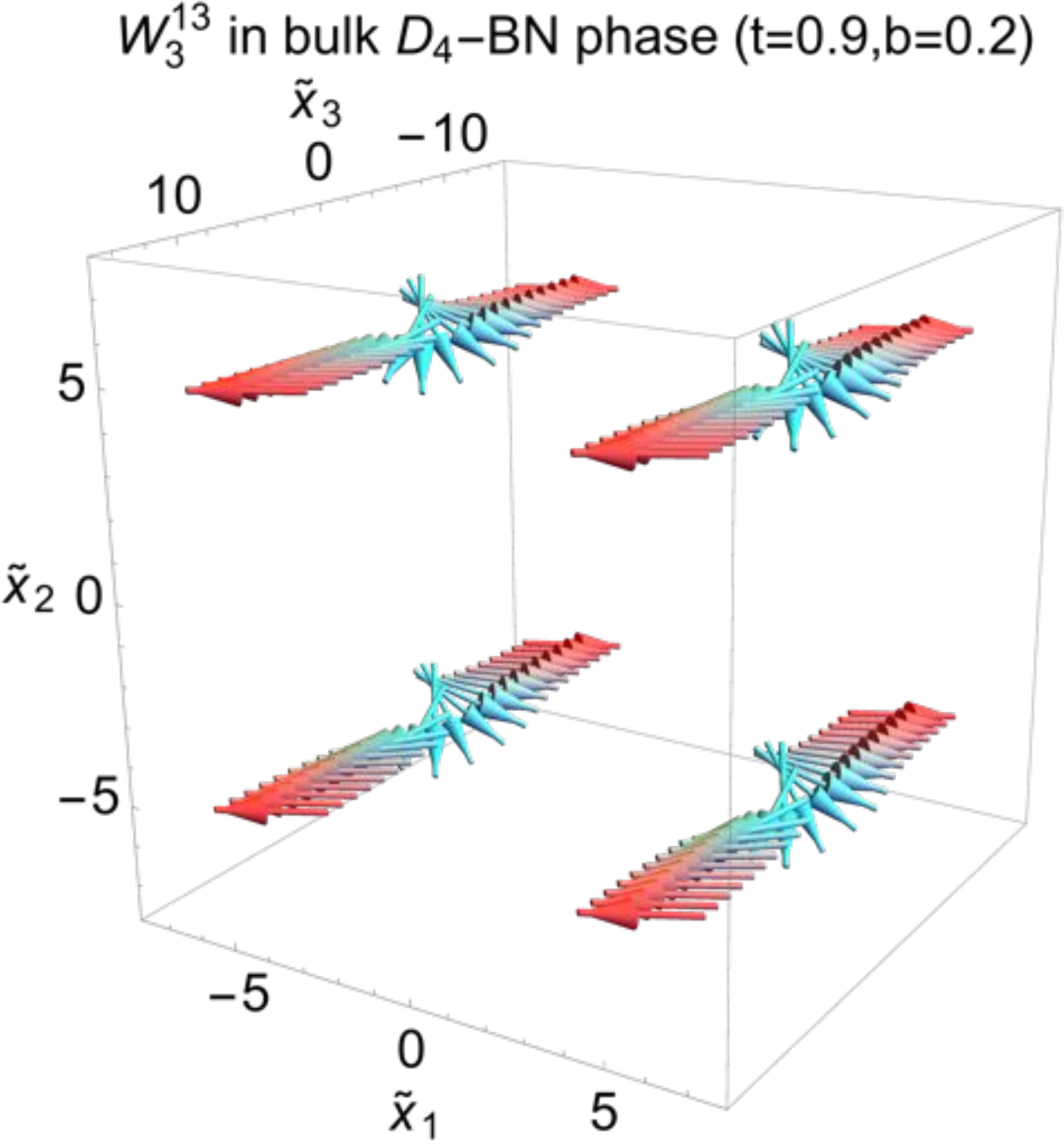}
\caption{The plots of the profile functions as the three-dimensional vectors $\vec{f}(\tilde{d}) = \bigl(f_{1}(\tilde{d};\vec{n}),f_{2}(\tilde{d};\vec{n}),f_{3}(\tilde{d};\vec{n})\bigr)$ with $f_{1}(\tilde{d};\vec{n})+f_{2}(\tilde{d};\vec{n})+f_{3}(\tilde{d};\vec{n})=0$ for the domain walls $W^{13}_{i}$ ($i=1$, $2$, $3$ the directions along $\tilde{x}_{1}$, $\tilde{x}_{2}$, and $\tilde{x}_{3}$ direction) in the bulk D$_{4}$-BN phase.}
\label{fig:190521_soliton_3P2_f1f2_D4BN}
\end{center}
\end{figure}

We present the plots of the three-dimensional vector $\vec{f}(\tilde{d}) = \big(f_{1}(\tilde{d};\vec{n}),f_{2}(\tilde{d};\vec{n}),f_{3}(\tilde{d};\vec{n}))$ in Eq.~\eqref{eq:A_matrix_diagonal}, as functions of $\tilde{d}$ for a fixed $\vec{n}$, in the three-dimensional coordinate $(\tilde{x}_{1},\tilde{x}_{2},\tilde{x}_{3})$ as the solutions of the EL equation \eqref{eq:EL_f} for the domain walls:
$W^{\alpha}_{i}(\mathrm{UN})$ with $\alpha=1, 2, 3$ in Fig.~\ref{fig:190521_soliton_3P2_f1f2_UN}, $W^{\alpha}_{i}(\mathrm{D}_{2}\mathrm{BN})$ with $\alpha=13, 2$ in Fig.~\ref{fig:190521_soliton_3P2_f1f2_D2BN}, and $W^{13}_{i}(\mathrm{D}_{4}\mathrm{BN})$ in Fig.~\ref{fig:190521_soliton_3P2_f1f2_D4BN}.
The subscripts $i=1, 2, 3$ denote the direction of the domain walls, $\tilde{x}_{1}$, $\tilde{x}_{2}$, and $\tilde{x}_{3}$ directions, respectively.

%%%%%%%%%%%%%%%%%%%%%%%%%%%%%%
%%%%%%%%%%%%%%%%%%%%%%%%%%%%%%

%%%%%%%%%%%%%%%%%%%%%%%%%%%%%%
%%%%%%%%%%%%%%%%%%%%%%%%%%%%%%

%%%%%%%%%%%%%%%%%%%%%%%%%%%%%%
%%%%%%%%%%%%%%%%%%%%%%%%%%%%%%

\newpage

\bibliography{neutronstar}

\end{document}